\documentclass[aps,prb,superscriptaddress,twocolumn,showpacs,10pt]{revtex4-2}
\usepackage{amsmath}
\usepackage{amssymb}
\usepackage{graphicx}
\usepackage{array}
\usepackage{dcolumn}
\usepackage{subfigure}
\usepackage{url}
\usepackage{color}
\usepackage{float}
\usepackage{xr-hyper}
\usepackage{multirow}
\usepackage[hidelinks=true]{hyperref}
\hypersetup{
  colorlinks   = true, 
  urlcolor     = blue, 
  linkcolor    = blue, 
  citecolor   = red 
}

\begin{document}
 
\title{Designing Non-Relativistic Spin Splitting in Oxide Perovskites
}

\author{Subhadeep Bandyopadhyay}  
\email{subha.7491@gmail.com}
\affiliation{Consiglio Nazionale delle Ricerche (CNR-SPIN),  Unità di Ricerca presso Terzo di Chieti, c/o Università G. D'Annunzio, I-66100 Chieti, Italy}

\author{Silvia Picozzi}
\affiliation{Consiglio Nazionale delle Ricerche (CNR-SPIN),  Unità di Ricerca presso Terzo di Chieti, c/o Università G. D'Annunzio, I-66100 Chieti, Italy}
\affiliation{Department of Materials Science, University of Milan-Bicocca, Via Roberto Cozzi 55, 20125 Milan, Italy}

\author{Sayantika Bhowal}
\email{sbhowal@iitb.ac.in}
\affiliation{Department of Physics, Indian Institute of Technology Bombay, Mumbai 400076, India}

\begin{abstract}
We investigate the role of atomic distortions in non-relativistic spin splitting in perovskite oxides with $Pbnm$ symmetry. Using LaMnO$_3$ as a representative material, we analyze its non-relativistic spin splitting through a combined phonon and multipolar analysis. Our study provides key insights into how structural distortions and magnetic ordering drive ferroically ordered magnetic multipoles, which, in turn, give rise to non-relativistic spin splitting. Based on these findings, we propose three strategies for engineering non-relativistic spin splitting: modifying the $A$-site cation size, strain engineering, and electric field control in superlattice structures. Our work establishes a framework for designing non-relativistic spin splitting in the Brillouin zone of oxide perovskites.  
\end{abstract}

\maketitle

\section{Introduction}

Recently, the phenomenon of the splitting of spin polarised bands  in the absence of spin-orbit interaction in  compensated antiferromagnetic materials  
has gained significant attention \cite{Hayami2019, Yuan2020, Yuan2021, Smejkal2022PRX, YuanZunger2023, Guo2023, Zeng2024, Lee2024PRL, Krempask2024, Reimers2024, Aoyama2024, Lin2024, Kyo-Hoon2019, Libor2020, Libor2022Review, Paul2024, Bai2024}. The symmetry conditions necessary for realizing such non-relativistic spin splitting (NRSS) have been identified, leading to the theoretical recognition of numerous previously known materials exhibiting NRSS \cite{Yuan2021, Smejkal2022PRX, Guo2023, Zeng2024}. This phenomenon enables the realization of unconventional physical properties by combining the spin splitting typically associated with ferromagnets with the zero magnetization characteristic of conventional antiferromagnets. These properties include efficient spin-current generation \cite{Naka2019, Hernandez2021, Shao2021, Bose2022, Hu2024}, spin-splitting torque \cite{Bai2022, Karube2021}, giant magnetoresistance \cite{Libor2022}, and a spontaneous Hall effect \cite{Libor2020, Libor2022NatRev, Helena2020, Feng2020, Betancourt2021, SmejkalAHE2022, Cheong2024, Sato2024} in the absence of net magnetization, as well as unconventional superconducting properties \cite{Mazin2022, Zhu2023, Banerjee2024, Chakraborty2024, Zhang2024, Lee2024}, enhanced thermal transport \cite{Zhou2024, Yershov2024}, chiral magnons \cite{Libor2023, McClarty2024, Liu2024, Morano2024, Bandyopadhyay2024}, and the piezomagnetic effect \cite{BhowalSpaldin2024, Radaelli2024}.  

Initially explored through theoretical studies, NRSS has recently been confirmed experimentally via angle-resolved photoemission spectroscopy (ARPES) measurements in several candidate materials \cite{Lee2024, Dale2024, Reimers2024, Yang2025}. Consequently, controlling NRSS has emerged as a growing research direction \cite{Smejkal2024arxiv, Gu2025, Bandyopadhyay2024, Duan2025}. In this context, the development of multipolar theory, which correlates the magnitude of NRSS with that of a specific ferroic magnetic multipole, offers a crucial framework for tuning NRSS. This approach enables the control of NRSS by manipulating a particular ferroic magnetic multipole \cite{BhowalSpaldin2024, Verbeek2024, Nag2024}. Interestingly, these ferroic magnetic multipoles emerge from a combination of antiferroic charge multipoles, and antiferroic magnetic dipoles with matching spatial patterns. This intrinsic relationship suggests that the magnitude of the ferroic magnetic multipole—and thus NRSS—is strongly influenced by structural distortions, which are quantified by antiferroic charge multipoles. Consequently, this insight provides an opportunity for the controlled design and optimization of NRSS in real materials.  

In this work, we integrate a phonon-assisted approach with the multipolar framework to reveal the role of distinct atomic distortion modes and magnetic dipolar order in driving NRSS. By establishing the contribution of specific distortion modes, we propose strategies for designing and controlling NRSS, which we further validate through first-principles calculations. To illustrate our approach, we focus on LaMnO$_3$ (LMO), a representative perovskite material, known to exhibit NRSS \cite{Yuan2021}. 
Perovskite oxides, with the general chemical formula $ABO_3$, constitute a vast family of materials exhibiting a diverse range of fascinating physical properties, including ferroelectricity, magnetoelectric coupling, superconductivity, metal-insulator transitions, anomalous Hall effects, and photovoltaic effects \cite{ABO3_Ghosez_ferroelectricity,ABO3_Eric_magnetoelectric,ABO3_Takagi_superconductivity,ABO3_Tokura_MIT,naka_ABO3,ABO3_RSC_photovolatics}. Our choice of perovskite oxides for designing NRSS is motivated by their versatility, which enables precise engineering and manipulation through epitaxial strain, chemical doping, or layer thickness control.

  GdFeO$_3$-type $Pbnm$ phase is  common  among the oxide Perovskite family.
 Many $Pbnm$ perovskites, including LMO, have recently been identified as hosting NRSS in their electronic band structures \cite{Smejkal2022, YuanZunger2023, naka_ABO3, Rooj_ganguly_Pnma}. LMO has long been of interest due to its metal-to-insulator transition occurring below 750 K \cite{LMO_TM}, which arises from a Jahn-Teller (JT) distortion that stabilizes a $C$-type orbital ordering of the Mn $d$ states \cite{LMnO_OO}. Notably, this transition does not break any symmetries, as the material retains its orthorhombic $Pbnm$ structure across all temperatures. Upon further cooling below 140 K, LMO undergoes an antiferromagnetic (AFM)  transition, stabilizing an $A$-type AFM spin alignment in its ground state \cite{LMO_TN}. Our study focuses on this magnetic ground state of LMO and presents a method to control the NRSS within it.

The remainder of the manuscript is structured as follows. Section \ref{method} provides details of our calculations, with a particular focus on the multipole analysis. In Section \ref{results}, we present and discuss our findings, beginning with the magnetic ground state of LMO. We then investigate the influence of magnetic ordering by considering alternative magnetic configurations, commonly observed in perovskites, viz., $C$- and $G$-type AFM order. Subsequently, we analyze the impact of various atomic distortions in the $Pbnm$ structure on the NRSS of LMO. Based on these insights, we propose three strategies for controlling NRSS: tuning the A-site cation radius, applying strain engineering, and utilizing superlattice structures to manipulate ferroelectric domains via an external electric field. Finally, in Section \ref{summary}, we summarize our findings and outline potential directions for future research.

\section{Computational details} \label{method}
Electronic structure calculations are carried out
using the plane wave-based projector augmented wave (PAW) \cite{Bloch1994, Kresse1999} method as implemented in the Vienna ab initio simulation package (VASP) \cite{Kresse1993, Kresse1996}, relying on the PBESol exchange correlation functional. On site Coulomb interaction is introduced by including Hubbard interaction $U$ and Hund's coupling $J$ to the Mn-$3d$
orbital \cite{Liechtenstein_U}. We use $U=$ 5 eV and $J=$ 1 eV in our calculations \cite{LaMnO3_marcus}. All the calculations, presented in the work, are performed in the absence of spin-orbit interaction. For electronic convergence, we use an energy cutoff of 520 eV and $12\times12\times8$ $k$ mesh for the sampling of the Brilliouin zone (BZ). Structural optimizations are carried out until the Hellman-Feynman forces on each atom become less than 0.005 eV/\AA. Atomic distortions have been identified employing symmetry mode analysis, as implemented in ISODISTORT software \cite{Isodistort,Isodistort1}.

\subsection{Multipole analysis}

To identify the existence of NRSS of a given structure and the momentum direction in reciprocal space along which the NRSS occurs, we adopted the framework discussed in Ref. \cite{BhowalSpaldin2024} and performed a multipole analysis. Before detailing the methodology used in this work to compute higher-order magnetic multipoles, we first provide an introduction to these concepts.  

The magnetic interaction energy, $\cal{E}^{\rm int}_{\rm mag}$, for a system with a magnetization density $\vec{\mu}(\Vec{r})$ in the presence of an applied magnetic field $\vec{H}(\vec{r})$, is given by:  

\begin{widetext}
\begin{align}\nonumber \label{int}
    \cal{E}^{\rm int}_{\rm mag} = & -\underbrace{\Big( \int \mu_{i} \,d\Vec{r} \Big)}_\text{magnetic dipole} H_i({\vec{r}=0}) \thickspace - \thickspace
    \underbrace{\Big( \int \mu_{i} r_{j} \,d\Vec{r} \Big)}_{\substack{\text{magnetic quadrupole} \\ \text{moment}}}
    \partial_{j} H_{i}({\vec{r}=0}) 
    \thickspace - \thickspace \underbrace{\Big( \int \mu_{i} r_{j} r_{k} \,d\Vec{r} \Big)}_\text{magnetic octupole}  \partial_{j} \partial_{k} H_{i}({\vec{r}=0}) \\ 
    & \thickspace - \thickspace {\underbrace{\Big( \int \mu_{i} r_{j} r_{k} r_{l} \,d\Vec{r} \Big)}_\text{magnetic hexadecapole}  \partial_{j} \partial_{k} \partial_{l} H_{i}({\vec{r}=0}) \thickspace - \thickspace \underbrace{\Big( \int \mu_{i} r_{j} r_{k} r_{l} r_{m} \,d\Vec{r} \Big)}_\text{magnetic triakontadipole}  \partial_{j} \partial_{k} \partial_{l} \partial_{m} H_{i}({\vec{r}=0})} \thickspace - \thickspace ... ,
\end{align}
\end{widetext}

The magnetic dipole, magnetic quadrupole, magnetic octupole, magnetic hexadecapole, and magnetic triakontadipole represent the magnetic multipoles of different orders, characterizing the asymmetries and anisotropies in the magnetization density. Their ranks range from 1 to 5, respectively. In the Taylor series expansion of Eq. (\ref{int}), we have explicitly shown magnetic multipoles up to rank 5, as higher-order multipoles are not relevant to the discussion of the present work.
It is important to point out that the magnetic quadrupole moments in Eq. (\ref{int}) are also referred to as magnetoelectric multipoles, as they are related to the linear magnetoelectric effect \cite{Spaldin2013}.

We compute the angular part of these different magnetic multipoles around a sphere at every Mn atomic site. For this, we first compute the density matrix $\rho_{lm,l'm'}$ within the density functional theory (DFT) method as implemented in VASP using the noncollinear setting, followed by its decomposition into irreducible (IR) spherical tensor components $w^{k p r}_t$ \cite{Cricchio2010, Spaldin2013}. 
Here, $k$, $p$, and $r$ represent the spatial index, spin index, and the tensor rank respectively, while $t$ labels the tensor components. The $p = 0, 1$ represent respectively the charge and the magnetic multipoles; 
rank $r \in { | k - p |, | k - p | + 1, \dots, k + p }$, 
and $t \in { -r, -r + 1, \dots, r}$.
For example, the atomic-site magnetic octupoles correspond to $k=2$, $p=1$, and $r=1, 2, 3$, representing the non-symmetric ($r=1$ with 3 components, also known as moment of toroidal moment $t_i^{(\tau)}$ and $r=2$ with 5 components, known as toroidal quadrupole moment ${\cal Q}^{(\tau)}_{ij}$) and the totally symmetric ($r=3$ with 7 components) components of the magnetic octupoles ${\cal O}_{3m}$, where $m= -3,...+3$ represent the seven components \cite{Urru2022}. 
$M$ and $R$-type structural distortionsSimilarly, magnetic triakontadipoles are defined by $k=4$, $p=1$, and $r=3, 4, 5$. Throughout the manuscript, we will use the $kpr$ notation to represent the multipole components. Since, both magnetic octupoles and triakontadipoles are inversion symmetric, only $l+l'=$ even terms in the density matrix $\rho_{lm,l'm'}$ contribute to these multipoles.

\section{Results and discussions} \label{results}

We now employ the multipolar framework, as discussed above, to analyze the NRSS in our example perovskite oxide, LMO. Using this approach, we first examine the influence of different magnetic dipolar orders and structural distortions, characterized by phonon modes. Building on this understanding, we propose strategies to tune NRSS in perovskite oxide systems, with LMO as our model material.

\begin{figure}[t]
\centering
\includegraphics[width=\columnwidth]{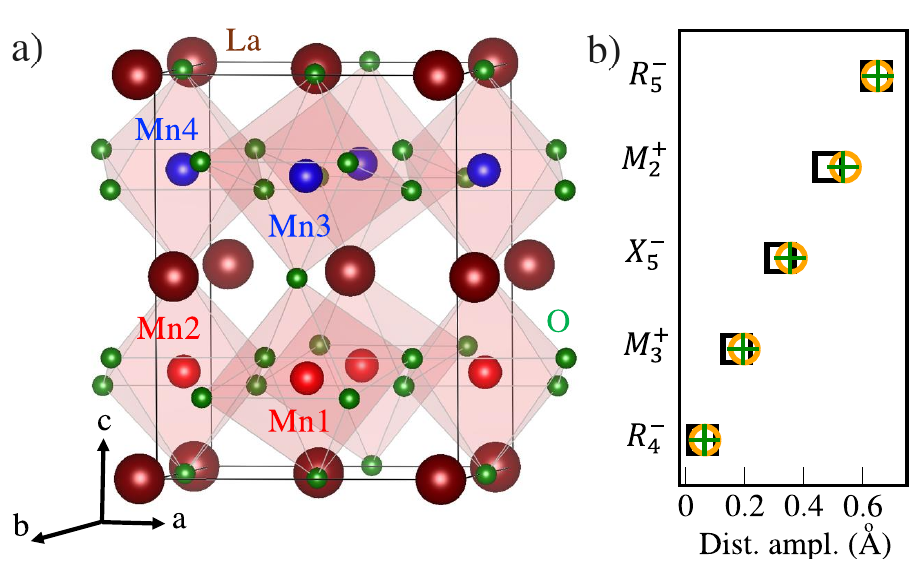}
\caption{(a) $Pbnm$ crystal structure of LMO with $A$-type AFM ordering. Mn atoms with up/down spin are shown with red/blue colors.  
(b) Amplitude of the $Pbnm$ distortions relative to the cubic $Pm\overline3m$ structure. Black open sqaure represents  distortions of optimized structures, whereas, open orange circle and green + represent distortions of the experimental  $Pbnm$ structures, reported in Refs. \cite{LMO_dist1} and \cite{LMO_dist2} respectively. }
\label{Gs_modes}
\end{figure}

\subsection{Effect of different magnetic dipolar ordering in the NRSS}\label{magorder}

We begin by discussing the ground state of LMO.
 LMO crystallizes in the orthorhombic $Pbnm$ structure below 140 K and exhibits an A-type AFM ordering \cite{LMO_dist1} as shown in Fig. \ref{Gs_modes}a. In the A-type AFM ordering, the Mn spins within the plane are ferromagnetically aligned while interacting antiferromagnetically along the out-of-plane direction (see Fig. \ref{Gs_modes}a). 
We carry out the electronic structure calculations for the optimized $Pbnm$ structure of LMO with the $A$-type AFM order. Our calculations show that LMO is an insulator with a band gap of 1.12 eV and a magnetic moment of 3.8 $\mu_B$/Mn as expected for Mn$^{3+} (3d^4)$ ions.

\begin{figure}[t]
\centering
\includegraphics[width=\columnwidth]{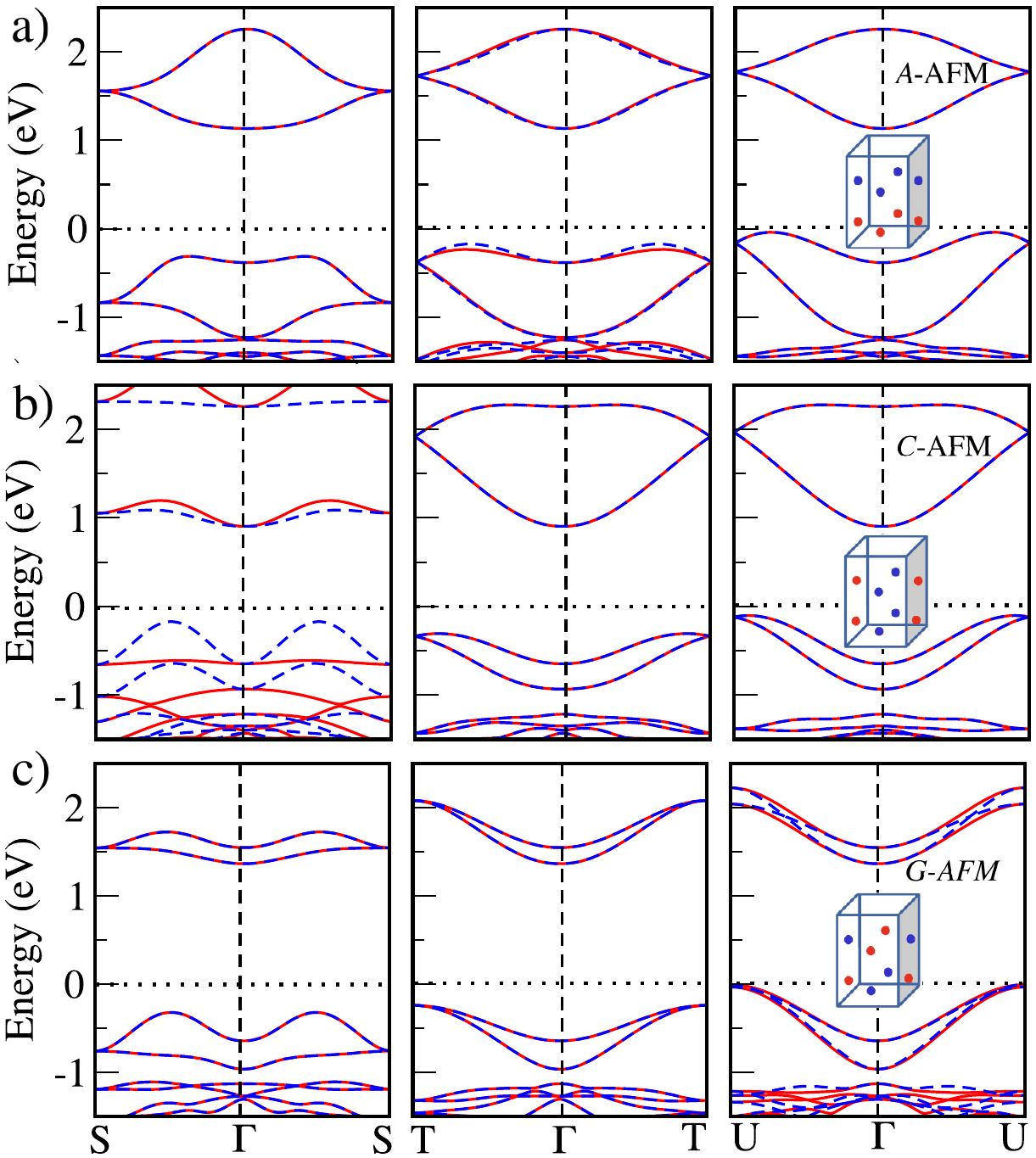}
\caption{ Electronic bands of LMO with the  $Pbnm$ structure for (a) A-, (b) C-,  and (c) G-type AFM configurations. The insets depict the corresponding magnetic order with  red and blue balls denoting the up and down Mn spins.  Bands are plotted along $S (-0.5,-0.5,~0) \rightarrow\Gamma (0,0,0)\rightarrow S (0.5,0.5,0)$ (left panel), $T (0,-0.5,-0.5)\rightarrow\Gamma (0,0,0)\rightarrow T (0, 0.5, 0.5)$ (middle panel), and $U (-0.5, 0,-0.5)\rightarrow\Gamma (0, 0, 0)\rightarrow U (0.5, 0, 0.5)$ (right panel) high symmetry $k$ path. Solid red and dashed blue lines denote the up and down spin-polarized bands respectively. Zero in the energy axis represents the Fermi energy of the system.}
\label{bands}
\end{figure}

In order to see if the NRSS is present in LMO, we further perform the multipolar calculations in the absence of spin-orbit coupling for the magnetic ordering shown in Fig. \ref{Gs_modes}a with the spin polarization along $\hat z$ to analyze the charge and magnetic multipoles at the Mn sites. Our calculations show that there are some magnetic octupole components present at the Mn sites that have ferroic ordering. These components are $w^{213}_{-1}$, $w^{212}_{1}$, and $w^{211}_{-1}$ (see config. 7 in Table \ref{tab1}), 
that describe an anisotropic magnetization density on the $xy$ plane. We find that these octupoles form the lowest-order ferroic magnetic multipole in LMO. 

In addition to the magnetic multipoles, we also analyze the charge multipoles in the system. Our analysis shows the presence of antiferroically ordered charge quadrupole components, $w^{202}_{-2}$, $w^{202}_{-1}$, and $w^{202}_{1}$ (see config. 7 in Table \ref{tab1}), that characterize the locally anisotropic charge density at the Mn ions on the $xy$, $yz$, and $xz$ planes respectively. The antiferroic pattern of the charge quadrupole moments is listed in Table \ref{tab1}. We note that out of these three charge quadrupole moments, the antiferroic pattern of the $w^{202}_{-1}$ quadrupole moment is the same as that of the antiferroic pattern of the magnetic dipole moment ($w^{011}_{0}$) for the A-type AFM order, which in turn leads to a ferroic ordering of $w^{213}_{-1}$ and other magnetic octupole components.

We next analyze the $k$-space representation of the existing ferroically ordered magnetic octupole components to determine the momentum direction of NRSS \cite{BhowalSpaldin2024}. The $k$-space representation of the existing magnetic octupoles in this case is $k_yk_zm_z$ \cite{BhowalSpaldin2024,Urru2022}. This suggests the presence of a splitting between spin-polarized bands for any direction in the BZ with $k_y \ne 0$ and $k_z \ne 0$. To verify this, we analyze the band structure of LMO in the absence of spin-orbit coupling. The computed band structure, as shown in Fig. \ref{bands}a, indeed depicts a splitting along the high-symmetry $\Gamma-T$ path, while it is absent along $\Gamma-S$ and $\Gamma-U$, consistent with the $k$-space representation.

In order to investigate if the other existing antiferroic charge quadrupole components,  $w^{202}_{-2}$ and $w^{202}_{1}$, could also lead to ferroically ordered magnetic octupoles for different antiferroic patterns of magnetic dipoles, we further investigate the multipoles and the band structures for the 
C- and G-type collinear AFM ordering with spin polarization along $\hat z$.  
Unlike the A-type AFM ordering, discussed previously, the C- and G-type AFM ordering depicts an AFM alignment of the in-plane Mn spins, whereas the out-of-plane Mn spins are ferromagnetically and antiferromagnetically aligned, respectively (see the insets of Fig. \ref{bands} b and c). For this analysis, we consider the same LMO structure but with different magnetic ordering. Although, in LMO the C- and G-type collinear AFM configurations are respectively 45  and 49 meV/f.u. higher in energy compared to the ground state A-type AFM configuration, there are many other $Pbnm$ perovskite materials with a C- or G-type AFM ground state \cite{CMO_matar, LFO_wood, NdFeO_2005, Subhadeep_BiNiO3, CaCrO3_Khomskii, AVO_1, AVO_2}.

Interestingly, we find that the magnetic dipolar ordering in the C-type AFM configuration has the same antiferroic pattern as the charge quadrupole component $w^{202}_{-2}$ (see Table \ref{tab1}). This consequently leads to ferroically ordered magnetic octupole components $w^{213}_{-2}$ and $w^{212}_{2}$. While the existing charge quadrupoles characterize an anisotropic charge density on the $xy$ plane, the magnetic octupoles describe the anisotropic magnetization density on the same plane. The $k$-space representation of the existing magnetic octupole components is $k_xk_ym_z$ and accordingly, we find a spin splitting along the $\Gamma-S$ direction (see Fig.~\ref{bands}b). 

Similarly, the anti-ferromagnetic dipolar ordering in the G-type AFM has the same pattern as that for the charge quadrupole component $w^{202}_{1}$, leading to ferroically ordered magnetic octupole components $w^{213}_{1}$, $w^{212}_{-1}$ and $w^{211}_{1}$ (see Table~\ref{tab1}). Once again, consistent with the 
$k$-space representation $k_xk_zm_z$ of the existing magnetic octupole components, the non-relativistic spin splitting occurs along the $\Gamma-U$ high symmetry $k$ path (see Fig.~\ref{bands}c).

Interestingly, the C-type AFM order exhibits the largest spin splitting among these three magnetic configurations, as evident, for example, from the comparison of the spin splitting for the topmost valence bands, shown in Fig. \ref{bands}, even though the crystal structure is identical for the three magnetic configurations. We attribute this behavior to the computed largest ferroic magnetic octupole components in the C-type AFM order among the three considered magnetic configurations (e.g., $w^{213}_{-2} = -0.49 \mu_{\rm B}$ in C-type AFM vs. $w^{213}_{-1} = -0.15 \mu_{\rm B}$ in A-type AFM and $w^{213}_1 = 0.03\mu_{\rm B}$ in G-type AFM). Even though the crystal structure remains the same for all three magnetic configurations, different AFM magnetic dipolar ordering couples to different existing charge quadrupole moments of the structure. Interestingly, the charge quadrupole moment $w^{202}_{-2}$ is the largest in magnitude among the existing antiferroic charge quadrupole moments, which, however, only is in action for the C-type AFM order. This consequently leads to the largest ferroic magnetic octupole component and, hence, the largest spin splitting in the C-type AFM order.   

Our multipole analysis, described so far, highlights the combined roles of the structural distortion, characterized by the charge quadrupole moments, and the AFM ordering in dictating the ferroically ordered magnetic octupole component and hence the spin splitting. While so far we discussed the role of the AFM ordering in driving the NRSS along different directions in the reciprocal space, in the following we investigate the role of atomic distortion by explicitly analyzing different phonon modes in LMO. 

\begin{table*}[t]
\caption{ Different atomic distortions of the $Pbnm$ structure, the corresponding atomic-site antiferroic (AF) charge multipoles, and the lowest ferroically (F) ordered magnetic multipoles at the Mn atoms for various antiferromagnetic (AFM) orderings, along with the resulting non-relativistic spin splitting (NRSS). The `+' and `-' signs denote the relative orientation of a particular multipole on the Mn sites. Mn1–Mn4 atoms are labeled in Fig. \ref{Gs_modes}. The spin magnetic dipole moments are oriented along $\hat z$ ($w^{011}_0$). The presence and absence of NRSS are indicated in the last column as `Yes' and `No', respectively. Configurations 1–6 are considered within a  $\sqrt2\times\sqrt2\times2$ supercell of the cubic structure, while lattice parameters are optimized for the $Pbnm$ structure in configuration 7.}
\centering
\resizebox{\linewidth}{!}
{\begin{tabular}{|c|c|c|c|c|c|c|c|c|}
    \hline
     Config. &Distortion& Space gr. &\multicolumn{2}{|c|}{AF charge Multipole} &  \multicolumn{2}{|c|}{Spin magnetic dipole ($w^{011}_0$)} & Lowest F-ordered & NRSS \\
      &     &   & Multipole ($w^{kpr}_t$) & AF pattern & Order & AF pattern & magnetic multipole & \\
    \hline\hline
      &  & &  &  & A-AFM &  + + - -   & --  & No \\ 
       &  & &  &  &  &        & & \\
     1 &  $R_5^-(\phi_{xy}^-)$ & $Imma$ & Quadrupole ($w^{202}_1$) & + - - + & C-AFM & + - + -    & --  & No \\
       &  & &  &  &  &        &  &\\
      &   & &  &  & G-AFM & + - - +    & Octupole - $w^{213}_1$,  & Yes\\
      &   & &  &  & &     &  $w^{212}_{-1}, w^{212}_{2}, w^{211}_{1}$. & \\ 
    \hline
    &  & &  &  & A-AFM &   + + - -   & --  & No \\
     &  & &  &  &  &        & & \\
   2 & $M_2^+(\phi_{z}^+)$ &  $P4/mbm$ &Hexadecapole ($w^{404}_{-4}$)  & + - + -& C-AFM & + - + - & Triakontadipole & Yes \\
      &  & &  &  &  &     & $w^{415}_{-4}, w^{414}_{4}$    & \\
    &  & &  &  & G-AFM & + - - +    & --  & No \\ 
    &  & &  &  &  &        & & \\
     \hline
     &  & &  &  & A-AFM &  + + - -    & Octupole-$w^{213}_{-1}$  & Yes\\
     &  & &  &  &  &     & $w^{212}_{1}, w^{211}_{-1}$  & \\
    3  & $X_5^-$ &  $Cmcm$  & Quadrupole ($w^{202}_1$) & - - + + &C-AFM & + - + - & -- & No \\
      &  & &  &  &  &        & &\\
    &  & &  &  & G-AFM & + - - +    & --  & No \\ 
    &  & &  &  &  &        & & \\
     \hline
     &  & &  &  & A-AFM &  + + - -    & --  & No\\
       &  & &  &  &  &     &   & \\
     4 & $M_3^+$ &  $P4/mbm$ &   Quadrupole ($w^{202}_{-2}$) & + - + - &C-AFM & + - + - & Octupole  & Yes  \\
       &  & &  &  &  &    & $w^{213}_{-2}, w^{212}_{2}$  &  \\
       &  & &  &  & G-AFM &  + - - +   & --  & No \\ 
       &  & &  &  &  &        & & \\
      \hline 
      &  & &  &  & A-AFM &  + + - -    & --  & No\\
       &  & &  &  &  &        & &\\
      5 & $R_4^-$ &  $Imma$ &  Quadrupole ($w^{202}_{1}$) &  - + + - & C-AFM &  + - + - & -- & No \\
      &  & &  &  &  &     &   & \\
    &  & &  &  & G-AFM &  + - - +   & Octupole- $w^{213}_{1}$ & Yes \\ 
    &  & &  &  &  & & $w^{212}_{1}, w^{211}_{-1}$ & \\
     \hline 
     &  & &Quadrupole $w^{202}_{-1}$  & + + - -  & A-AFM &  + + - - & Octupole- $w^{213}_{-1}$    &   Yes\\
       &  & &   &  & & & $w^{212}_{1}, w^{211}_{-1}$ &        \\
    6 & $R_5^-+M_2^+$ &  $Pbnm$  & $w^{202}_{-2}$  & + - + - & C-AFM & + - + - & Octupole- $w^{213}_{-2}$& Yes \\
     &  & &  & &  &     & $w^{212}_{2}$  & \\
    &  & & $w^{202}_{1}$ &  + - - +  & G-AFM & + - - +    & Octupole- $w^{213}_{1}$ & Yes \\
    &  & &  & &  &       & $w^{212}_{-1}, w^{211}_{1}$  & \\
     \hline 
     &  & &Quadrupole $w^{202}_{-1}$  & + + - -  & A-AFM &  + + - - & Octupole- $w^{213}_{-1}$    &   Yes\\
         & & & &   &  & &  $w^{212}_{1}, w^{211}_{-1}$ &   \\
    7 & $R_5^-$+$M_2^+$ &  $Pbnm$   & $w^{202}_{-2}$  & + - + - & C-AFM & + - + - & Octupole- $w^{213}_{-2}$& Yes \\
     & +$X_5^-$+$R_4^-$ & &  & &  &     & $w^{212}_{2}$  & \\
    & $M_3^+$ & & $w^{202}_{1}$ &  + - - +  & G-AFM & + - - +    & Octupole- $w^{213}_{1}$ & Yes \\
 &  & &  & &  &     &  $w^{212}_{-1}$, $w^{211}_{1}$ & \\
\hline \hline 
\end{tabular}}
\label{tab1}
\end{table*}

\subsection{Atomic distortions of the {\it Pbnm} phase and their role in NRSS} \label{dist}

\begin{figure*}[t]
\centering
\includegraphics[width=\linewidth]{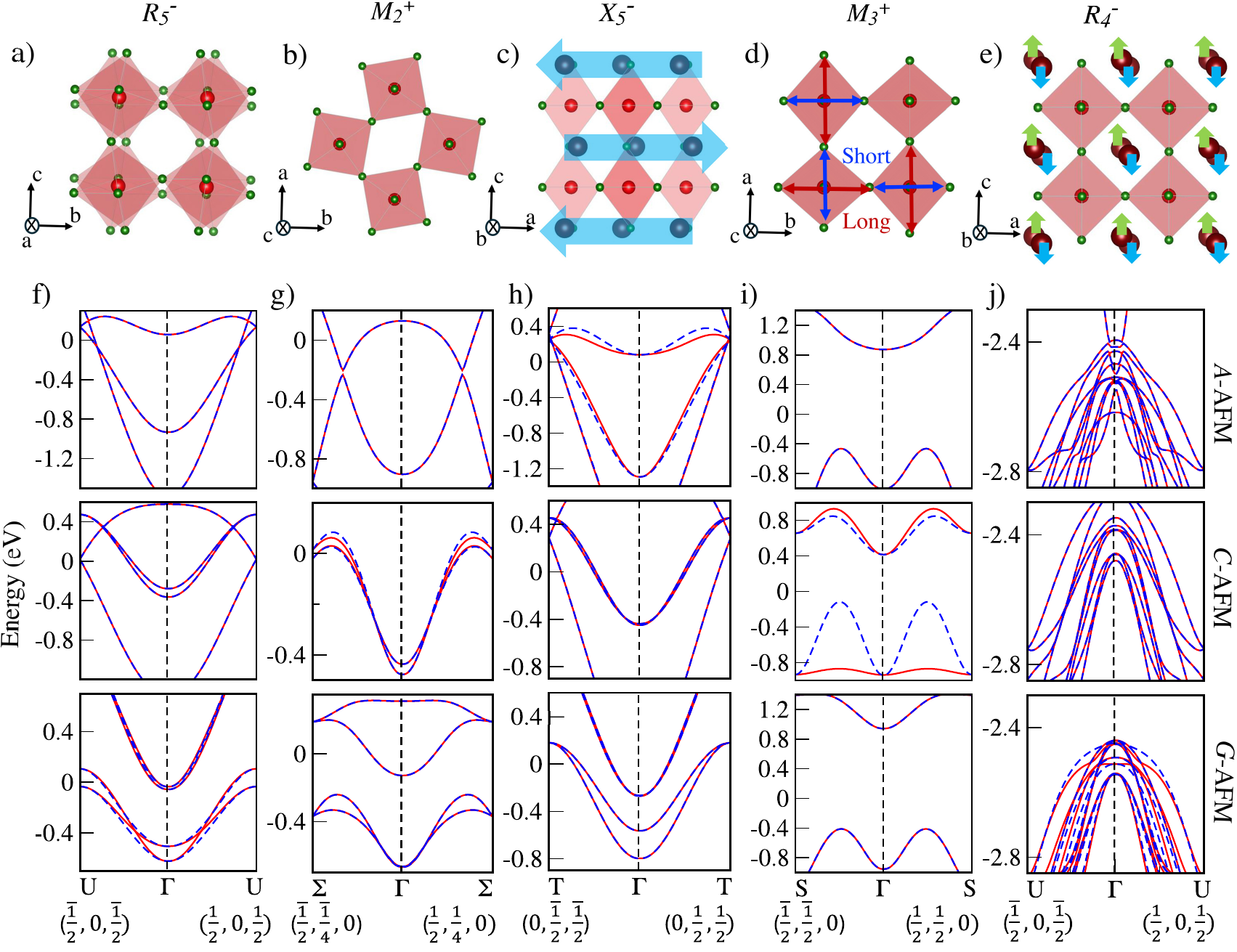}
\caption{Schematic representation of the atomic distortions of the $Pbnm$ structure and the corresponding band structure in the absence of spin-orbit coupling.  (a) Out-of-phase octahderal rotation: $\phi_{xy}^-$ ($R_5^-$), (b) In-phase octahderal rotation: $\phi_{z}^+$ ($M_2^+$), (c) Antipolar motion of La atoms: $X_5^-$, (d) JT distortion: $M_3^+$, and (e) Antipolar motion of La atoms: $R_4^-$ distortion. Crystallographic $a, b, c$ axes are aligned with the Cartesian $x, y,$ and $ z$ directions, respectively. Bottom panel: Electronic band structure for individual $Pbnm$ distortions: (f) $R_5^-$, (g) $M_2^+$, (h) $X_5^-$, (i) $M_3^+$, and (j) $R_4^-$ with A-type AFM ({\it top panel}), C-type AFM ({\it middle panel}), and G-type AFM ({\it bottom panel}) configurations. Up and down spin-polarized bands are plotted with solid red and dashed blue lines, respectively. Energies are plotted relative to the Fermi energy of the system such that zero in the energy corresponds to the Fermi energy.
}
\label{distortion_vs_band}
\end{figure*}

To investigate the role of atomic distortions in the NRSS, we first perform the  
 symmetry mode analysis of the 
 the optimized $Pbnm$ structure
 with reference to the cubic $Pm\overline3m$ structure using ISODISTORT~\cite{Isodistort, Isodistort1}. We note that although LMO does not crystallize in $Pm\overline3m$ structure at high temperature,  it provides a simple description of the individual atomic distortions, forming thereby a good choice of reference for symmetry mode analysis.  Our symmetry mode analysis
 shows the presence of  atomic distortions corresponding to the irreducible representations (IRs)
 $R_5^-$,  $M_2^+$, $X_5^-$, $M_3^+$, and $R_4^-$  of the $Pm\overline3m$ structure~\cite{Varignon_titanates, Subhadeep_BiNiO3, TMO_subhadeep}. 
 The computed amplitudes of the atomic distortions, as shown in Fig. \ref{Gs_modes}b, are in reasonable agreement with the values calculated for the experimentally reported $Pbnm$ structures, confirming the accuracy of our optimized structure.

We find that in the absence of these distortions, $i.e.$ in the $Pm\overline3m$ structure, there is no antiferroically ordered charge quadrupole moment in the system and accordingly the NRSS is absent irrespective of the magnetic ordering. 

We now proceed to illustrate each of the computed atomic distortions, the corresponding charge and magnetic multipoles, and their effect on the NRSS in the electronic band structure,  considering all three, A-, C- and G-type AFM orders. The results of our findings are summarized in Table \ref{tab1}. In the following discussion, we consider a similar amplitude of distortions to that present in the ground state $Pbnm$ structure while condensing individual distortions.

We start by considering the $R_5^-$ phonon mode, associated with an unstable phonon at the $R$ point in the BZ of the phonon band structure of a cubic $Pm\overline3m$ perovskite with a small tolerance factor ($t <$ 1) \cite{Goldschmidt_1926,Benedek_2013,TMO_subhadeep,YNO_Mercy}.
$R_5^-$ induces an out-of-phase (i.e., alternating clockwise and anti-clockwise rotation) octahedral rotation  
along $x$ and $y$ directions, as shown in Fig. \ref{distortion_vs_band}a, abbreviated as  $\phi_{xy}^-$, stabilizing an $a^-a^-c^0$ octahedral rotation pattern, following the Glazer's notation.  
Since $\phi_{xy}^-$  involves atomic distortions in all three dimensions (3D),
we refer to it as a $R$-type structural distortion.

Condensation of $\phi_{xy}^-$ lowers the symmetry of the structure to $Imma$ and induces an antiferroic charge quadrupole  component $w^{202}_1$, as listed in Table \ref{tab1}. Interestingly, this antiferroic pattern is identical to that of AFM dipolar ordering for the G-type AFM configuration. Consequently, it leads to ferroically ordered magnetic octupole components $w^{213}_1$, $w^{212}_{-1}$, $w^{212}_2$, and $w^{211}_1$ for G-type AFM configuration, while no ferroic magnetic octupole component exists for the A- and C-type AFM (see Table \ref{tab1}). The $k$-space representation, $k_xk_zm_z$, of the existing magnetic octupole components in G-type AFM dictates the NRSS along any direction in momentum space with $k_x \ne 0$ and $k_z \ne 0$. This is also confirmed by our band structure calculations  in Fig. \ref{distortion_vs_band}f, showing the presence of the NRSS along $\Gamma$-$U$.  
In contrast to the G-type AFM configuration, the absence of a ferroic magnetic octupole in A and C-type AFM order leads to degenerate up and down spin-polarized bands (see Table \ref{tab1}).

$M_2^+$ is another dominant
atomic distortion that spontaneously appears in perovskites with $t < 1$ \cite{Goldschmidt_1926,Benedek_2013,TMO_subhadeep,YNO_Mercy}. The $M_2^+$ distortion
corresponds to an unstable phonon mode at the $M$-point in the phonon band structure of the cubic $Pm\overline3m$ structure. It induces an in-phase octahedral rotation along $\hat z$, $\phi_{z}^+$, as shown in Fig. \ref{distortion_vs_band}b, in which the O atoms are displaced only on the $ab$ plane, i.e., in two dimensions in contrast to the $R_5^{-}$ distortion, discussed above.  The condensation of $\phi_{z}^+$ reduces the symmetry to $P4/mbm$, stabilizing an octahedral rotation pattern corresponding to the Glazer notation $a^0a^0c^+$. 

Interestingly, the $M_2^+$ distortion does not lead to any antiferroic charge quadrupole moment but rather it leads to an antiferroic pattern of higher-order rank-4 charge multipole, viz., charge hexadecapole component $w^{404}_{-4}$, as listed in Table \ref{tab1}. We find that the antiferroic pattern of $w^{404}_{-4}$ is identical to the antiferroic magnetic dipolar ordering of the  C-type AFM configuration. Consequently, we get ferroically ordered magnetic triakontadipole components, $w^{415}_{-4}$ and $w^{414}_{4}$. In contrast to the case of the $R_5^{-}$ distortion, here the magnetic triakontadipoles form the lowest ferroic magnetic multipole, and the corresponding $k$-space representation, $(k_x^2-k_y^2)k_xk_ym_z$, dictates a $g$-wave splitting. This is consistent with the previous report \cite{Verbeek2024} that also correlates magnetic triakontadipoles to the $g$-wave splitting. According to the $k$-space representation of the existing magnetic triakontadipoles, we expect a NRSS along $k_xk_y$ direction only if $k_x\ne k_y$. This indeed becomes evident from our computed band structure, showing the splitting along $\Gamma-\Sigma$($k_x=0.5, k_y=0.25, k_z=0$) in Fig. \ref{distortion_vs_band}g.

For the A and G-type AFM order, however, such a splitting is absent due to the absence of any net ferroically ordered magnetic traikontadipole moment.

Moving on to the $X_5^-$ distortion, which is an $X$-point phonon mode of the reference $Pm\overline 3m$ structure, introduces the anti-polar motion of the La atoms as indicated in Fig. \ref{distortion_vs_band}c. $X_5^-$ is coupled trilinearly with the $R_5^-$ and $M_2^+$ octaherdral rotations as discussed later in Eq. (\ref{Fpbnm}), and therefore appears naturally in all $Pbnm$ perovskite structure \cite{Varignon_titanates,varignon_AVO}.
  
 Condensation of only $X_5^-$ distortion induces a $Cmcm$ structure, which hosts a
 charge quadrupole component $w^{202}_{-1}$ with an antiferroic pattern identical to the antiferromagnetic dipolar ordering of the A-type AFM order (see Table \ref{tab1}). As a result, it leads to ferroically ordered magnetic octupole components, $w^{213}_{-1}$, $w^{212}_{1}$, and $w^{211}_{-1}$. Consequently, they lead to a NRSS along $\Gamma-T$ direction (see Fig. \ref{distortion_vs_band}h) consistent with the $k$-space representation of the existing magnetic octupole components.

$M_3^+$ phonon mode induces a Jahn-Teller (JT) distortion at the MnO$_6$ octahedra, resulting in alternating long and short Mn-O bonds, as indicated in Fig. \ref{distortion_vs_band}d. Such a JT distortion, also known as $Q_{2z}^M$-type JT distortion following canonical notation in Refs. ~\cite{LaMnO3_marcus,TMO_subhadeep}, stabilizes a C-type orbital ordering in LMO~\cite{LMnO_OO}. Condensation of the $M_3^+$ distortion leads to a $P4/mbm$ structure, similar to the $M_2^+$ phonon mode, as discussed above. Interestingly, however, we note that the existing multipoles are different. In this case, we have an antiferroic charge quadrupole  component $w^{202}_{-2}$. The antiferroic pattern of $w^{202}_{-2}$, as listed in Table \ref{tab1}, coincides with that of the magnetic dipolar ordering of C-type AFM, leading to ferroic magnetic octupole components $w^{213}_{-2}$ and $w^{212}_{2}$. For the  A- and C-type AFM order, however, there is no ferroic magnetic octupole component. The $k$-space representation, $k_xk_ym_z$, of the existing magnetic octupole components in the C-type AFM accordingly leads to a NRSS along any direction in momentum space with $k_x \ne 0$ and $k_y \ne 0$. Consistently, a NRSS of the bands emerges along $\Gamma-S$ direction in the bandstructure shown in Fig. \ref{distortion_vs_band}i.  Thus the  nature of the spin splitting is found to be very different 
for the $M_3^+$ distortion compared to the $M_2^+ (\phi_{z}^+)$ distortion ($d$-wave Vs. $g$-wave).
This emphasizes the crucial roles of the structural distortion, and consequent multipoles in describing the NRSS.

Finally, $R_4^-$  is also a $R$-type structural distortion that distorts La atoms as shown in Fig. \ref{distortion_vs_band}e.
Condensing $R_4^-$ lowers the symmetry of the system to $Imma$, which consequently induces a charge quadrupole  component $w^{202}_1$ with an antiferroic pattern, identical to that of AFM dipolar ordering for the G-type AFM configuration (see Table \ref{tab1}). Consequently, it leads to ferroically ordered magnetic octupole components $w^{213}_1$, $w^{212}_{-1}$, $w^{212}_2$, and $w^{211}_1$ for the G-type AFM configuration, while no ferroic magnetic octupole component exists for the A- and C-type AFM.  Similarly to $R_5^-$, the existing ferroic magnetic octupoles for the $R_4^-$ distortion lead to an NRSS along $\Gamma-U$ direction for the G-type AFM order. 

The multipolar analysis of the individual atomic distortions highlights the importance of the combined effect of the atomic distortions and the type of AFM order in inducing the NRSS. For example,  $X_5^-, M_3^+,$ and $R_4^-$ distortions lead to NRSS for  
A-, C-, and G-type AFM order, respectively. In the $Pbnm$ phase, all these atomic distortions are present and coupled to each other. We note that, however, starting from the reference $Pm\overline3m$ structure, joint action of $R_5^-$ ($\phi_{xy}^-$) and $M_2^+ (\phi_z^+)$ already give rise to the $Pbnm$ phase with $a^-a^-c^+$ octahedral rotation pattern, corresponding to the configuration 6 in Table \ref{tab1}.

Our multipole calculation shows that the crystal structure in this case has three antiferroic charge quadrupole moment components, $w^{202}_{-2}, w^{202}_{-1},$ and $w^{202}_{1}$ with antiferroic patterns identical to that of C-, A-, and G-type AFM dipolar order respectively. Consequently, the $w^{202}_{-2}$ quadrupole moment in combination with AFM dipolar order give rise to ferroically ordered magnetic octupole components $w^{213}_{-2}$ and $w^{212}_{2}$ in the C-type AFM oder. Similarly, we have ferroically ordered magnetic octupole components $w^{213}_{-1}$, $w^{212}_{1}$ and $w^{211}_{-1}$ in the A-type AFM order, while the G-type AFM order hosts ferroic octupole components $w^{213}_{1}$, $w^{212}_{-1}$ and $w^{211}_{1}$. The presence of ferroic magnetic octupole components in A-, C-, and G-type AFM gives rise to NRSS for all cases but along different directions, dictated by the $k$-space representation of the existing magnetic octupole components. 

A fully optimized $Pbnm$ structure additionally incorporates $X_5^-$, $R_4^-$, and $M_3^+$ distortions. In the presence of these additional distortions, the relevant multipoles and the direction of NRSS remain the same, as discussed earlier in Section \ref{magorder}, and also evident from Table \ref{tab1}. We note that the magnitudes of the multipoles and, hence, the NRSS are, however, different for the optimized $Pbnm$ structure compared to configuration 6 in Table \ref{tab1}.

\subsection{Engineering NRSS} \label{design}

Building on our understanding of the role of different phonon modes in NRSS discussed thus far, we explore in this section various approaches to tuning specific phonon modes through chemical substitution, strain, and interface engineering. These strategies consequently enable the design and control of NRSS in the perovskite structure.

\begin{figure}[t]
\centering
\includegraphics[width=\columnwidth]{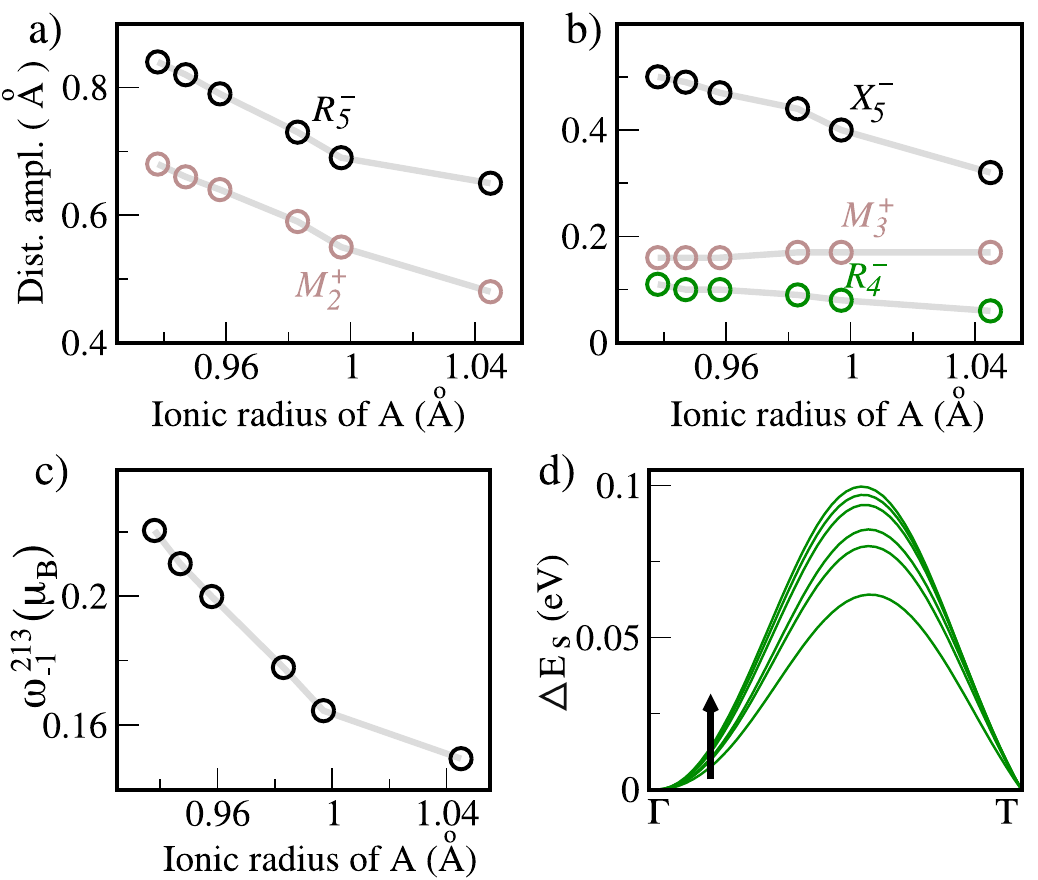}
\caption{ Effect of chemical engineering. (a) Amplitudes of distortions for $R_5^{-}$ and $M_2^{+}$ phonon modes of the $Pbnm$ structure as a function of the ionic radius of the $A$-site cation. (b) The same variations for $X_5^{-}$, $M_3^{+}$, and $R_4^{-}$ phonon modes. (c) Variation of the lowest-order ferroic magnetic octupole component $w^{213}_{-1}$ as a function of the  $A$-site ionic radius. (d) Corresponding variation in the magnitude of the NRSS energy $\Delta E_{\rm s}$ at the VBM along $\Gamma$-$T$ for different $A$-site ionic radii. The arrow denotes the direction of decreasing $A$-site ionic radius.}
\label{tolerance}
\end{figure}

 \subsubsection{Chemical engineering} \label{chemical}
As discussed in Section \ref{dist}, octahedral rotations, $\phi_{xy}^-$ and $\phi_z^+$, play a crucial role in inducing NRSS. Notably, these rotations are governed by the tolerance factor ($t$) in ABO$_3$ perovskites, which can be modulated by applying chemical pressure through variations in the A-site cation radius. Specifically, a reduction in the $A$-site atomic radius enhances the $R_5^-$ ($\phi_{xy}^{-}$) and $M_2^+$ ($\phi_{z}^{+}$) octahedral rotations, which, in turn, is expected to result in a larger NRSS.

To illustrate this idea, we consider $Pbnm$-$A$MnO$_3$ systems with different rare earth $A$-site cations \cite{Radii}: La (1.045 \AA), Pr (0.997 \AA), Nd (0.983 \AA), Sm (0.958 \AA), Eu (0.947 \AA), and Gd (0.938 \AA), where the values in parentheses indicate the ionic radii of the A-site cations.  As shown in Fig. \ref{tolerance}a and b, changing the $A$-site cation changes the amplitude of different phonon mode distortions of the $Pbnm$ structure. The most significant changes occur in the octahedral rotational distortions $R_5^-$ ($\phi_{xy}^{-}$) and $M_2^+$ ($\phi_{z}^{+}$) (see Fig. \ref{tolerance}a). The smaller the $A^{3+}$ cation, the stronger the octahedral rotational distortions $R_5^-$ and $M_2^+$. The changes in the anti-polar motion $X_5^-$ are rather moderate, while those in the $M_3^+$ and $R_4^-$ distortions are relatively weak. 

These weak changes in other distortion modes can be understood from the symmetry-allowed tri-linear anaharmonic coupling between different distortion modes in the free energy $F_{Pbnm}$ of the $Pbnm$ structure \cite{varignon_AVO,Invariants,Invariants2},

\begin{equation}\label{Fpbnm}
    F_{Pbnm} \propto \phi_{xy}^-\phi_{z}^+X_5^- + \phi_{xy}^-X_5^-M_3^+ + R_4^-X_5^-\phi_{z}^+ + R_4^-X_5^-M_3^+
\end{equation}

To investigate the effect of the octahedral rotations, we analyze the magnetic multipoles and the resulting NRSS in $A$MnO$_3$ structures with different $A$-site cations for the A-type AFM order, corresponding to the ground state of LMO.  The computed multipoles show that the magnitudes of the ferroic magnetic octupole components increase with the decrease in the ionic radius of the $A$-site ion.  The corresponding variation for one of the ferroically ordered magnetic octupole components $w^{213}_{-1}$ is shown in Fig. \ref{tolerance}c. Consequently, we find that the magnitude of the NRSS energy $\Delta E_{\rm s} = E_{\uparrow}-E_{\downarrow}$  also increases as the $A$-site ion radius decreases, 
as shown in Fig. \ref{tolerance}d.  

\subsubsection{Strain engineering}
\begin{figure}[t]
\centering
\includegraphics[width=\columnwidth]{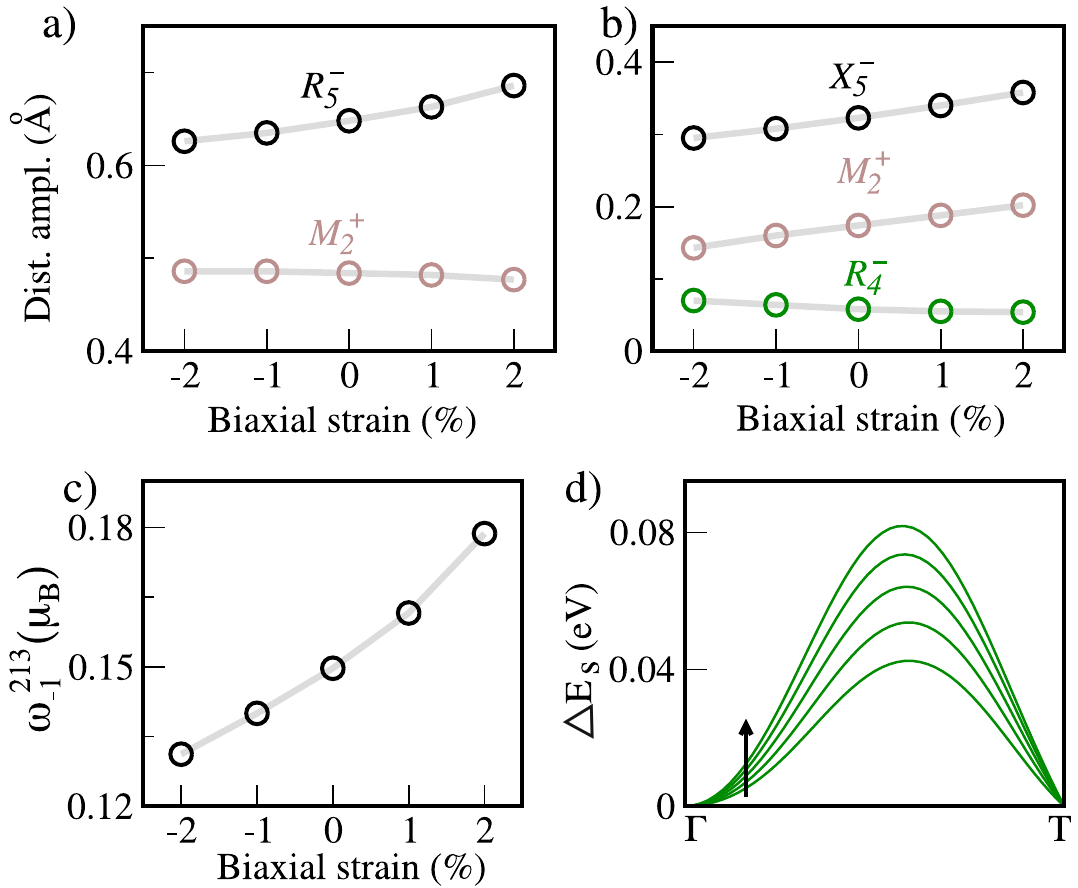}
\caption{Strain engineering of NRSS. (a) Amplitudes of distortions for $R_5^{-}$ and $M_2^{+}$ phonon modes of the $Pbnm$ structure with varying biaxial strain. The +ve and -ve values denote the compressive and tensile strain, respectively. (b) The same variation for $X_5^{-}$, $M_3^{+}$, and $R_4^{-}$ phonon modes. (c) The magnitude of the lowest-order ferroic magnetic octupole component $w^{213}_{-1}$ as a function of the biaxial strain. (d) Corresponding variation in the magnitude of the NRSS energy $\Delta E_{\rm s}$ at the VBM along $\Gamma$-$T$ for different biaxial strains. The arrow denotes the direction of increasing biaxial strain. }
\label{strain}
\end{figure}
An alternative approach to tuning octahedral rotation for a given material is through strain engineering or the application of external pressure. Experimentally, epitaxial strain can be controlled during the growth of thin films on different substrates.  
Here, we consider a biaxial strain to LMO,  ranging from -2\% (tensile) to 2\% (compressive), which falls within the experimentally achieved range for LMO \cite{LaMnO3_ceramics_2023, LaMnO3_on_LaAlO3, LaMnO3_marton, LaMnO3_Roqueta}. 
We construct the strained structures by modifying the in-plane lattice parameters (here, $ab$ in the pseudocubic reference frame), while the out-of-plane lattice parameter is adjusted to keep the system-volume invariant. This is followed by optimization of the interatomic coordinates.  As shown in Fig. \ref{strain}a, a tensile (compressive) strain progressively increases (decreases) the $\phi_{xy}^-$ ($R_5^{-}$) octahedral rotation. In contrast, the $\phi_z^+$ ($M_2^{+}$) octahedral rotation exhibits the opposite trend, although the change is minimal. Additionally, due to anharmonic couplings among different phonon modes, as discussed earlier, the $X_5^-$ and $M_3^+$ distortions also undergo modifications (see Fig. \ref{strain}b).  
Our multipole calculations for the A-type AFM order show that the magnitude of the ferroically ordered magnetic octupole increases with compressive strain, as depicted in Fig. \ref{strain}c. The corresponding spin splitting energy $\Delta E_s$, {\it i.e.,} the energy difference between the up- and down-spin bands,  along the $\Gamma$-T is shown in Fig. \ref{strain}d, demonstrating a consistent trend — larger compressive (tensile) strain results in larger (smaller) NRSS- demonstrating the strain engineering of the NRSS.

\subsubsection{Superlattice engineering and electric field tuning }
We now discuss the NRSS engineering by designing superlattice structures and its electric-field-driven switchability. While $Pbnm$ perovskites are non-polar, a spontaneous polar distortion, required for the electric field control, can be induced by stacking  $Pbnm$ structures of two different materials along the (001) lattice plane \cite{HIF_Pnma_adv_mat_2012}. Here, the difference in the octahedral rotation amplitudes between two different perovskites give rise to an uncompensated anti-polar motion of the A-site cations ($X_5^-$), which in turn induces a sizable in-plane electric polarization ($\vec P$) \cite{HIF_Pnma_adv_mat_2012}, leading to {\it hybrid~ improper ferroelectricity} \cite{Eric_2008,HIF_Pnma_adv_mat_2012,BENEDEK2012}. The polar distortion, and the spontaneous electric polarization can also be induced in superlattices made of $Pbnm$ and non-$Pbnm$ perovskites \cite{HIF_Pnma_adv_mat_2012,Subhdeep_APL}.

\begin{figure}[t]
\centering
\includegraphics[width=\columnwidth]{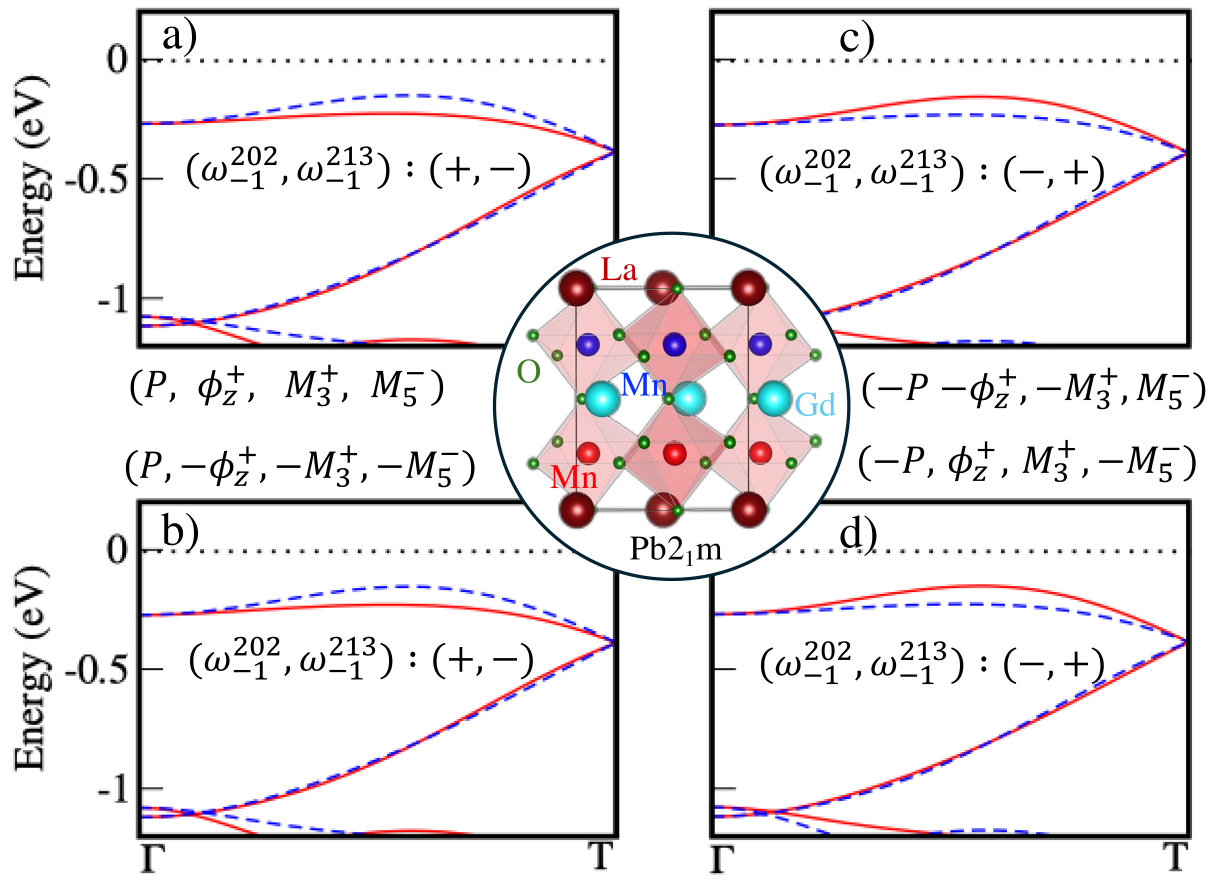}
\caption{ Electronic bands of the superlattice structures (a) S1: ($P$, $\phi_z^+$, $M_3^+$, $M_5^-$),
(b) S2: ($P$, $-\phi_z^+$, $-M_3^+$, $-M_5^-$),
(c) S3: ($-P$, $-\phi_z^+$, $-M_3^+$, $M_5^-$), and (d) S4: ($-P$, $\phi_z^+$, $M_3^+$, $-M_5^-$) in the presence of $A$-type AFM order (see text for details). Spin-up and spin-down bands are plotted with solid red and dashed blue lines, respectively. Energies are plotted relative to the Fermi energy of the system. The inset shows the crystal structure of the LMO-GdMnO$_3$ superlattice with the space group symmetry $Pb2_1m$.
}
\label{superlattice_band}
\end{figure}
To illustrate our idea, we construct the LMO-GdMnO$_3$ superlattice (see inset of  Fig. \ref{superlattice_band}).
 LMO and GdMnO$_3$ are chosen as they have the largest and smallest tolerance factor $t$ among the $A$MnO$_3$ compounds as discussed previously in Section \ref{chemical}, inducing thereby the largest uncompensated  $X_5^-$ distortion. We optimize both the lattice parameters and the internal atomic coordinates of the LMO-GdMnO$_3$ superlattice, keeping an $A$-type AFM order, which lowers the symmetry to $Pb2_1m$ space group. We note that the $Pb2_1m$ structure has $M_5^-$ and $X_1^+$ anti-polar distortions of the La/Gd cations, and  polar $\Gamma_4^-$ and $\Gamma_5^-$ distortions in addition to the existing distortions of the $Pbnm$ structure, discussed in Section \ref{dist}. We denote the combined distortion of $\Gamma_4^-$ and $\Gamma_5^-$ as $P$ in the rest of the paper, which gives rise to an in-plane electric polarization $\vec{P}$ in the $Pb2_1m$ structure.  
The free energy of the $Pb2_1m$ structure allows for the following anharmonic coupling terms 
in addition to the anharmonic couplings presented in Eq. (\ref{Fpbnm}),

\begin{equation} \label{FPb2_1m}
    F_{Pb2_1m} \propto  P\phi_{z}^+M_5^-
    + PM_3^+M_5^-.
\end{equation}

Guided by Eq. (\ref{FPb2_1m}), we obtain a four-fold degenerate ground state for the 
$Pb2_1m$ superlattice structure. These are S1: ($P$, $\phi_z^+$, $M_3^+$, $M_5^-$),
S2: ($P$, $-\phi_z^+$, $-M_3^+$, $-M_5^-$),
S3: ($-P$, $-\phi_z^+$, $-M_3^+$, $M_5^-$), and S4: ($-P$, $\phi_z^+$, $M_3^+$, $-M_5^-$),
 where ``$-$''  sign indicates condensation of a distortion in the reverse direction but with equal amplitude. We note that condensing any other combination of these modes is energetically unfavorable and gets back to one of the aforementioned four structures during structural optimization.

 To analyze the NRSS, we carry out both the electronic structure calculations and the multipole analysis for the ground state structures, S1-S4, of the  $Pb2_1m$ phase. Our calculations for the A-type AFM order show that for all the structures, S1 to S4, there is a net non-zero atomic-site electric dipole moment at the Mn sites, which has opposite signs for S1, S2 vs. S3, S4 structures. We find that with the reversal in the sign of the electric dipole moment, the antiferroic charge quadrupole moment component $w^{202}_{-1}$ also switches sign due to the anharmonic coupling described in Eq. (\ref{FPb2_1m}). Since the antiferroic charge quadrupole moment component $w^{202}_{-1}$ couples to the antiferroic magnetic dipole moment at the Mn sites to give rise to the ferroic magnetic octupole component, the sign change in the antiferroic charge quadrupole also leads to a sign change in the ferroic magnetic octupole. 
 
 Consequently, our computed band structure shows a reversal in the NRSS, as depicted in Fig. \ref{superlattice_band}a-d. As expected, the spin polarization of the split bands is opposite for S3 and S4 as compared to S1 and S2, demonstrating the reversal of the NRSS upon switching the polar distortion. 
 As a characteristic feature of ferroelectric materials, switching of the polar distortion that is associated with the structural transition between (S1, S2) and (S3, S4)  and concomitant electric polarization should be achieved by applying an external electric field.  Eq. (\ref{FPb2_1m}) reveals two possible pathways for this switching, viz.,  path I: transforms S1 (S2) to S3 (S4)  and path II: transforms S1 (S2) to S4 (S3). Path I reverses $\phi_z^{-}$  and $M_3^+$, while Path II reverses $M_5^-$ during the electric polarization switching. For a given material, the path with a lower energy barrier specific to that system will be favored for the ferroelectric switching. Interestingly, however, our analysis shows that both ferroic magnetic octupoles and the resulting NRSS reverse irrespective of the path it may follow for the ferroelectric switching.
\par

\section{Summary and Outlook} \label{summary}

In this work, we demonstrate the crucial interplay between structural symmetry and magnetic order in determining the presence of NRSS. Our analysis reveals that for a given crystal structure, NRSS may or may not emerge depending on the specific magnetic order, as summarized in Table \ref{tab1}. We focus on three common AFM orders in perovskite oxides—A-type, C-type, and G-type—highlighting how their distinct magnetic configurations influence the NRSS. 
We note that other AFM orderings, such as E'-type and E-type AFM \footnote{E and E'-type AFM describe an in-plane up-up-down-down spin order that couples ferroically and antiferroically along the out of plane direction respectively.}, which are relevant to the observed ground state of HoMnO$_3$ \cite{Dagotto_PRL, Silvia_HMnO3} and a low-energy magnetic state in rare-earth nickelate perovskites \cite{Varignon_2017}, preserve time-reversal symmetry and therefore cannot host NRSS. 

Similarly, for a given magnetic order, variations in structural symmetry can either enable or suppress NRSS. This combined effect of structure and magnetism is effectively captured by ferroically ordered magnetic multipoles, such as magnetic octupoles or triakontadipoles.  

Through a detailed examination of the atomic distortions, we gain key insights into the role of structure in the emergence of NRSS. We find that a necessary condition for NRSS is a matching pattern between the antiferroic charge multipole (which quantifies structural distortions) and the underlying AFM dipolar order. For instance, in a G-type AFM order, where magnetic dipoles are antiferroically aligned in all three spatial directions, the emergence of a ferroically ordered magnetic multipole, and hence NRSS, requires a three-dimensional distortion, i.e., an $R$-type structural distortion.
However, for a C- (A-) type AFM order with magnetic dipoles antiferroically aligned in two (one) spatial directions requires two (one) dimensional  distortion, i.e., an $M$- ($X$-) type structural distortion to induce the NRSS.
This insight is crucial in designing NRSS in real materials. 
Our analysis also applies to the NRSS in orbital-ordered systems, as proposed recently for SrCrO$_3$\cite{Cano_claude_2025}, in which the NRSS with the C-type AFM order survives only with a C-type orbital order and vanishes for the ground-state G-type orbital order.   

Furthermore, we clarify that, while the direction of spin polarization does not affect the momentum direction of the NRSS, it does influence the specific components of the ferroically ordered magnetic multipole. For example, in the $Pbnm$ structure (Table \ref{tab1}), an A-type AFM order with spin polarization along $\hat{z}$ results in a ferroically ordered magnetic octupole component $w^{213}_{-1}$, while spin polarization along $\hat{y}$ instead leads to $w^{213}_{2}$. The former corresponds to a $k$-space representation of $k_y k_z m_z$, while the latter is represented by $k_y k_z m_y$. In both cases, NRSS emerges along momentum directions with nonzero $k_y$ and $k_z$, but the spin polarization of the bands differs, aligning along $\hat{z}$ in the former case and $\hat{y}$ in the latter.

Based on our understanding of the role of various atomic distortions, we propose three approaches for engineering NRSS: chemical substitution, strain engineering, and the formation of superlattices. While the first two approaches primarily offer control over the magnitude of NRSS, the superlattice approach introduces 
improper ferroelectricity, providing a platform for tuning the  NRSS using an external electric field.  This phenomenon is not limited to only  perovskite systems, and can be seen in other family of materials. For instance, our recent report \cite{Bandyopadhyay2024} shows, polar distortion coupling with Raman active phonon mode gives rise to change in the ligand environment and thereby controls the NRSS in rultile MnF$_2$.

Furthermore, our analysis shows that if switching the polar distortion switches some of the coupled distortions that directly influence the NRSS, then the sign of the NRSS can also be switched.
As discussed explicitly for the $Pb2_1m$ superlattice, the electric field can modulate the polar distortion, which is coupled to rotational modes, thereby altering NRSS. 
We highlight that our proposed framework serves as a universal theory to explain recent reports on electric field tuning of NRSS \cite{Libor_2025_BaCuF4, Stroppa_2025}. 
For example, the coupled octahedral rotation and deformation modes reverse upon reversing the polar mode in certain ferroelectrics, resulting in the switching of the NRSS \cite{Libor_2025_BaCuF4, Stroppa_2025}.
A similar situation also arises during the antiferroelectric (AFE)-ferroelectric (FE) transition, as recently discussed in Ref.~\cite{Duan_2025}. In the AFE phase, antipolar displacements are coupled to the rotational symmetry that connects the two magnetic sublattices that get suppressed during the transition to the FE phase in the presence of an external electric field. Accordingly, the NRSS is present only in the AFE phase and is absent in the presence of an external electric field when the system transforms to a FE phase.
Conversely, in multiferroic systems, the electric field may control magnetic domains, leading to a corresponding switch in NRSS. Notably, some multiferroic materials have recently been identified to exhibit NRSS, and their electric-field-driven control has also been explored \cite{Dong2025}. 
We hope that the general framework developed in the present work will inspire further research on manipulating NRSS in the near future.

\begin{acknowledgments}
S.Ba acknowledges computational resources that are provided by CINECA under the ISCRA initiative, specifically through the project ISCRA-B HP10BA00W3.
S.P. acknowledges partial financial support by the Next-Generation-EU program via the PRIN-2022 SORBET (Grant No. 2022ZY8HJY).
SBh thanks National Supercomputing Mission for providing computing resources
of ‘PARAM Porul’ at NIT Trichy, implemented by C-DAC and supported by the Ministry
of Electronics and Information Technology (MeitY) and Department of Science and
Technology, Government of India and acknowledges funding support from the Industrial Research and Consultancy Centre (IRCC) Seed Grant (RD/0523-IRCCSH0-018) and the INSPIRE research grant (project code RD/0124-DST0030-002). 
\end{acknowledgments}

\vspace{2 cm}

\bibliography{main}

\providecommand{\noopsort}[1]{}\providecommand{\singleletter}[1]{#1}
\begin{thebibliography}{111}%
\makeatletter
\providecommand \@ifxundefined [1]{%
 \@ifx{#1\undefined}
}%
\providecommand \@ifnum [1]{%
 \ifnum #1\expandafter \@firstoftwo
 \else \expandafter \@secondoftwo
 \fi
}%
\providecommand \@ifx [1]{%
 \ifx #1\expandafter \@firstoftwo
 \else \expandafter \@secondoftwo
 \fi
}%
\providecommand \natexlab [1]{#1}%
\providecommand \enquote  [1]{``#1''}%
\providecommand \bibnamefont  [1]{#1}%
\providecommand \bibfnamefont [1]{#1}%
\providecommand \citenamefont [1]{#1}%
\providecommand \href@noop [0]{\@secondoftwo}%
\providecommand \href [0]{\begingroup \@sanitize@url \@href}%
\providecommand \@href[1]{\@@startlink{#1}\@@href}%
\providecommand \@@href[1]{\endgroup#1\@@endlink}%
\providecommand \@sanitize@url [0]{\catcode `\\12\catcode `\$12\catcode
  `\&12\catcode `\#12\catcode `\^12\catcode `\_12\catcode `\%12\relax}%
\providecommand \@@startlink[1]{}%
\providecommand \@@endlink[0]{}%
\providecommand \url  [0]{\begingroup\@sanitize@url \@url }%
\providecommand \@url [1]{\endgroup\@href {#1}{\urlprefix }}%
\providecommand \urlprefix  [0]{URL }%
\providecommand \Eprint [0]{\href }%
\providecommand \doibase [0]{https://doi.org/}%
\providecommand \selectlanguage [0]{\@gobble}%
\providecommand \bibinfo  [0]{\@secondoftwo}%
\providecommand \bibfield  [0]{\@secondoftwo}%
\providecommand \translation [1]{[#1]}%
\providecommand \BibitemOpen [0]{}%
\providecommand \bibitemStop [0]{}%
\providecommand \bibitemNoStop [0]{.\EOS\space}%
\providecommand \EOS [0]{\spacefactor3000\relax}%
\providecommand \BibitemShut  [1]{\csname bibitem#1\endcsname}%
\let\auto@bib@innerbib\@empty
\bibitem [{\citenamefont {Hayami}\ \emph {et~al.}(2019)\citenamefont {Hayami},
  \citenamefont {Yanagi},\ and\ \citenamefont {Kusunose}}]{Hayami2019}%
  \BibitemOpen
  \bibfield  {author} {\bibinfo {author} {\bibfnamefont {S.}~\bibnamefont
  {Hayami}}, \bibinfo {author} {\bibfnamefont {Y.}~\bibnamefont {Yanagi}},\
  and\ \bibinfo {author} {\bibfnamefont {H.}~\bibnamefont {Kusunose}},\
  }\bibfield  {title} {\bibinfo {title} {Momentum-dependent spin splitting by
  collinear antiferromagnetic ordering},\ }\href
  {https://doi.org/10.7566/JPSJ.88.123702} {\bibfield  {journal} {\bibinfo
  {journal} {J. Phys. Soc. Jpn.}\ }\textbf {\bibinfo {volume} {88}},\ \bibinfo
  {pages} {123702} (\bibinfo {year} {2019})}\BibitemShut {NoStop}%
\bibitem [{\citenamefont {Yuan}\ \emph {et~al.}(2020)\citenamefont {Yuan},
  \citenamefont {Wang}, \citenamefont {Luo}, \citenamefont {Rashba},\ and\
  \citenamefont {Zunger}}]{Yuan2020}%
  \BibitemOpen
  \bibfield  {author} {\bibinfo {author} {\bibfnamefont {L.-D.}\ \bibnamefont
  {Yuan}}, \bibinfo {author} {\bibfnamefont {Z.}~\bibnamefont {Wang}}, \bibinfo
  {author} {\bibfnamefont {J.-W.}\ \bibnamefont {Luo}}, \bibinfo {author}
  {\bibfnamefont {E.~I.}\ \bibnamefont {Rashba}},\ and\ \bibinfo {author}
  {\bibfnamefont {A.}~\bibnamefont {Zunger}},\ }\bibfield  {title} {\bibinfo
  {title} {Giant momentum-dependent spin splitting in centrosymmetric low-$z$
  antiferromagnets},\ }\href {https://doi.org/10.1103/PhysRevB.102.014422}
  {\bibfield  {journal} {\bibinfo  {journal} {Phys. Rev. B}\ }\textbf {\bibinfo
  {volume} {102}},\ \bibinfo {pages} {014422} (\bibinfo {year}
  {2020})}\BibitemShut {NoStop}%
\bibitem [{\citenamefont {Yuan}\ \emph {et~al.}(2021)\citenamefont {Yuan},
  \citenamefont {Wang}, \citenamefont {Luo},\ and\ \citenamefont
  {Zunger}}]{Yuan2021}%
  \BibitemOpen
  \bibfield  {author} {\bibinfo {author} {\bibfnamefont {L.-D.}\ \bibnamefont
  {Yuan}}, \bibinfo {author} {\bibfnamefont {Z.}~\bibnamefont {Wang}}, \bibinfo
  {author} {\bibfnamefont {J.-W.}\ \bibnamefont {Luo}},\ and\ \bibinfo {author}
  {\bibfnamefont {A.}~\bibnamefont {Zunger}},\ }\bibfield  {title} {\bibinfo
  {title} {Prediction of low-z collinear and noncollinear antiferromagnetic
  compounds having momentum-dependent spin splitting even without spin-orbit
  coupling},\ }\href {https://doi.org/10.1103/PhysRevMaterials.5.014409}
  {\bibfield  {journal} {\bibinfo  {journal} {Phys. Rev. Materials}\ }\textbf
  {\bibinfo {volume} {5}},\ \bibinfo {pages} {014409} (\bibinfo {year}
  {2021})}\BibitemShut {NoStop}%
\bibitem [{\citenamefont {\ifmmode~\check{S}\else \v{S}\fi{}mejkal}\ \emph
  {et~al.}(2022{\natexlab{a}})\citenamefont {\ifmmode~\check{S}\else
  \v{S}\fi{}mejkal}, \citenamefont {Sinova},\ and\ \citenamefont
  {Jungwirth}}]{Smejkal2022PRX}%
  \BibitemOpen
  \bibfield  {author} {\bibinfo {author} {\bibfnamefont {L.}~\bibnamefont
  {\ifmmode~\check{S}\else \v{S}\fi{}mejkal}}, \bibinfo {author} {\bibfnamefont
  {J.}~\bibnamefont {Sinova}},\ and\ \bibinfo {author} {\bibfnamefont
  {T.}~\bibnamefont {Jungwirth}},\ }\bibfield  {title} {\bibinfo {title}
  {Beyond conventional ferromagnetism and antiferromagnetism: A phase with
  nonrelativistic spin and crystal rotation symmetry},\ }\href
  {https://doi.org/10.1103/PhysRevX.12.031042} {\bibfield  {journal} {\bibinfo
  {journal} {Phys. Rev. X}\ }\textbf {\bibinfo {volume} {12}},\ \bibinfo
  {pages} {031042} (\bibinfo {year} {2022}{\natexlab{a}})}\BibitemShut
  {NoStop}%
\bibitem [{\citenamefont {Yuan}\ and\ \citenamefont
  {Zunger}(2023)}]{YuanZunger2023}%
  \BibitemOpen
  \bibfield  {author} {\bibinfo {author} {\bibfnamefont {L.-D.}\ \bibnamefont
  {Yuan}}\ and\ \bibinfo {author} {\bibfnamefont {A.}~\bibnamefont {Zunger}},\
  }\bibfield  {title} {\bibinfo {title} {Degeneracy removal of spin bands in
  collinear antiferromagnets with non-interconvertible spin-structure motif
  pair},\ }\href {https://doi.org/https://doi.org/10.1002/adma.202211966}
  {\bibfield  {journal} {\bibinfo  {journal} {Advanced Materials}\ }\textbf
  {\bibinfo {volume} {35}},\ \bibinfo {pages} {2211966} (\bibinfo {year}
  {2023})}\BibitemShut {NoStop}%
\bibitem [{\citenamefont {Guo}\ \emph {et~al.}(2023)\citenamefont {Guo},
  \citenamefont {Liu}, \citenamefont {Janson}, \citenamefont {Fulga},
  \citenamefont {{van den Brink}},\ and\ \citenamefont {Facio}}]{Guo2023}%
  \BibitemOpen
  \bibfield  {author} {\bibinfo {author} {\bibfnamefont {Y.}~\bibnamefont
  {Guo}}, \bibinfo {author} {\bibfnamefont {H.}~\bibnamefont {Liu}}, \bibinfo
  {author} {\bibfnamefont {O.}~\bibnamefont {Janson}}, \bibinfo {author}
  {\bibfnamefont {I.~C.}\ \bibnamefont {Fulga}}, \bibinfo {author}
  {\bibfnamefont {J.}~\bibnamefont {{van den Brink}}},\ and\ \bibinfo {author}
  {\bibfnamefont {J.~I.}\ \bibnamefont {Facio}},\ }\bibfield  {title} {\bibinfo
  {title} {Spin-split collinear antiferromagnets: A large-scale ab-initio
  study},\ }\href
  {https://doi.org/https://doi.org/10.1016/j.mtphys.2023.100991} {\bibfield
  {journal} {\bibinfo  {journal} {Mater. Today Phys.}\ }\textbf {\bibinfo
  {volume} {32}},\ \bibinfo {pages} {100991} (\bibinfo {year}
  {2023})}\BibitemShut {NoStop}%
\bibitem [{\citenamefont {Zeng}\ and\ \citenamefont {Zhao}(2024)}]{Zeng2024}%
  \BibitemOpen
  \bibfield  {author} {\bibinfo {author} {\bibfnamefont {S.}~\bibnamefont
  {Zeng}}\ and\ \bibinfo {author} {\bibfnamefont {Y.-J.}\ \bibnamefont
  {Zhao}},\ }\bibfield  {title} {\bibinfo {title} {Description of
  two-dimensional altermagnetism: Categorization using spin group theory},\
  }\href {https://doi.org/10.1103/PhysRevB.110.054406} {\bibfield  {journal}
  {\bibinfo  {journal} {Phys. Rev. B}\ }\textbf {\bibinfo {volume} {110}},\
  \bibinfo {pages} {054406} (\bibinfo {year} {2024})}\BibitemShut {NoStop}%
\bibitem [{\citenamefont {Lee}\ \emph {et~al.}(2024{\natexlab{a}})\citenamefont
  {Lee}, \citenamefont {Lee}, \citenamefont {Jung}, \citenamefont {Jung},
  \citenamefont {Kim}, \citenamefont {Lee}, \citenamefont {Seok}, \citenamefont
  {Kim}, \citenamefont {Park}, \citenamefont {\ifmmode~\check{S}\else
  \v{S}\fi{}mejkal}, \citenamefont {Kang},\ and\ \citenamefont
  {Kim}}]{Lee2024PRL}%
  \BibitemOpen
  \bibfield  {author} {\bibinfo {author} {\bibfnamefont {S.}~\bibnamefont
  {Lee}}, \bibinfo {author} {\bibfnamefont {S.}~\bibnamefont {Lee}}, \bibinfo
  {author} {\bibfnamefont {S.}~\bibnamefont {Jung}}, \bibinfo {author}
  {\bibfnamefont {J.}~\bibnamefont {Jung}}, \bibinfo {author} {\bibfnamefont
  {D.}~\bibnamefont {Kim}}, \bibinfo {author} {\bibfnamefont {Y.}~\bibnamefont
  {Lee}}, \bibinfo {author} {\bibfnamefont {B.}~\bibnamefont {Seok}}, \bibinfo
  {author} {\bibfnamefont {J.}~\bibnamefont {Kim}}, \bibinfo {author}
  {\bibfnamefont {B.~G.}\ \bibnamefont {Park}}, \bibinfo {author}
  {\bibfnamefont {L.}~\bibnamefont {\ifmmode~\check{S}\else \v{S}\fi{}mejkal}},
  \bibinfo {author} {\bibfnamefont {C.-J.}\ \bibnamefont {Kang}},\ and\
  \bibinfo {author} {\bibfnamefont {C.}~\bibnamefont {Kim}},\ }\bibfield
  {title} {\bibinfo {title} {Broken {K}ramers degeneracy in altermagnetic
  {M}n{T}e},\ }\href {https://doi.org/10.1103/PhysRevLett.132.036702}
  {\bibfield  {journal} {\bibinfo  {journal} {Phys. Rev. Lett.}\ }\textbf
  {\bibinfo {volume} {132}},\ \bibinfo {pages} {036702} (\bibinfo {year}
  {2024}{\natexlab{a}})}\BibitemShut {NoStop}%
\bibitem [{\citenamefont {Krempask{\'y}}\ \emph {et~al.}(2024)\citenamefont
  {Krempask{\'y}}, \citenamefont {{\v S}mejkal}, \citenamefont {D'Souza},
  \citenamefont {Hajlaoui}, \citenamefont {Springholz}, \citenamefont
  {Uhl{\'\i}{\v r}ov{\'a}}, \citenamefont {Alarab}, \citenamefont
  {Constantinou}, \citenamefont {Strocov}, \citenamefont {Usanov},
  \citenamefont {Pudelko}, \citenamefont {Gonz{\'a}lez-Hern{\'a}ndez},
  \citenamefont {Birk~Hellenes}, \citenamefont {Jansa}, \citenamefont
  {Reichlov{\'a}}, \citenamefont {{\v S}ob{\'a}{\v n}}, \citenamefont
  {Gonzalez~Betancourt}, \citenamefont {Wadley}, \citenamefont {Sinova},
  \citenamefont {Kriegner}, \citenamefont {Min{\'a}r}, \citenamefont {Dil},\
  and\ \citenamefont {Jungwirth}}]{Krempask2024}%
  \BibitemOpen
  \bibfield  {author} {\bibinfo {author} {\bibfnamefont {J.}~\bibnamefont
  {Krempask{\'y}}}, \bibinfo {author} {\bibfnamefont {L.}~\bibnamefont {{\v
  S}mejkal}}, \bibinfo {author} {\bibfnamefont {S.~W.}\ \bibnamefont
  {D'Souza}}, \bibinfo {author} {\bibfnamefont {M.}~\bibnamefont {Hajlaoui}},
  \bibinfo {author} {\bibfnamefont {G.}~\bibnamefont {Springholz}}, \bibinfo
  {author} {\bibfnamefont {K.}~\bibnamefont {Uhl{\'\i}{\v r}ov{\'a}}}, \bibinfo
  {author} {\bibfnamefont {F.}~\bibnamefont {Alarab}}, \bibinfo {author}
  {\bibfnamefont {P.~C.}\ \bibnamefont {Constantinou}}, \bibinfo {author}
  {\bibfnamefont {V.}~\bibnamefont {Strocov}}, \bibinfo {author} {\bibfnamefont
  {D.}~\bibnamefont {Usanov}}, \bibinfo {author} {\bibfnamefont {W.~R.}\
  \bibnamefont {Pudelko}}, \bibinfo {author} {\bibfnamefont {R.}~\bibnamefont
  {Gonz{\'a}lez-Hern{\'a}ndez}}, \bibinfo {author} {\bibfnamefont
  {A.}~\bibnamefont {Birk~Hellenes}}, \bibinfo {author} {\bibfnamefont
  {Z.}~\bibnamefont {Jansa}}, \bibinfo {author} {\bibfnamefont
  {H.}~\bibnamefont {Reichlov{\'a}}}, \bibinfo {author} {\bibfnamefont
  {Z.}~\bibnamefont {{\v S}ob{\'a}{\v n}}}, \bibinfo {author} {\bibfnamefont
  {R.~D.}\ \bibnamefont {Gonzalez~Betancourt}}, \bibinfo {author}
  {\bibfnamefont {P.}~\bibnamefont {Wadley}}, \bibinfo {author} {\bibfnamefont
  {J.}~\bibnamefont {Sinova}}, \bibinfo {author} {\bibfnamefont
  {D.}~\bibnamefont {Kriegner}}, \bibinfo {author} {\bibfnamefont
  {J.}~\bibnamefont {Min{\'a}r}}, \bibinfo {author} {\bibfnamefont {J.~H.}\
  \bibnamefont {Dil}},\ and\ \bibinfo {author} {\bibfnamefont {T.}~\bibnamefont
  {Jungwirth}},\ }\bibfield  {title} {\bibinfo {title} {Altermagnetic lifting
  of {K}ramers spin degeneracy},\ }\href
  {https://doi.org/10.1038/s41586-023-06907-7} {\bibfield  {journal} {\bibinfo
  {journal} {Nature}\ }\textbf {\bibinfo {volume} {626}},\ \bibinfo {pages}
  {517} (\bibinfo {year} {2024})}\BibitemShut {NoStop}%
\bibitem [{\citenamefont {Reimers}\ \emph {et~al.}(2024)\citenamefont
  {Reimers}, \citenamefont {Odenbreit}, \citenamefont {Šmejkal}, \citenamefont
  {Strocov}, \citenamefont {Constantinou}, \citenamefont {Hellenes},
  \citenamefont {Jaeschke~Ubiergo}, \citenamefont {Campos}, \citenamefont
  {Bharadwaj}, \citenamefont {Chakraborty}, \citenamefont {Denneulin},
  \citenamefont {Shi}, \citenamefont {Dunin-Borkowski}, \citenamefont {Das},
  \citenamefont {Kläui}, \citenamefont {Sinova},\ and\ \citenamefont
  {Jourdan}}]{Reimers2024}%
  \BibitemOpen
  \bibfield  {author} {\bibinfo {author} {\bibfnamefont {S.}~\bibnamefont
  {Reimers}}, \bibinfo {author} {\bibfnamefont {L.}~\bibnamefont {Odenbreit}},
  \bibinfo {author} {\bibfnamefont {L.}~\bibnamefont {Šmejkal}}, \bibinfo
  {author} {\bibfnamefont {V.~N.}\ \bibnamefont {Strocov}}, \bibinfo {author}
  {\bibfnamefont {P.}~\bibnamefont {Constantinou}}, \bibinfo {author}
  {\bibfnamefont {A.~B.}\ \bibnamefont {Hellenes}}, \bibinfo {author}
  {\bibfnamefont {R.}~\bibnamefont {Jaeschke~Ubiergo}}, \bibinfo {author}
  {\bibfnamefont {W.~H.}\ \bibnamefont {Campos}}, \bibinfo {author}
  {\bibfnamefont {V.~K.}\ \bibnamefont {Bharadwaj}}, \bibinfo {author}
  {\bibfnamefont {A.}~\bibnamefont {Chakraborty}}, \bibinfo {author}
  {\bibfnamefont {T.}~\bibnamefont {Denneulin}}, \bibinfo {author}
  {\bibfnamefont {W.}~\bibnamefont {Shi}}, \bibinfo {author} {\bibfnamefont
  {R.~E.}\ \bibnamefont {Dunin-Borkowski}}, \bibinfo {author} {\bibfnamefont
  {S.}~\bibnamefont {Das}}, \bibinfo {author} {\bibfnamefont {M.}~\bibnamefont
  {Kläui}}, \bibinfo {author} {\bibfnamefont {J.}~\bibnamefont {Sinova}},\
  and\ \bibinfo {author} {\bibfnamefont {M.}~\bibnamefont {Jourdan}},\
  }\bibfield  {title} {\bibinfo {title} {Direct observation of altermagnetic
  band splitting in {CrSb} thin films},\ }\href
  {https://doi.org/10.1038/s41467-024-46476-5} {\bibfield  {journal} {\bibinfo
  {journal} {Nat. Commun.}\ }\textbf {\bibinfo {volume} {15}},\ \bibinfo
  {pages} {2116} (\bibinfo {year} {2024})}\BibitemShut {NoStop}%
\bibitem [{\citenamefont {Aoyama}\ and\ \citenamefont
  {Ohgushi}(2024)}]{Aoyama2024}%
  \BibitemOpen
  \bibfield  {author} {\bibinfo {author} {\bibfnamefont {T.}~\bibnamefont
  {Aoyama}}\ and\ \bibinfo {author} {\bibfnamefont {K.}~\bibnamefont
  {Ohgushi}},\ }\bibfield  {title} {\bibinfo {title} {Piezomagnetic properties
  in altermagnetic {M}n{T}e},\ }\href
  {https://doi.org/10.1103/PhysRevMaterials.8.L041402} {\bibfield  {journal}
  {\bibinfo  {journal} {Phys. Rev. Mater.}\ }\textbf {\bibinfo {volume} {8}},\
  \bibinfo {pages} {L041402} (\bibinfo {year} {2024})}\BibitemShut {NoStop}%
\bibitem [{\citenamefont {Lin}\ \emph {et~al.}(2024)\citenamefont {Lin},
  \citenamefont {Chen}, \citenamefont {Lu}, \citenamefont {Liang},
  \citenamefont {Feng}, \citenamefont {Yamagami}, \citenamefont {Osiecki},
  \citenamefont {Leandersson}, \citenamefont {Thiagarajan}, \citenamefont
  {Liu}, \citenamefont {Felser},\ and\ \citenamefont {Ma}}]{Lin2024}%
  \BibitemOpen
  \bibfield  {author} {\bibinfo {author} {\bibfnamefont {Z.}~\bibnamefont
  {Lin}}, \bibinfo {author} {\bibfnamefont {D.}~\bibnamefont {Chen}}, \bibinfo
  {author} {\bibfnamefont {W.}~\bibnamefont {Lu}}, \bibinfo {author}
  {\bibfnamefont {X.}~\bibnamefont {Liang}}, \bibinfo {author} {\bibfnamefont
  {S.}~\bibnamefont {Feng}}, \bibinfo {author} {\bibfnamefont {K.}~\bibnamefont
  {Yamagami}}, \bibinfo {author} {\bibfnamefont {J.}~\bibnamefont {Osiecki}},
  \bibinfo {author} {\bibfnamefont {M.}~\bibnamefont {Leandersson}}, \bibinfo
  {author} {\bibfnamefont {B.}~\bibnamefont {Thiagarajan}}, \bibinfo {author}
  {\bibfnamefont {J.}~\bibnamefont {Liu}}, \bibinfo {author} {\bibfnamefont
  {C.}~\bibnamefont {Felser}},\ and\ \bibinfo {author} {\bibfnamefont
  {J.}~\bibnamefont {Ma}},\ }\href@noop {} {\bibinfo {title} {Observation of
  giant spin splitting and d-wave spin texture in room temperature altermagnet
  {RuO$_2$}}} (\bibinfo {year} {2024}),\ \Eprint
  {https://arxiv.org/abs/2402.04995} {arXiv:2402.04995 [cond-mat.mtrl-sci]}
  \BibitemShut {NoStop}%
\bibitem [{\citenamefont {Ahn}\ \emph {et~al.}(2019)\citenamefont {Ahn},
  \citenamefont {Hariki}, \citenamefont {Lee},\ and\ \citenamefont
  {Kune\ifmmode~\check{s}\else \v{s}\fi{}}}]{Kyo-Hoon2019}%
  \BibitemOpen
  \bibfield  {author} {\bibinfo {author} {\bibfnamefont {K.-H.}\ \bibnamefont
  {Ahn}}, \bibinfo {author} {\bibfnamefont {A.}~\bibnamefont {Hariki}},
  \bibinfo {author} {\bibfnamefont {K.-W.}\ \bibnamefont {Lee}},\ and\ \bibinfo
  {author} {\bibfnamefont {J.}~\bibnamefont {Kune\ifmmode~\check{s}\else
  \v{s}\fi{}}},\ }\bibfield  {title} {\bibinfo {title} {Antiferromagnetism in
  {RuO}$_{2}$ as $d$-wave {Pomeranchuk} instability},\ }\href
  {https://doi.org/10.1103/PhysRevB.99.184432} {\bibfield  {journal} {\bibinfo
  {journal} {Phys. Rev. B}\ }\textbf {\bibinfo {volume} {99}},\ \bibinfo
  {pages} {184432} (\bibinfo {year} {2019})}\BibitemShut {NoStop}%
\bibitem [{\citenamefont {Šmejkal}\ \emph {et~al.}(2020)\citenamefont
  {Šmejkal}, \citenamefont {González-Hernández}, \citenamefont {Jungwirth},\
  and\ \citenamefont {Sinova}}]{Libor2020}%
  \BibitemOpen
  \bibfield  {author} {\bibinfo {author} {\bibfnamefont {L.}~\bibnamefont
  {Šmejkal}}, \bibinfo {author} {\bibfnamefont {R.}~\bibnamefont
  {González-Hernández}}, \bibinfo {author} {\bibfnamefont {T.}~\bibnamefont
  {Jungwirth}},\ and\ \bibinfo {author} {\bibfnamefont {J.}~\bibnamefont
  {Sinova}},\ }\bibfield  {title} {\bibinfo {title} {Crystal time-reversal
  symmetry breaking and spontaneous {H}all effect in collinear
  antiferromagnets},\ }\href {https://doi.org/10.1126/sciadv.aaz8809}
  {\bibfield  {journal} {\bibinfo  {journal} {Sci. Adv.}\ }\textbf {\bibinfo
  {volume} {6}},\ \bibinfo {pages} {eaaz8809} (\bibinfo {year} {2020})},\
  \Eprint
  {https://arxiv.org/abs/https://www.science.org/doi/pdf/10.1126/sciadv.aaz8809}
  {https://www.science.org/doi/pdf/10.1126/sciadv.aaz8809} \BibitemShut
  {NoStop}%
\bibitem [{\citenamefont {\ifmmode~\check{S}\else \v{S}\fi{}mejkal}\ \emph
  {et~al.}(2022{\natexlab{b}})\citenamefont {\ifmmode~\check{S}\else
  \v{S}\fi{}mejkal}, \citenamefont {Sinova},\ and\ \citenamefont
  {Jungwirth}}]{Libor2022Review}%
  \BibitemOpen
  \bibfield  {author} {\bibinfo {author} {\bibfnamefont {L.}~\bibnamefont
  {\ifmmode~\check{S}\else \v{S}\fi{}mejkal}}, \bibinfo {author} {\bibfnamefont
  {J.}~\bibnamefont {Sinova}},\ and\ \bibinfo {author} {\bibfnamefont
  {T.}~\bibnamefont {Jungwirth}},\ }\bibfield  {title} {\bibinfo {title}
  {Emerging research landscape of altermagnetism},\ }\href
  {https://doi.org/10.1103/PhysRevX.12.040501} {\bibfield  {journal} {\bibinfo
  {journal} {Phys. Rev. X}\ }\textbf {\bibinfo {volume} {12}},\ \bibinfo
  {pages} {040501} (\bibinfo {year} {2022}{\natexlab{b}})}\BibitemShut
  {NoStop}%
\bibitem [{\citenamefont {McClarty}\ and\ \citenamefont
  {Rau}(2024)}]{Paul2024}%
  \BibitemOpen
  \bibfield  {author} {\bibinfo {author} {\bibfnamefont {P.~A.}\ \bibnamefont
  {McClarty}}\ and\ \bibinfo {author} {\bibfnamefont {J.~G.}\ \bibnamefont
  {Rau}},\ }\bibfield  {title} {\bibinfo {title} {Landau theory of
  altermagnetism},\ }\href {https://doi.org/10.1103/PhysRevLett.132.176702}
  {\bibfield  {journal} {\bibinfo  {journal} {Phys. Rev. Lett.}\ }\textbf
  {\bibinfo {volume} {132}},\ \bibinfo {pages} {176702} (\bibinfo {year}
  {2024})}\BibitemShut {NoStop}%
\bibitem [{\citenamefont {Bai}\ \emph {et~al.}(2024)\citenamefont {Bai},
  \citenamefont {Feng}, \citenamefont {Liu}, \citenamefont {Šmejkal},
  \citenamefont {Mokrousov},\ and\ \citenamefont {Yao}}]{Bai2024}%
  \BibitemOpen
  \bibfield  {author} {\bibinfo {author} {\bibfnamefont {L.}~\bibnamefont
  {Bai}}, \bibinfo {author} {\bibfnamefont {W.}~\bibnamefont {Feng}}, \bibinfo
  {author} {\bibfnamefont {S.}~\bibnamefont {Liu}}, \bibinfo {author}
  {\bibfnamefont {L.}~\bibnamefont {Šmejkal}}, \bibinfo {author}
  {\bibfnamefont {Y.}~\bibnamefont {Mokrousov}},\ and\ \bibinfo {author}
  {\bibfnamefont {Y.}~\bibnamefont {Yao}},\ }\href
  {https://arxiv.org/abs/2406.02123} {\bibinfo {title} {Altermagnetism:
  Exploring new frontiers in magnetism and spintronics}} (\bibinfo {year}
  {2024}),\ \Eprint {https://arxiv.org/abs/2406.02123} {arXiv:2406.02123
  [cond-mat.mtrl-sci]} \BibitemShut {NoStop}%
\bibitem [{\citenamefont {Naka}\ \emph {et~al.}(2019)\citenamefont {Naka},
  \citenamefont {Hayami}, \citenamefont {Kusunose}, \citenamefont {Yanagi},
  \citenamefont {Motome},\ and\ \citenamefont {Seo}}]{Naka2019}%
  \BibitemOpen
  \bibfield  {author} {\bibinfo {author} {\bibfnamefont {M.}~\bibnamefont
  {Naka}}, \bibinfo {author} {\bibfnamefont {S.}~\bibnamefont {Hayami}},
  \bibinfo {author} {\bibfnamefont {H.}~\bibnamefont {Kusunose}}, \bibinfo
  {author} {\bibfnamefont {Y.}~\bibnamefont {Yanagi}}, \bibinfo {author}
  {\bibfnamefont {Y.}~\bibnamefont {Motome}},\ and\ \bibinfo {author}
  {\bibfnamefont {H.}~\bibnamefont {Seo}},\ }\bibfield  {title} {\bibinfo
  {title} {Spin current generation in organic antiferromagnets},\ }\href
  {https://doi.org/10.1038/s41467-019-12229-y} {\bibfield  {journal} {\bibinfo
  {journal} {Nat. Commun.}\ }\textbf {\bibinfo {volume} {10}},\ \bibinfo
  {pages} {4305} (\bibinfo {year} {2019})}\BibitemShut {NoStop}%
\bibitem [{\citenamefont {Gonz\'alez-Hern\'andez}\ \emph
  {et~al.}(2021)\citenamefont {Gonz\'alez-Hern\'andez}, \citenamefont
  {\ifmmode~\check{S}\else \v{S}\fi{}mejkal}, \citenamefont {V\'yborn\'y},
  \citenamefont {Yahagi}, \citenamefont {Sinova}, \citenamefont {Jungwirth},\
  and\ \citenamefont {\ifmmode~\check{Z}\else
  \v{Z}\fi{}elezn\'y}}]{Hernandez2021}%
  \BibitemOpen
  \bibfield  {author} {\bibinfo {author} {\bibfnamefont {R.}~\bibnamefont
  {Gonz\'alez-Hern\'andez}}, \bibinfo {author} {\bibfnamefont {L.}~\bibnamefont
  {\ifmmode~\check{S}\else \v{S}\fi{}mejkal}}, \bibinfo {author} {\bibfnamefont
  {K.}~\bibnamefont {V\'yborn\'y}}, \bibinfo {author} {\bibfnamefont
  {Y.}~\bibnamefont {Yahagi}}, \bibinfo {author} {\bibfnamefont
  {J.}~\bibnamefont {Sinova}}, \bibinfo {author} {\bibfnamefont
  {T.}~\bibnamefont {Jungwirth}},\ and\ \bibinfo {author} {\bibfnamefont
  {J.}~\bibnamefont {\ifmmode~\check{Z}\else \v{Z}\fi{}elezn\'y}},\ }\bibfield
  {title} {\bibinfo {title} {Efficient electrical spin splitter based on
  nonrelativistic collinear antiferromagnetism},\ }\href
  {https://doi.org/10.1103/PhysRevLett.126.127701} {\bibfield  {journal}
  {\bibinfo  {journal} {Phys. Rev. Lett.}\ }\textbf {\bibinfo {volume} {126}},\
  \bibinfo {pages} {127701} (\bibinfo {year} {2021})}\BibitemShut {NoStop}%
\bibitem [{\citenamefont {Shao}\ \emph {et~al.}(2021)\citenamefont {Shao},
  \citenamefont {Zhang}, \citenamefont {Li}, \citenamefont {Eom},\ and\
  \citenamefont {Tsymbal}}]{Shao2021}%
  \BibitemOpen
  \bibfield  {author} {\bibinfo {author} {\bibfnamefont {D.-F.}\ \bibnamefont
  {Shao}}, \bibinfo {author} {\bibfnamefont {S.-H.}\ \bibnamefont {Zhang}},
  \bibinfo {author} {\bibfnamefont {M.}~\bibnamefont {Li}}, \bibinfo {author}
  {\bibfnamefont {C.-B.}\ \bibnamefont {Eom}},\ and\ \bibinfo {author}
  {\bibfnamefont {E.~Y.}\ \bibnamefont {Tsymbal}},\ }\bibfield  {title}
  {\bibinfo {title} {Spin-neutral currents for spintronics},\ }\href
  {https://doi.org/10.1038/s41467-021-26915-3} {\bibfield  {journal} {\bibinfo
  {journal} {Nat. Commun.}\ }\textbf {\bibinfo {volume} {12}},\ \bibinfo
  {pages} {7061} (\bibinfo {year} {2021})}\BibitemShut {NoStop}%
\bibitem [{\citenamefont {Bose}\ \emph {et~al.}(2022)\citenamefont {Bose},
  \citenamefont {Schreiber}, \citenamefont {Jain}, \citenamefont {Shao},
  \citenamefont {Nair}, \citenamefont {Sun}, \citenamefont {Zhang},
  \citenamefont {Muller}, \citenamefont {Tsymbal}, \citenamefont {Schlom},\
  and\ \citenamefont {Ralph}}]{Bose2022}%
  \BibitemOpen
  \bibfield  {author} {\bibinfo {author} {\bibfnamefont {A.}~\bibnamefont
  {Bose}}, \bibinfo {author} {\bibfnamefont {N.~J.}\ \bibnamefont {Schreiber}},
  \bibinfo {author} {\bibfnamefont {R.}~\bibnamefont {Jain}}, \bibinfo {author}
  {\bibfnamefont {D.-F.}\ \bibnamefont {Shao}}, \bibinfo {author}
  {\bibfnamefont {H.~P.}\ \bibnamefont {Nair}}, \bibinfo {author}
  {\bibfnamefont {J.}~\bibnamefont {Sun}}, \bibinfo {author} {\bibfnamefont
  {X.~S.}\ \bibnamefont {Zhang}}, \bibinfo {author} {\bibfnamefont {D.~A.}\
  \bibnamefont {Muller}}, \bibinfo {author} {\bibfnamefont {E.~Y.}\
  \bibnamefont {Tsymbal}}, \bibinfo {author} {\bibfnamefont {D.~G.}\
  \bibnamefont {Schlom}},\ and\ \bibinfo {author} {\bibfnamefont {D.~C.}\
  \bibnamefont {Ralph}},\ }\bibfield  {title} {\bibinfo {title} {Tilted spin
  current generated by the collinear antiferromagnet ruthenium dioxide},\
  }\href {https://doi.org/10.1038/s41928-022-00744-8} {\bibfield  {journal}
  {\bibinfo  {journal} {Nat. Electron.}\ }\textbf {\bibinfo {volume} {5}},\
  \bibinfo {pages} {267} (\bibinfo {year} {2022})}\BibitemShut {NoStop}%
\bibitem [{\citenamefont {Hu}\ \emph {et~al.}(2024)\citenamefont {Hu},
  \citenamefont {Janson}, \citenamefont {Felser}, \citenamefont {McClarty},
  \citenamefont {van~den Brink},\ and\ \citenamefont {Vergniory}}]{Hu2024}%
  \BibitemOpen
  \bibfield  {author} {\bibinfo {author} {\bibfnamefont {M.}~\bibnamefont
  {Hu}}, \bibinfo {author} {\bibfnamefont {O.}~\bibnamefont {Janson}}, \bibinfo
  {author} {\bibfnamefont {C.}~\bibnamefont {Felser}}, \bibinfo {author}
  {\bibfnamefont {P.}~\bibnamefont {McClarty}}, \bibinfo {author}
  {\bibfnamefont {J.}~\bibnamefont {van~den Brink}},\ and\ \bibinfo {author}
  {\bibfnamefont {M.~G.}\ \bibnamefont {Vergniory}},\ }\href
  {https://arxiv.org/abs/2410.17993} {\bibinfo {title} {Spin {H}all and
  edelstein effects in novel chiral noncollinear altermagnets}} (\bibinfo
  {year} {2024}),\ \Eprint {https://arxiv.org/abs/2410.17993} {arXiv:2410.17993
  [cond-mat.mtrl-sci]} \BibitemShut {NoStop}%
\bibitem [{\citenamefont {Bai}\ \emph {et~al.}(2022)\citenamefont {Bai},
  \citenamefont {Han}, \citenamefont {Feng}, \citenamefont {Zhou},
  \citenamefont {Su}, \citenamefont {Wang}, \citenamefont {Liao}, \citenamefont
  {Zhu}, \citenamefont {Chen}, \citenamefont {Pan}, \citenamefont {Fan},\ and\
  \citenamefont {Song}}]{Bai2022}%
  \BibitemOpen
  \bibfield  {author} {\bibinfo {author} {\bibfnamefont {H.}~\bibnamefont
  {Bai}}, \bibinfo {author} {\bibfnamefont {L.}~\bibnamefont {Han}}, \bibinfo
  {author} {\bibfnamefont {X.~Y.}\ \bibnamefont {Feng}}, \bibinfo {author}
  {\bibfnamefont {Y.~J.}\ \bibnamefont {Zhou}}, \bibinfo {author}
  {\bibfnamefont {R.~X.}\ \bibnamefont {Su}}, \bibinfo {author} {\bibfnamefont
  {Q.}~\bibnamefont {Wang}}, \bibinfo {author} {\bibfnamefont {L.~Y.}\
  \bibnamefont {Liao}}, \bibinfo {author} {\bibfnamefont {W.~X.}\ \bibnamefont
  {Zhu}}, \bibinfo {author} {\bibfnamefont {X.~Z.}\ \bibnamefont {Chen}},
  \bibinfo {author} {\bibfnamefont {F.}~\bibnamefont {Pan}}, \bibinfo {author}
  {\bibfnamefont {X.~L.}\ \bibnamefont {Fan}},\ and\ \bibinfo {author}
  {\bibfnamefont {C.}~\bibnamefont {Song}},\ }\bibfield  {title} {\bibinfo
  {title} {Observation of spin splitting torque in a collinear antiferromagnet
  {RuO}$_{2}$},\ }\href {https://doi.org/10.1103/PhysRevLett.128.197202}
  {\bibfield  {journal} {\bibinfo  {journal} {Phys. Rev. Lett.}\ }\textbf
  {\bibinfo {volume} {128}},\ \bibinfo {pages} {197202} (\bibinfo {year}
  {2022})}\BibitemShut {NoStop}%
\bibitem [{\citenamefont {Karube}\ \emph {et~al.}(2022)\citenamefont {Karube},
  \citenamefont {Tanaka}, \citenamefont {Sugawara}, \citenamefont {Kadoguchi},
  \citenamefont {Kohda},\ and\ \citenamefont {Nitta}}]{Karube2021}%
  \BibitemOpen
  \bibfield  {author} {\bibinfo {author} {\bibfnamefont {S.}~\bibnamefont
  {Karube}}, \bibinfo {author} {\bibfnamefont {T.}~\bibnamefont {Tanaka}},
  \bibinfo {author} {\bibfnamefont {D.}~\bibnamefont {Sugawara}}, \bibinfo
  {author} {\bibfnamefont {N.}~\bibnamefont {Kadoguchi}}, \bibinfo {author}
  {\bibfnamefont {M.}~\bibnamefont {Kohda}},\ and\ \bibinfo {author}
  {\bibfnamefont {J.}~\bibnamefont {Nitta}},\ }\bibfield  {title} {\bibinfo
  {title} {Observation of spin-splitter torque in collinear antiferromagnetic
  {RuO}$_{2}$},\ }\href {https://doi.org/10.1103/PhysRevLett.129.137201}
  {\bibfield  {journal} {\bibinfo  {journal} {Phys. Rev. Lett.}\ }\textbf
  {\bibinfo {volume} {129}},\ \bibinfo {pages} {137201} (\bibinfo {year}
  {2022})}\BibitemShut {NoStop}%
\bibitem [{\citenamefont {\ifmmode~\check{S}\else \v{S}\fi{}mejkal}\ \emph
  {et~al.}(2022{\natexlab{c}})\citenamefont {\ifmmode~\check{S}\else
  \v{S}\fi{}mejkal}, \citenamefont {Hellenes}, \citenamefont
  {Gonz\'alez-Hern\'andez}, \citenamefont {Sinova},\ and\ \citenamefont
  {Jungwirth}}]{Libor2022}%
  \BibitemOpen
  \bibfield  {author} {\bibinfo {author} {\bibfnamefont {L.}~\bibnamefont
  {\ifmmode~\check{S}\else \v{S}\fi{}mejkal}}, \bibinfo {author} {\bibfnamefont
  {A.~B.}\ \bibnamefont {Hellenes}}, \bibinfo {author} {\bibfnamefont
  {R.}~\bibnamefont {Gonz\'alez-Hern\'andez}}, \bibinfo {author} {\bibfnamefont
  {J.}~\bibnamefont {Sinova}},\ and\ \bibinfo {author} {\bibfnamefont
  {T.}~\bibnamefont {Jungwirth}},\ }\bibfield  {title} {\bibinfo {title} {Giant
  and tunneling magnetoresistance in unconventional collinear antiferromagnets
  with nonrelativistic spin-momentum coupling},\ }\href
  {https://doi.org/10.1103/PhysRevX.12.011028} {\bibfield  {journal} {\bibinfo
  {journal} {Phys. Rev. X}\ }\textbf {\bibinfo {volume} {12}},\ \bibinfo
  {pages} {011028} (\bibinfo {year} {2022}{\natexlab{c}})}\BibitemShut
  {NoStop}%
\bibitem [{\citenamefont {{\v S}mejkal}\ \emph
  {et~al.}(2022{\natexlab{a}})\citenamefont {{\v S}mejkal}, \citenamefont
  {MacDonald}, \citenamefont {Sinova}, \citenamefont {Nakatsuji},\ and\
  \citenamefont {Jungwirth}}]{Libor2022NatRev}%
  \BibitemOpen
  \bibfield  {author} {\bibinfo {author} {\bibfnamefont {L.}~\bibnamefont {{\v
  S}mejkal}}, \bibinfo {author} {\bibfnamefont {A.~H.}\ \bibnamefont
  {MacDonald}}, \bibinfo {author} {\bibfnamefont {J.}~\bibnamefont {Sinova}},
  \bibinfo {author} {\bibfnamefont {S.}~\bibnamefont {Nakatsuji}},\ and\
  \bibinfo {author} {\bibfnamefont {T.}~\bibnamefont {Jungwirth}},\ }\bibfield
  {title} {\bibinfo {title} {Anomalous {H}all antiferromagnets},\ }\href
  {https://doi.org/10.1038/s41578-022-00430-3} {\bibfield  {journal} {\bibinfo
  {journal} {Nature Reviews Materials}\ }\textbf {\bibinfo {volume} {7}},\
  \bibinfo {pages} {482} (\bibinfo {year} {2022}{\natexlab{a}})}\BibitemShut
  {NoStop}%
\bibitem [{\citenamefont {Reichlová}\ \emph {et~al.}(2020)\citenamefont
  {Reichlová}, \citenamefont {Seeger}, \citenamefont {González-Hernández},
  \citenamefont {Kounta}, \citenamefont {Schlitz}, \citenamefont {Kriegner},
  \citenamefont {Ritzinger}, \citenamefont {Lammel}, \citenamefont {Leiviskä},
  \citenamefont {Petříček}, \citenamefont {Doležal}, \citenamefont
  {Schmoranzerová}, \citenamefont {Bad'ura}, \citenamefont {Thomas},
  \citenamefont {Baltz}, \citenamefont {Michez}, \citenamefont {Sinova},
  \citenamefont {Goennenwein}, \citenamefont {Jungwirth},\ and\ \citenamefont
  {Šmejkal}}]{Helena2020}%
  \BibitemOpen
  \bibfield  {author} {\bibinfo {author} {\bibfnamefont {H.}~\bibnamefont
  {Reichlová}}, \bibinfo {author} {\bibfnamefont {R.~L.}\ \bibnamefont
  {Seeger}}, \bibinfo {author} {\bibfnamefont {R.}~\bibnamefont
  {González-Hernández}}, \bibinfo {author} {\bibfnamefont {I.}~\bibnamefont
  {Kounta}}, \bibinfo {author} {\bibfnamefont {R.}~\bibnamefont {Schlitz}},
  \bibinfo {author} {\bibfnamefont {D.}~\bibnamefont {Kriegner}}, \bibinfo
  {author} {\bibfnamefont {P.}~\bibnamefont {Ritzinger}}, \bibinfo {author}
  {\bibfnamefont {M.}~\bibnamefont {Lammel}}, \bibinfo {author} {\bibfnamefont
  {M.}~\bibnamefont {Leiviskä}}, \bibinfo {author} {\bibfnamefont
  {V.}~\bibnamefont {Petříček}}, \bibinfo {author} {\bibfnamefont
  {P.}~\bibnamefont {Doležal}}, \bibinfo {author} {\bibfnamefont
  {E.}~\bibnamefont {Schmoranzerová}}, \bibinfo {author} {\bibfnamefont
  {A.}~\bibnamefont {Bad'ura}}, \bibinfo {author} {\bibfnamefont
  {A.}~\bibnamefont {Thomas}}, \bibinfo {author} {\bibfnamefont
  {V.}~\bibnamefont {Baltz}}, \bibinfo {author} {\bibfnamefont
  {L.}~\bibnamefont {Michez}}, \bibinfo {author} {\bibfnamefont
  {J.}~\bibnamefont {Sinova}}, \bibinfo {author} {\bibfnamefont {S.~T.~B.}\
  \bibnamefont {Goennenwein}}, \bibinfo {author} {\bibfnamefont
  {T.}~\bibnamefont {Jungwirth}},\ and\ \bibinfo {author} {\bibfnamefont
  {L.}~\bibnamefont {Šmejkal}},\ }\bibfield  {title} {\bibinfo {title}
  {Macroscopic time reversal symmetry breaking by staggered spin-momentum
  interaction},\ }\href {https://arxiv.org/abs/2012.15651} {\bibfield
  {journal} {\bibinfo  {journal} {arXiv 2012.15651}\ } (\bibinfo {year}
  {2020})}\BibitemShut {NoStop}%
\bibitem [{\citenamefont {Feng}\ \emph {et~al.}(2022)\citenamefont {Feng},
  \citenamefont {Zhou}, \citenamefont {{\v S}mejkal}, \citenamefont {Wu},
  \citenamefont {Zhu}, \citenamefont {Guo}, \citenamefont
  {Gonz{\'a}lez-Hern{\'a}ndez}, \citenamefont {Wang}, \citenamefont {Yan},
  \citenamefont {Qin}, \citenamefont {Zhang}, \citenamefont {Wu}, \citenamefont
  {Chen}, \citenamefont {Meng}, \citenamefont {Liu}, \citenamefont {Xia},
  \citenamefont {Sinova}, \citenamefont {Jungwirth},\ and\ \citenamefont
  {Liu}}]{Feng2020}%
  \BibitemOpen
  \bibfield  {author} {\bibinfo {author} {\bibfnamefont {Z.}~\bibnamefont
  {Feng}}, \bibinfo {author} {\bibfnamefont {X.}~\bibnamefont {Zhou}}, \bibinfo
  {author} {\bibfnamefont {L.}~\bibnamefont {{\v S}mejkal}}, \bibinfo {author}
  {\bibfnamefont {L.}~\bibnamefont {Wu}}, \bibinfo {author} {\bibfnamefont
  {Z.}~\bibnamefont {Zhu}}, \bibinfo {author} {\bibfnamefont {H.}~\bibnamefont
  {Guo}}, \bibinfo {author} {\bibfnamefont {R.}~\bibnamefont
  {Gonz{\'a}lez-Hern{\'a}ndez}}, \bibinfo {author} {\bibfnamefont
  {X.}~\bibnamefont {Wang}}, \bibinfo {author} {\bibfnamefont {H.}~\bibnamefont
  {Yan}}, \bibinfo {author} {\bibfnamefont {P.}~\bibnamefont {Qin}}, \bibinfo
  {author} {\bibfnamefont {X.}~\bibnamefont {Zhang}}, \bibinfo {author}
  {\bibfnamefont {H.}~\bibnamefont {Wu}}, \bibinfo {author} {\bibfnamefont
  {H.}~\bibnamefont {Chen}}, \bibinfo {author} {\bibfnamefont {Z.}~\bibnamefont
  {Meng}}, \bibinfo {author} {\bibfnamefont {L.}~\bibnamefont {Liu}}, \bibinfo
  {author} {\bibfnamefont {Z.}~\bibnamefont {Xia}}, \bibinfo {author}
  {\bibfnamefont {J.}~\bibnamefont {Sinova}}, \bibinfo {author} {\bibfnamefont
  {T.}~\bibnamefont {Jungwirth}},\ and\ \bibinfo {author} {\bibfnamefont
  {Z.}~\bibnamefont {Liu}},\ }\bibfield  {title} {\bibinfo {title} {An
  anomalous {H}all effect in altermagnetic ruthenium dioxide},\ }\href
  {https://doi.org/10.1038/s41928-022-00866-z} {\bibfield  {journal} {\bibinfo
  {journal} {Nat. Electron.}\ }\textbf {\bibinfo {volume} {5}},\ \bibinfo
  {pages} {735} (\bibinfo {year} {2022})}\BibitemShut {NoStop}%
\bibitem [{\citenamefont {Gonzalez~Betancourt}\ \emph
  {et~al.}(2023)\citenamefont {Gonzalez~Betancourt}, \citenamefont
  {Zub\'a\ifmmode~\check{c}\else \v{c}\fi{}}, \citenamefont
  {Gonzalez-Hernandez}, \citenamefont {Geishendorf}, \citenamefont {\ifmmode
  \check{S}\else \v{S}\fi{}ob\'a\ifmmode~\check{n}\else \v{n}\fi{}},
  \citenamefont {Springholz}, \citenamefont {Olejn\'{\i}k}, \citenamefont
  {\ifmmode~\check{S}\else \v{S}\fi{}mejkal}, \citenamefont {Sinova},
  \citenamefont {Jungwirth}, \citenamefont {Goennenwein}, \citenamefont
  {Thomas}, \citenamefont {Reichlov\'a}, \citenamefont {\ifmmode~\check{Z}\else
  \v{Z}\fi{}elezn\'y},\ and\ \citenamefont {Kriegner}}]{Betancourt2021}%
  \BibitemOpen
  \bibfield  {author} {\bibinfo {author} {\bibfnamefont {R.~D.}\ \bibnamefont
  {Gonzalez~Betancourt}}, \bibinfo {author} {\bibfnamefont {J.}~\bibnamefont
  {Zub\'a\ifmmode~\check{c}\else \v{c}\fi{}}}, \bibinfo {author} {\bibfnamefont
  {R.}~\bibnamefont {Gonzalez-Hernandez}}, \bibinfo {author} {\bibfnamefont
  {K.}~\bibnamefont {Geishendorf}}, \bibinfo {author} {\bibfnamefont
  {Z.}~\bibnamefont {\ifmmode \check{S}\else
  \v{S}\fi{}ob\'a\ifmmode~\check{n}\else \v{n}\fi{}}}, \bibinfo {author}
  {\bibfnamefont {G.}~\bibnamefont {Springholz}}, \bibinfo {author}
  {\bibfnamefont {K.}~\bibnamefont {Olejn\'{\i}k}}, \bibinfo {author}
  {\bibfnamefont {L.}~\bibnamefont {\ifmmode~\check{S}\else \v{S}\fi{}mejkal}},
  \bibinfo {author} {\bibfnamefont {J.}~\bibnamefont {Sinova}}, \bibinfo
  {author} {\bibfnamefont {T.}~\bibnamefont {Jungwirth}}, \bibinfo {author}
  {\bibfnamefont {S.~T.~B.}\ \bibnamefont {Goennenwein}}, \bibinfo {author}
  {\bibfnamefont {A.}~\bibnamefont {Thomas}}, \bibinfo {author} {\bibfnamefont
  {H.}~\bibnamefont {Reichlov\'a}}, \bibinfo {author} {\bibfnamefont
  {J.}~\bibnamefont {\ifmmode~\check{Z}\else \v{Z}\fi{}elezn\'y}},\ and\
  \bibinfo {author} {\bibfnamefont {D.}~\bibnamefont {Kriegner}},\ }\bibfield
  {title} {\bibinfo {title} {Spontaneous anomalous {H}all effect arising from
  an unconventional compensated magnetic phase in a semiconductor},\ }\href
  {https://doi.org/10.1103/PhysRevLett.130.036702} {\bibfield  {journal}
  {\bibinfo  {journal} {Phys. Rev. Lett.}\ }\textbf {\bibinfo {volume} {130}},\
  \bibinfo {pages} {036702} (\bibinfo {year} {2023})}\BibitemShut {NoStop}%
\bibitem [{\citenamefont {{\v S}mejkal}\ \emph
  {et~al.}(2022{\natexlab{b}})\citenamefont {{\v S}mejkal}, \citenamefont
  {MacDonald}, \citenamefont {Sinova}, \citenamefont {Nakatsuji},\ and\
  \citenamefont {Jungwirth}}]{SmejkalAHE2022}%
  \BibitemOpen
  \bibfield  {author} {\bibinfo {author} {\bibfnamefont {L.}~\bibnamefont {{\v
  S}mejkal}}, \bibinfo {author} {\bibfnamefont {A.~H.}\ \bibnamefont
  {MacDonald}}, \bibinfo {author} {\bibfnamefont {J.}~\bibnamefont {Sinova}},
  \bibinfo {author} {\bibfnamefont {S.}~\bibnamefont {Nakatsuji}},\ and\
  \bibinfo {author} {\bibfnamefont {T.}~\bibnamefont {Jungwirth}},\ }\bibfield
  {title} {\bibinfo {title} {Anomalous {H}all antiferromagnets},\ }\href
  {https://doi.org/10.1038/s41578-022-00430-3} {\bibfield  {journal} {\bibinfo
  {journal} {Nat. Rev. Mater.}\ }\textbf {\bibinfo {volume} {7}},\ \bibinfo
  {pages} {482} (\bibinfo {year} {2022}{\natexlab{b}})}\BibitemShut {NoStop}%
\bibitem [{\citenamefont {Cheong}\ and\ \citenamefont
  {Huang}(2024)}]{Cheong2024}%
  \BibitemOpen
  \bibfield  {author} {\bibinfo {author} {\bibfnamefont {S.-W.}\ \bibnamefont
  {Cheong}}\ and\ \bibinfo {author} {\bibfnamefont {F.-T.}\ \bibnamefont
  {Huang}},\ }\bibfield  {title} {\bibinfo {title} {Altermagnetism with
  non-collinear spins},\ }\href {https://doi.org/10.1038/s41535-024-00626-6}
  {\bibfield  {journal} {\bibinfo  {journal} {npj Quantum Materials}\ }\textbf
  {\bibinfo {volume} {9}},\ \bibinfo {pages} {13} (\bibinfo {year}
  {2024})}\BibitemShut {NoStop}%
\bibitem [{\citenamefont {Sato}\ \emph {et~al.}(2024)\citenamefont {Sato},
  \citenamefont {Haddad}, \citenamefont {Fulga}, \citenamefont {Assaad},\ and\
  \citenamefont {van~den Brink}}]{Sato2024}%
  \BibitemOpen
  \bibfield  {author} {\bibinfo {author} {\bibfnamefont {T.}~\bibnamefont
  {Sato}}, \bibinfo {author} {\bibfnamefont {S.}~\bibnamefont {Haddad}},
  \bibinfo {author} {\bibfnamefont {I.~C.}\ \bibnamefont {Fulga}}, \bibinfo
  {author} {\bibfnamefont {F.~F.}\ \bibnamefont {Assaad}},\ and\ \bibinfo
  {author} {\bibfnamefont {J.}~\bibnamefont {van~den Brink}},\ }\bibfield
  {title} {\bibinfo {title} {Altermagnetic anomalous {H}all effect emerging
  from electronic correlations},\ }\href
  {https://doi.org/10.1103/PhysRevLett.133.086503} {\bibfield  {journal}
  {\bibinfo  {journal} {Phys. Rev. Lett.}\ }\textbf {\bibinfo {volume} {133}},\
  \bibinfo {pages} {086503} (\bibinfo {year} {2024})}\BibitemShut {NoStop}%
\bibitem [{\citenamefont {Mazin}(2022)}]{Mazin2022}%
  \BibitemOpen
  \bibfield  {author} {\bibinfo {author} {\bibfnamefont {I.~I.}\ \bibnamefont
  {Mazin}},\ }\bibfield  {title} {\bibinfo {title} {Notes on altermagnetism and
  superconductivity},\ }\href {https://arxiv.org/abs/2203.05000} {\bibfield
  {journal} {\bibinfo  {journal} {arXiv 2203.05000}\ } (\bibinfo {year}
  {2022})}\BibitemShut {NoStop}%
\bibitem [{\citenamefont {Zhu}\ \emph {et~al.}(2023)\citenamefont {Zhu},
  \citenamefont {Zhuang}, \citenamefont {Wu},\ and\ \citenamefont
  {Yan}}]{Zhu2023}%
  \BibitemOpen
  \bibfield  {author} {\bibinfo {author} {\bibfnamefont {D.}~\bibnamefont
  {Zhu}}, \bibinfo {author} {\bibfnamefont {Z.-Y.}\ \bibnamefont {Zhuang}},
  \bibinfo {author} {\bibfnamefont {Z.}~\bibnamefont {Wu}},\ and\ \bibinfo
  {author} {\bibfnamefont {Z.}~\bibnamefont {Yan}},\ }\bibfield  {title}
  {\bibinfo {title} {Topological superconductivity in two-dimensional
  altermagnetic metals},\ }\href {https://doi.org/10.1103/PhysRevB.108.184505}
  {\bibfield  {journal} {\bibinfo  {journal} {Phys. Rev. B}\ }\textbf {\bibinfo
  {volume} {108}},\ \bibinfo {pages} {184505} (\bibinfo {year}
  {2023})}\BibitemShut {NoStop}%
\bibitem [{\citenamefont {Banerjee}\ and\ \citenamefont
  {Scheurer}(2024)}]{Banerjee2024}%
  \BibitemOpen
  \bibfield  {author} {\bibinfo {author} {\bibfnamefont {S.}~\bibnamefont
  {Banerjee}}\ and\ \bibinfo {author} {\bibfnamefont {M.~S.}\ \bibnamefont
  {Scheurer}},\ }\bibfield  {title} {\bibinfo {title} {Altermagnetic
  superconducting diode effect},\ }\href
  {https://doi.org/10.1103/PhysRevB.110.024503} {\bibfield  {journal} {\bibinfo
   {journal} {Phys. Rev. B}\ }\textbf {\bibinfo {volume} {110}},\ \bibinfo
  {pages} {024503} (\bibinfo {year} {2024})}\BibitemShut {NoStop}%
\bibitem [{\citenamefont {Chakraborty}\ and\ \citenamefont
  {Black-Schaffer}(2024)}]{Chakraborty2024}%
  \BibitemOpen
  \bibfield  {author} {\bibinfo {author} {\bibfnamefont {D.}~\bibnamefont
  {Chakraborty}}\ and\ \bibinfo {author} {\bibfnamefont {A.~M.}\ \bibnamefont
  {Black-Schaffer}},\ }\bibfield  {title} {\bibinfo {title} {Zero-field
  finite-momentum and field-induced superconductivity in altermagnets},\ }\href
  {https://doi.org/10.1103/PhysRevB.110.L060508} {\bibfield  {journal}
  {\bibinfo  {journal} {Phys. Rev. B}\ }\textbf {\bibinfo {volume} {110}},\
  \bibinfo {pages} {L060508} (\bibinfo {year} {2024})}\BibitemShut {NoStop}%
\bibitem [{\citenamefont {Zhang}\ \emph {et~al.}(2024)\citenamefont {Zhang},
  \citenamefont {Hu},\ and\ \citenamefont {Neupert}}]{Zhang2024}%
  \BibitemOpen
  \bibfield  {author} {\bibinfo {author} {\bibfnamefont {S.-B.}\ \bibnamefont
  {Zhang}}, \bibinfo {author} {\bibfnamefont {L.-H.}\ \bibnamefont {Hu}},\ and\
  \bibinfo {author} {\bibfnamefont {T.}~\bibnamefont {Neupert}},\ }\bibfield
  {title} {\bibinfo {title} {Finite-momentum cooper pairing in proximitized
  altermagnets},\ }\href {https://doi.org/10.1038/s41467-024-45951-3}
  {\bibfield  {journal} {\bibinfo  {journal} {Nat. Commun.}\ }\textbf {\bibinfo
  {volume} {15}},\ \bibinfo {pages} {1801} (\bibinfo {year}
  {2024})}\BibitemShut {NoStop}%
\bibitem [{\citenamefont {Lee}\ \emph {et~al.}(2024{\natexlab{b}})\citenamefont
  {Lee}, \citenamefont {Qian},\ and\ \citenamefont {Yang}}]{Lee2024}%
  \BibitemOpen
  \bibfield  {author} {\bibinfo {author} {\bibfnamefont {S.~H.}\ \bibnamefont
  {Lee}}, \bibinfo {author} {\bibfnamefont {Y.}~\bibnamefont {Qian}},\ and\
  \bibinfo {author} {\bibfnamefont {B.-J.}\ \bibnamefont {Yang}},\ }\bibfield
  {title} {\bibinfo {title} {Fermi surface spin texture and topological
  superconductivity in spin-orbit free noncollinear antiferromagnets},\ }\href
  {https://doi.org/10.1103/PhysRevLett.132.196602} {\bibfield  {journal}
  {\bibinfo  {journal} {Phys. Rev. Lett.}\ }\textbf {\bibinfo {volume} {132}},\
  \bibinfo {pages} {196602} (\bibinfo {year} {2024}{\natexlab{b}})}\BibitemShut
  {NoStop}%
\bibitem [{\citenamefont {Zhou}\ \emph {et~al.}(2024)\citenamefont {Zhou},
  \citenamefont {Feng}, \citenamefont {Zhang}, \citenamefont
  {\ifmmode~\check{S}\else \v{S}\fi{}mejkal}, \citenamefont {Sinova},
  \citenamefont {Mokrousov},\ and\ \citenamefont {Yao}}]{Zhou2024}%
  \BibitemOpen
  \bibfield  {author} {\bibinfo {author} {\bibfnamefont {X.}~\bibnamefont
  {Zhou}}, \bibinfo {author} {\bibfnamefont {W.}~\bibnamefont {Feng}}, \bibinfo
  {author} {\bibfnamefont {R.-W.}\ \bibnamefont {Zhang}}, \bibinfo {author}
  {\bibfnamefont {L.}~\bibnamefont {\ifmmode~\check{S}\else \v{S}\fi{}mejkal}},
  \bibinfo {author} {\bibfnamefont {J.}~\bibnamefont {Sinova}}, \bibinfo
  {author} {\bibfnamefont {Y.}~\bibnamefont {Mokrousov}},\ and\ \bibinfo
  {author} {\bibfnamefont {Y.}~\bibnamefont {Yao}},\ }\bibfield  {title}
  {\bibinfo {title} {Crystal thermal transport in altermagnetic
  ${\mathrm{ruo}}_{2}$},\ }\href
  {https://doi.org/10.1103/PhysRevLett.132.056701} {\bibfield  {journal}
  {\bibinfo  {journal} {Phys. Rev. Lett.}\ }\textbf {\bibinfo {volume} {132}},\
  \bibinfo {pages} {056701} (\bibinfo {year} {2024})}\BibitemShut {NoStop}%
\bibitem [{\citenamefont {Yershov}\ \emph {et~al.}(2024)\citenamefont
  {Yershov}, \citenamefont {Kravchuk}, \citenamefont {Daghofer},\ and\
  \citenamefont {van~den Brink}}]{Yershov2024}%
  \BibitemOpen
  \bibfield  {author} {\bibinfo {author} {\bibfnamefont {K.~V.}\ \bibnamefont
  {Yershov}}, \bibinfo {author} {\bibfnamefont {V.~P.}\ \bibnamefont
  {Kravchuk}}, \bibinfo {author} {\bibfnamefont {M.}~\bibnamefont {Daghofer}},\
  and\ \bibinfo {author} {\bibfnamefont {J.}~\bibnamefont {van~den Brink}},\
  }\bibfield  {title} {\bibinfo {title} {Fluctuation-induced piezomagnetism in
  local moment altermagnets},\ }\href
  {https://doi.org/10.1103/PhysRevB.110.144421} {\bibfield  {journal} {\bibinfo
   {journal} {Phys. Rev. B}\ }\textbf {\bibinfo {volume} {110}},\ \bibinfo
  {pages} {144421} (\bibinfo {year} {2024})}\BibitemShut {NoStop}%
\bibitem [{\citenamefont {\ifmmode~\check{S}\else \v{S}\fi{}mejkal}\ \emph
  {et~al.}(2023)\citenamefont {\ifmmode~\check{S}\else \v{S}\fi{}mejkal},
  \citenamefont {Marmodoro}, \citenamefont {Ahn}, \citenamefont
  {Gonz\'alez-Hern\'andez}, \citenamefont {Turek}, \citenamefont {Mankovsky},
  \citenamefont {Ebert}, \citenamefont {D'Souza}, \citenamefont
  {\ifmmode~\check{S}\else \v{S}\fi{}ipr}, \citenamefont {Sinova},\ and\
  \citenamefont {Jungwirth}}]{Libor2023}%
  \BibitemOpen
  \bibfield  {author} {\bibinfo {author} {\bibfnamefont {L.}~\bibnamefont
  {\ifmmode~\check{S}\else \v{S}\fi{}mejkal}}, \bibinfo {author} {\bibfnamefont
  {A.}~\bibnamefont {Marmodoro}}, \bibinfo {author} {\bibfnamefont {K.-H.}\
  \bibnamefont {Ahn}}, \bibinfo {author} {\bibfnamefont {R.}~\bibnamefont
  {Gonz\'alez-Hern\'andez}}, \bibinfo {author} {\bibfnamefont {I.}~\bibnamefont
  {Turek}}, \bibinfo {author} {\bibfnamefont {S.}~\bibnamefont {Mankovsky}},
  \bibinfo {author} {\bibfnamefont {H.}~\bibnamefont {Ebert}}, \bibinfo
  {author} {\bibfnamefont {S.~W.}\ \bibnamefont {D'Souza}}, \bibinfo {author}
  {\bibfnamefont {O.~c.~v.}\ \bibnamefont {\ifmmode~\check{S}\else
  \v{S}\fi{}ipr}}, \bibinfo {author} {\bibfnamefont {J.}~\bibnamefont
  {Sinova}},\ and\ \bibinfo {author} {\bibfnamefont {T.~c.~v.}\ \bibnamefont
  {Jungwirth}},\ }\bibfield  {title} {\bibinfo {title} {Chiral magnons in
  altermagnetic {RuO}$_{2}$},\ }\href
  {https://doi.org/10.1103/PhysRevLett.131.256703} {\bibfield  {journal}
  {\bibinfo  {journal} {Phys. Rev. Lett.}\ }\textbf {\bibinfo {volume} {131}},\
  \bibinfo {pages} {256703} (\bibinfo {year} {2023})}\BibitemShut {NoStop}%
\bibitem [{\citenamefont {McClarty}\ \emph {et~al.}(2024)\citenamefont
  {McClarty}, \citenamefont {Gukasov},\ and\ \citenamefont
  {Rau}}]{McClarty2024}%
  \BibitemOpen
  \bibfield  {author} {\bibinfo {author} {\bibfnamefont {P.~A.}\ \bibnamefont
  {McClarty}}, \bibinfo {author} {\bibfnamefont {A.}~\bibnamefont {Gukasov}},\
  and\ \bibinfo {author} {\bibfnamefont {J.~G.}\ \bibnamefont {Rau}},\ }\href
  {https://arxiv.org/abs/2410.10771} {\bibinfo {title} {Observing
  altermagnetism using polarized neutrons}} (\bibinfo {year} {2024}),\ \Eprint
  {https://arxiv.org/abs/2410.10771} {arXiv:2410.10771 [cond-mat.str-el]}
  \BibitemShut {NoStop}%
\bibitem [{\citenamefont {Liu}\ \emph {et~al.}(2024)\citenamefont {Liu},
  \citenamefont {Ozeki}, \citenamefont {Asai}, \citenamefont {Itoh},\ and\
  \citenamefont {Masuda}}]{Liu2024}%
  \BibitemOpen
  \bibfield  {author} {\bibinfo {author} {\bibfnamefont {Z.}~\bibnamefont
  {Liu}}, \bibinfo {author} {\bibfnamefont {M.}~\bibnamefont {Ozeki}}, \bibinfo
  {author} {\bibfnamefont {S.}~\bibnamefont {Asai}}, \bibinfo {author}
  {\bibfnamefont {S.}~\bibnamefont {Itoh}},\ and\ \bibinfo {author}
  {\bibfnamefont {T.}~\bibnamefont {Masuda}},\ }\bibfield  {title} {\bibinfo
  {title} {Chiral split magnon in altermagnetic {MnTe}},\ }\href
  {https://doi.org/10.1103/PhysRevLett.133.156702} {\bibfield  {journal}
  {\bibinfo  {journal} {Phys. Rev. Lett.}\ }\textbf {\bibinfo {volume} {133}},\
  \bibinfo {pages} {156702} (\bibinfo {year} {2024})}\BibitemShut {NoStop}%
\bibitem [{\citenamefont {Morano}\ \emph {et~al.}(2024)\citenamefont {Morano},
  \citenamefont {Maesen}, \citenamefont {Nikitin}, \citenamefont {Lass},
  \citenamefont {Mazzone},\ and\ \citenamefont {Zaharko}}]{Morano2024}%
  \BibitemOpen
  \bibfield  {author} {\bibinfo {author} {\bibfnamefont {V.~C.}\ \bibnamefont
  {Morano}}, \bibinfo {author} {\bibfnamefont {Z.}~\bibnamefont {Maesen}},
  \bibinfo {author} {\bibfnamefont {S.~E.}\ \bibnamefont {Nikitin}}, \bibinfo
  {author} {\bibfnamefont {J.}~\bibnamefont {Lass}}, \bibinfo {author}
  {\bibfnamefont {D.~G.}\ \bibnamefont {Mazzone}},\ and\ \bibinfo {author}
  {\bibfnamefont {O.}~\bibnamefont {Zaharko}},\ }\href
  {https://arxiv.org/abs/2412.03545} {\bibinfo {title} {Absence of
  altermagnetic magnon band splitting in {MnF$_2$}}} (\bibinfo {year} {2024}),\
  \Eprint {https://arxiv.org/abs/2412.03545} {arXiv:2412.03545
  [cond-mat.str-el]} \BibitemShut {NoStop}%
\bibitem [{\citenamefont {Bandyopadhyay}\ \emph {et~al.}(2024)\citenamefont
  {Bandyopadhyay}, \citenamefont {Raj}, \citenamefont {Ghosez}, \citenamefont
  {Pujari},\ and\ \citenamefont {Bhowal}}]{Bandyopadhyay2024}%
  \BibitemOpen
  \bibfield  {author} {\bibinfo {author} {\bibfnamefont {S.}~\bibnamefont
  {Bandyopadhyay}}, \bibinfo {author} {\bibfnamefont {A.}~\bibnamefont {Raj}},
  \bibinfo {author} {\bibfnamefont {P.}~\bibnamefont {Ghosez}}, \bibinfo
  {author} {\bibfnamefont {S.}~\bibnamefont {Pujari}},\ and\ \bibinfo {author}
  {\bibfnamefont {S.}~\bibnamefont {Bhowal}},\ }\href
  {https://arxiv.org/abs/2412.04934} {\bibinfo {title} {Phonon-assisted control
  of magnonic and electronic band splitting}} (\bibinfo {year} {2024}),\
  \Eprint {https://arxiv.org/abs/2412.04934} {arXiv:2412.04934
  [cond-mat.mtrl-sci]} \BibitemShut {NoStop}%
\bibitem [{\citenamefont {Bhowal}\ and\ \citenamefont
  {Spaldin}(2024)}]{BhowalSpaldin2024}%
  \BibitemOpen
  \bibfield  {author} {\bibinfo {author} {\bibfnamefont {S.}~\bibnamefont
  {Bhowal}}\ and\ \bibinfo {author} {\bibfnamefont {N.~A.}\ \bibnamefont
  {Spaldin}},\ }\bibfield  {title} {\bibinfo {title} {Ferroically ordered
  magnetic octupoles in $d$-wave altermagnets},\ }\href
  {https://doi.org/10.1103/PhysRevX.14.011019} {\bibfield  {journal} {\bibinfo
  {journal} {Phys. Rev. X}\ }\textbf {\bibinfo {volume} {14}},\ \bibinfo
  {pages} {011019} (\bibinfo {year} {2024})}\BibitemShut {NoStop}%
\bibitem [{\citenamefont {Radaelli}(2024)}]{Radaelli2024}%
  \BibitemOpen
  \bibfield  {author} {\bibinfo {author} {\bibfnamefont {P.~G.}\ \bibnamefont
  {Radaelli}},\ }\bibfield  {title} {\bibinfo {title} {Tensorial approach to
  altermagnetism},\ }\href {https://doi.org/10.1103/PhysRevB.110.214428}
  {\bibfield  {journal} {\bibinfo  {journal} {Phys. Rev. B}\ }\textbf {\bibinfo
  {volume} {110}},\ \bibinfo {pages} {214428} (\bibinfo {year}
  {2024})}\BibitemShut {NoStop}%
\bibitem [{\citenamefont {Dale}\ \emph {et~al.}(2024)\citenamefont {Dale},
  \citenamefont {Ashour}, \citenamefont {Vila}, \citenamefont {Regmi},
  \citenamefont {Fox}, \citenamefont {Johnson}, \citenamefont {Fedorov},
  \citenamefont {Stibor}, \citenamefont {Ghimire},\ and\ \citenamefont
  {Griffin}}]{Dale2024}%
  \BibitemOpen
  \bibfield  {author} {\bibinfo {author} {\bibfnamefont {N.}~\bibnamefont
  {Dale}}, \bibinfo {author} {\bibfnamefont {O.~A.}\ \bibnamefont {Ashour}},
  \bibinfo {author} {\bibfnamefont {M.}~\bibnamefont {Vila}}, \bibinfo {author}
  {\bibfnamefont {R.~B.}\ \bibnamefont {Regmi}}, \bibinfo {author}
  {\bibfnamefont {J.}~\bibnamefont {Fox}}, \bibinfo {author} {\bibfnamefont
  {C.~W.}\ \bibnamefont {Johnson}}, \bibinfo {author} {\bibfnamefont
  {A.}~\bibnamefont {Fedorov}}, \bibinfo {author} {\bibfnamefont
  {A.}~\bibnamefont {Stibor}}, \bibinfo {author} {\bibfnamefont {N.~J.}\
  \bibnamefont {Ghimire}},\ and\ \bibinfo {author} {\bibfnamefont {S.~M.}\
  \bibnamefont {Griffin}},\ }\href {https://arxiv.org/abs/2411.18761} {\bibinfo
  {title} {Non-relativistic spin splitting above and below the fermi level in a
  $g$-wave altermagnet}} (\bibinfo {year} {2024}),\ \Eprint
  {https://arxiv.org/abs/2411.18761} {arXiv:2411.18761 [cond-mat.mtrl-sci]}
  \BibitemShut {NoStop}%
\bibitem [{\citenamefont {Yang}\ \emph {et~al.}(2025)\citenamefont {Yang},
  \citenamefont {Li}, \citenamefont {Yang}, \citenamefont {Li}, \citenamefont
  {Zheng}, \citenamefont {Zhu}, \citenamefont {Pan}, \citenamefont {Xu},
  \citenamefont {Cao}, \citenamefont {Zhao}, \citenamefont {Jana},
  \citenamefont {Zhang}, \citenamefont {Ye}, \citenamefont {Song},
  \citenamefont {Hu}, \citenamefont {Yang}, \citenamefont {Fujii},
  \citenamefont {Vobornik}, \citenamefont {Shi}, \citenamefont {Yuan},
  \citenamefont {Zhang}, \citenamefont {Xu},\ and\ \citenamefont
  {Liu}}]{Yang2025}%
  \BibitemOpen
  \bibfield  {author} {\bibinfo {author} {\bibfnamefont {G.}~\bibnamefont
  {Yang}}, \bibinfo {author} {\bibfnamefont {Z.}~\bibnamefont {Li}}, \bibinfo
  {author} {\bibfnamefont {S.}~\bibnamefont {Yang}}, \bibinfo {author}
  {\bibfnamefont {J.}~\bibnamefont {Li}}, \bibinfo {author} {\bibfnamefont
  {H.}~\bibnamefont {Zheng}}, \bibinfo {author} {\bibfnamefont
  {W.}~\bibnamefont {Zhu}}, \bibinfo {author} {\bibfnamefont {Z.}~\bibnamefont
  {Pan}}, \bibinfo {author} {\bibfnamefont {Y.}~\bibnamefont {Xu}}, \bibinfo
  {author} {\bibfnamefont {S.}~\bibnamefont {Cao}}, \bibinfo {author}
  {\bibfnamefont {W.}~\bibnamefont {Zhao}}, \bibinfo {author} {\bibfnamefont
  {A.}~\bibnamefont {Jana}}, \bibinfo {author} {\bibfnamefont {J.}~\bibnamefont
  {Zhang}}, \bibinfo {author} {\bibfnamefont {M.}~\bibnamefont {Ye}}, \bibinfo
  {author} {\bibfnamefont {Y.}~\bibnamefont {Song}}, \bibinfo {author}
  {\bibfnamefont {L.-H.}\ \bibnamefont {Hu}}, \bibinfo {author} {\bibfnamefont
  {L.}~\bibnamefont {Yang}}, \bibinfo {author} {\bibfnamefont {J.}~\bibnamefont
  {Fujii}}, \bibinfo {author} {\bibfnamefont {I.}~\bibnamefont {Vobornik}},
  \bibinfo {author} {\bibfnamefont {M.}~\bibnamefont {Shi}}, \bibinfo {author}
  {\bibfnamefont {H.}~\bibnamefont {Yuan}}, \bibinfo {author} {\bibfnamefont
  {Y.}~\bibnamefont {Zhang}}, \bibinfo {author} {\bibfnamefont
  {Y.}~\bibnamefont {Xu}},\ and\ \bibinfo {author} {\bibfnamefont
  {Y.}~\bibnamefont {Liu}},\ }\bibfield  {title} {\bibinfo {title}
  {Three-dimensional mapping of the altermagnetic spin splitting in crsb},\
  }\href {https://doi.org/10.1038/s41467-025-56647-7} {\bibfield  {journal}
  {\bibinfo  {journal} {Nature Communications}\ }\textbf {\bibinfo {volume}
  {16}},\ \bibinfo {pages} {1442} (\bibinfo {year} {2025})}\BibitemShut
  {NoStop}%
\bibitem [{\citenamefont {Šmejkal}(2024{\natexlab{a}})}]{Smejkal2024arxiv}%
  \BibitemOpen
  \bibfield  {author} {\bibinfo {author} {\bibfnamefont {L.}~\bibnamefont
  {Šmejkal}},\ }\href {https://arxiv.org/abs/2411.19928} {\bibinfo {title}
  {Altermagnetic multiferroics and altermagnetoelectric effect}} (\bibinfo
  {year} {2024}{\natexlab{a}}),\ \Eprint {https://arxiv.org/abs/2411.19928}
  {arXiv:2411.19928 [cond-mat.mtrl-sci]} \BibitemShut {NoStop}%
\bibitem [{\citenamefont {Gu}\ \emph {et~al.}(2025{\natexlab{a}})\citenamefont
  {Gu}, \citenamefont {Liu}, \citenamefont {Zhu}, \citenamefont {Yananose},
  \citenamefont {Chen}, \citenamefont {Hu}, \citenamefont {Stroppa},\ and\
  \citenamefont {Liu}}]{Gu2025}%
  \BibitemOpen
  \bibfield  {author} {\bibinfo {author} {\bibfnamefont {M.}~\bibnamefont
  {Gu}}, \bibinfo {author} {\bibfnamefont {Y.}~\bibnamefont {Liu}}, \bibinfo
  {author} {\bibfnamefont {H.}~\bibnamefont {Zhu}}, \bibinfo {author}
  {\bibfnamefont {K.}~\bibnamefont {Yananose}}, \bibinfo {author}
  {\bibfnamefont {X.}~\bibnamefont {Chen}}, \bibinfo {author} {\bibfnamefont
  {Y.}~\bibnamefont {Hu}}, \bibinfo {author} {\bibfnamefont {A.}~\bibnamefont
  {Stroppa}},\ and\ \bibinfo {author} {\bibfnamefont {Q.}~\bibnamefont {Liu}},\
  }\href {https://arxiv.org/abs/2411.14216} {\bibinfo {title} {Ferroelectric
  switchable altermagnetism}} (\bibinfo {year} {2025}{\natexlab{a}}),\ \Eprint
  {https://arxiv.org/abs/2411.14216} {arXiv:2411.14216 [cond-mat.mtrl-sci]}
  \BibitemShut {NoStop}%
\bibitem [{\citenamefont {Duan}\ \emph
  {et~al.}(2025{\natexlab{a}})\citenamefont {Duan}, \citenamefont {Zhang},
  \citenamefont {Zhu}, \citenamefont {Liu}, \citenamefont {Zhang},
  \citenamefont {Zutic},\ and\ \citenamefont {Zhou}}]{Duan2025}%
  \BibitemOpen
  \bibfield  {author} {\bibinfo {author} {\bibfnamefont {X.}~\bibnamefont
  {Duan}}, \bibinfo {author} {\bibfnamefont {J.}~\bibnamefont {Zhang}},
  \bibinfo {author} {\bibfnamefont {Z.}~\bibnamefont {Zhu}}, \bibinfo {author}
  {\bibfnamefont {Y.}~\bibnamefont {Liu}}, \bibinfo {author} {\bibfnamefont
  {Z.}~\bibnamefont {Zhang}}, \bibinfo {author} {\bibfnamefont
  {I.}~\bibnamefont {Zutic}},\ and\ \bibinfo {author} {\bibfnamefont
  {T.}~\bibnamefont {Zhou}},\ }\href {https://arxiv.org/abs/2410.06071}
  {\bibinfo {title} {Antiferroelectric altermagnets: Antiferroelectricity
  alters magnets}} (\bibinfo {year} {2025}{\natexlab{a}}),\ \Eprint
  {https://arxiv.org/abs/2410.06071} {arXiv:2410.06071 [cond-mat.mtrl-sci]}
  \BibitemShut {NoStop}%
\bibitem [{\citenamefont {Verbeek}\ \emph {et~al.}(2024)\citenamefont
  {Verbeek}, \citenamefont {Voderholzer}, \citenamefont {Sch\"aren},
  \citenamefont {Gachnang}, \citenamefont {Spaldin},\ and\ \citenamefont
  {Bhowal}}]{Verbeek2024}%
  \BibitemOpen
  \bibfield  {author} {\bibinfo {author} {\bibfnamefont {X.~H.}\ \bibnamefont
  {Verbeek}}, \bibinfo {author} {\bibfnamefont {D.}~\bibnamefont
  {Voderholzer}}, \bibinfo {author} {\bibfnamefont {S.}~\bibnamefont
  {Sch\"aren}}, \bibinfo {author} {\bibfnamefont {Y.}~\bibnamefont {Gachnang}},
  \bibinfo {author} {\bibfnamefont {N.~A.}\ \bibnamefont {Spaldin}},\ and\
  \bibinfo {author} {\bibfnamefont {S.}~\bibnamefont {Bhowal}},\ }\bibfield
  {title} {\bibinfo {title} {Nonrelativistic ferromagnetotriakontadipolar order
  and spin splitting in hematite},\ }\href
  {https://doi.org/10.1103/PhysRevResearch.6.043157} {\bibfield  {journal}
  {\bibinfo  {journal} {Phys. Rev. Res.}\ }\textbf {\bibinfo {volume} {6}},\
  \bibinfo {pages} {043157} (\bibinfo {year} {2024})}\BibitemShut {NoStop}%
\bibitem [{\citenamefont {Nag}\ \emph {et~al.}(2024)\citenamefont {Nag},
  \citenamefont {Das}, \citenamefont {Bhowal}, \citenamefont {Nishioka},
  \citenamefont {Bandyopadhyay}, \citenamefont {Sarker}, \citenamefont {Kumar},
  \citenamefont {Kuroda}, \citenamefont {Gopalan}, \citenamefont {Kimura},
  \citenamefont {Suresh},\ and\ \citenamefont {Alam}}]{Nag2024}%
  \BibitemOpen
  \bibfield  {author} {\bibinfo {author} {\bibfnamefont {J.}~\bibnamefont
  {Nag}}, \bibinfo {author} {\bibfnamefont {B.}~\bibnamefont {Das}}, \bibinfo
  {author} {\bibfnamefont {S.}~\bibnamefont {Bhowal}}, \bibinfo {author}
  {\bibfnamefont {Y.}~\bibnamefont {Nishioka}}, \bibinfo {author}
  {\bibfnamefont {B.}~\bibnamefont {Bandyopadhyay}}, \bibinfo {author}
  {\bibfnamefont {S.}~\bibnamefont {Sarker}}, \bibinfo {author} {\bibfnamefont
  {S.}~\bibnamefont {Kumar}}, \bibinfo {author} {\bibfnamefont
  {K.}~\bibnamefont {Kuroda}}, \bibinfo {author} {\bibfnamefont
  {V.}~\bibnamefont {Gopalan}}, \bibinfo {author} {\bibfnamefont
  {A.}~\bibnamefont {Kimura}}, \bibinfo {author} {\bibfnamefont {K.~G.}\
  \bibnamefont {Suresh}},\ and\ \bibinfo {author} {\bibfnamefont
  {A.}~\bibnamefont {Alam}},\ }\bibfield  {title} {\bibinfo {title} {Gdalsi: An
  antiferromagnetic topological weyl semimetal with nonrelativistic spin
  splitting},\ }\href {https://doi.org/10.1103/PhysRevB.110.224436} {\bibfield
  {journal} {\bibinfo  {journal} {Phys. Rev. B}\ }\textbf {\bibinfo {volume}
  {110}},\ \bibinfo {pages} {224436} (\bibinfo {year} {2024})}\BibitemShut
  {NoStop}%
\bibitem [{\citenamefont {Ghosez}\ and\ \citenamefont
  {Junquera}(2022)}]{ABO3_Ghosez_ferroelectricity}%
  \BibitemOpen
  \bibfield  {author} {\bibinfo {author} {\bibfnamefont {P.}~\bibnamefont
  {Ghosez}}\ and\ \bibinfo {author} {\bibfnamefont {J.}~\bibnamefont
  {Junquera}},\ }\bibfield  {title} {\bibinfo {title} {Modeling of
  ferroelectric oxide perovskites: From first to second principles},\ }\href
  {https://doi.org/https://doi.org/10.1146/annurev-conmatphys-040220-045528}
  {\bibfield  {journal} {\bibinfo  {journal} {Annual Review of Condensed Matter
  Physics}\ }\textbf {\bibinfo {volume} {13}},\ \bibinfo {pages} {325}
  (\bibinfo {year} {2022})}\BibitemShut {NoStop}%
\bibitem [{\citenamefont {Bousquet}\ and\ \citenamefont
  {Spaldin}(2011)}]{ABO3_Eric_magnetoelectric}%
  \BibitemOpen
  \bibfield  {author} {\bibinfo {author} {\bibfnamefont {E.}~\bibnamefont
  {Bousquet}}\ and\ \bibinfo {author} {\bibfnamefont {N.}~\bibnamefont
  {Spaldin}},\ }\bibfield  {title} {\bibinfo {title} {Induced magnetoelectric
  response in $pnma$ perovskites},\ }\href
  {https://doi.org/10.1103/PhysRevLett.107.197603} {\bibfield  {journal}
  {\bibinfo  {journal} {Phys. Rev. Lett.}\ }\textbf {\bibinfo {volume} {107}},\
  \bibinfo {pages} {197603} (\bibinfo {year} {2011})}\BibitemShut {NoStop}%
\bibitem [{\citenamefont {Kim}\ \emph {et~al.}(2022)\citenamefont {Kim},
  \citenamefont {McNally}, \citenamefont {Kim}, \citenamefont {Oudah},
  \citenamefont {Gibbs}, \citenamefont {Manuel}, \citenamefont {Green},
  \citenamefont {Sutarto}, \citenamefont {Takayama}, \citenamefont {Yaresko},
  \citenamefont {Wedig}, \citenamefont {Isobe}, \citenamefont {Kremer},
  \citenamefont {Bonn}, \citenamefont {Keimer},\ and\ \citenamefont
  {Takagi}}]{ABO3_Takagi_superconductivity}%
  \BibitemOpen
  \bibfield  {author} {\bibinfo {author} {\bibfnamefont {M.}~\bibnamefont
  {Kim}}, \bibinfo {author} {\bibfnamefont {G.~M.}\ \bibnamefont {McNally}},
  \bibinfo {author} {\bibfnamefont {H.-H.}\ \bibnamefont {Kim}}, \bibinfo
  {author} {\bibfnamefont {M.}~\bibnamefont {Oudah}}, \bibinfo {author}
  {\bibfnamefont {A.~S.}\ \bibnamefont {Gibbs}}, \bibinfo {author}
  {\bibfnamefont {P.}~\bibnamefont {Manuel}}, \bibinfo {author} {\bibfnamefont
  {R.~J.}\ \bibnamefont {Green}}, \bibinfo {author} {\bibfnamefont
  {R.}~\bibnamefont {Sutarto}}, \bibinfo {author} {\bibfnamefont
  {T.}~\bibnamefont {Takayama}}, \bibinfo {author} {\bibfnamefont
  {A.}~\bibnamefont {Yaresko}}, \bibinfo {author} {\bibfnamefont
  {U.}~\bibnamefont {Wedig}}, \bibinfo {author} {\bibfnamefont
  {M.}~\bibnamefont {Isobe}}, \bibinfo {author} {\bibfnamefont {R.~K.}\
  \bibnamefont {Kremer}}, \bibinfo {author} {\bibfnamefont {D.~A.}\
  \bibnamefont {Bonn}}, \bibinfo {author} {\bibfnamefont {B.}~\bibnamefont
  {Keimer}},\ and\ \bibinfo {author} {\bibfnamefont {H.}~\bibnamefont
  {Takagi}},\ }\bibfield  {title} {\bibinfo {title} {Superconductivity in
  {(Ba,K)SbO$_3$}},\ }\href {https://doi.org/10.1038/s41563-022-01203-7}
  {\bibfield  {journal} {\bibinfo  {journal} {Nature Materials}\ }\textbf
  {\bibinfo {volume} {21}},\ \bibinfo {pages} {627} (\bibinfo {year}
  {2022})}\BibitemShut {NoStop}%
\bibitem [{\citenamefont {Imada}\ \emph {et~al.}(1998)\citenamefont {Imada},
  \citenamefont {Fujimori},\ and\ \citenamefont {Tokura}}]{ABO3_Tokura_MIT}%
  \BibitemOpen
  \bibfield  {author} {\bibinfo {author} {\bibfnamefont {M.}~\bibnamefont
  {Imada}}, \bibinfo {author} {\bibfnamefont {A.}~\bibnamefont {Fujimori}},\
  and\ \bibinfo {author} {\bibfnamefont {Y.}~\bibnamefont {Tokura}},\
  }\bibfield  {title} {\bibinfo {title} {Metal-insulator transitions},\ }\href
  {https://doi.org/10.1103/RevModPhys.70.1039} {\bibfield  {journal} {\bibinfo
  {journal} {Rev. Mod. Phys.}\ }\textbf {\bibinfo {volume} {70}},\ \bibinfo
  {pages} {1039} (\bibinfo {year} {1998})}\BibitemShut {NoStop}%
\bibitem [{\citenamefont {Naka}\ \emph {et~al.}(2025)\citenamefont {Naka},
  \citenamefont {Motome},\ and\ \citenamefont {Seo}}]{naka_ABO3}%
  \BibitemOpen
  \bibfield  {author} {\bibinfo {author} {\bibfnamefont {M.}~\bibnamefont
  {Naka}}, \bibinfo {author} {\bibfnamefont {Y.}~\bibnamefont {Motome}},\ and\
  \bibinfo {author} {\bibfnamefont {H.}~\bibnamefont {Seo}},\ }\bibfield
  {title} {\bibinfo {title} {Altermagnetic perovskites},\ }\href
  {https://doi.org/10.1038/s44306-024-00066-9} {\bibfield  {journal} {\bibinfo
  {journal} {npj Spintronics}\ }\textbf {\bibinfo {volume} {3}},\ \bibinfo
  {pages} {1} (\bibinfo {year} {2025})}\BibitemShut {NoStop}%
\bibitem [{\citenamefont {Fan}\ \emph {et~al.}(2015)\citenamefont {Fan},
  \citenamefont {Sun},\ and\ \citenamefont {Wang}}]{ABO3_RSC_photovolatics}%
  \BibitemOpen
  \bibfield  {author} {\bibinfo {author} {\bibfnamefont {Z.}~\bibnamefont
  {Fan}}, \bibinfo {author} {\bibfnamefont {K.}~\bibnamefont {Sun}},\ and\
  \bibinfo {author} {\bibfnamefont {J.}~\bibnamefont {Wang}},\ }\bibfield
  {title} {\bibinfo {title} {Perovskites for photovoltaics: a combined review
  of organic–inorganic halide perovskites and ferroelectric oxide
  perovskites},\ }\href {https://doi.org/10.1039/C5TA04235F} {\bibfield
  {journal} {\bibinfo  {journal} {J. Mater. Chem. A}\ }\textbf {\bibinfo
  {volume} {3}},\ \bibinfo {pages} {18809} (\bibinfo {year}
  {2015})}\BibitemShut {NoStop}%
\bibitem [{\citenamefont {\ifmmode~\check{S}\else \v{S}\fi{}mejkal}\ \emph
  {et~al.}(2022{\natexlab{d}})\citenamefont {\ifmmode~\check{S}\else
  \v{S}\fi{}mejkal}, \citenamefont {Sinova},\ and\ \citenamefont
  {Jungwirth}}]{Smejkal2022}%
  \BibitemOpen
  \bibfield  {author} {\bibinfo {author} {\bibfnamefont {L.}~\bibnamefont
  {\ifmmode~\check{S}\else \v{S}\fi{}mejkal}}, \bibinfo {author} {\bibfnamefont
  {J.}~\bibnamefont {Sinova}},\ and\ \bibinfo {author} {\bibfnamefont
  {T.}~\bibnamefont {Jungwirth}},\ }\bibfield  {title} {\bibinfo {title}
  {Emerging research landscape of altermagnetism},\ }\href
  {https://doi.org/10.1103/PhysRevX.12.040501} {\bibfield  {journal} {\bibinfo
  {journal} {Phys. Rev. X}\ }\textbf {\bibinfo {volume} {12}},\ \bibinfo
  {pages} {040501} (\bibinfo {year} {2022}{\natexlab{d}})}\BibitemShut
  {NoStop}%
\bibitem [{\citenamefont {Rooj}\ \emph {et~al.}(2025)\citenamefont {Rooj},
  \citenamefont {Saxena},\ and\ \citenamefont {Ganguli}}]{Rooj_ganguly_Pnma}%
  \BibitemOpen
  \bibfield  {author} {\bibinfo {author} {\bibfnamefont {S.}~\bibnamefont
  {Rooj}}, \bibinfo {author} {\bibfnamefont {S.}~\bibnamefont {Saxena}},\ and\
  \bibinfo {author} {\bibfnamefont {N.}~\bibnamefont {Ganguli}},\ }\bibfield
  {title} {\bibinfo {title} {Altermagnetism in the orthorhombic $pnma$
  structure through group theory and dft calculations},\ }\href
  {https://doi.org/10.1103/PhysRevB.111.014434} {\bibfield  {journal} {\bibinfo
   {journal} {Phys. Rev. B}\ }\textbf {\bibinfo {volume} {111}},\ \bibinfo
  {pages} {014434} (\bibinfo {year} {2025})}\BibitemShut {NoStop}%
\bibitem [{\citenamefont {Baldini}\ \emph {et~al.}(2011)\citenamefont
  {Baldini}, \citenamefont {Struzhkin}, \citenamefont {Goncharov},
  \citenamefont {Postorino},\ and\ \citenamefont {Mao}}]{LMO_TM}%
  \BibitemOpen
  \bibfield  {author} {\bibinfo {author} {\bibfnamefont {M.}~\bibnamefont
  {Baldini}}, \bibinfo {author} {\bibfnamefont {V.~V.}\ \bibnamefont
  {Struzhkin}}, \bibinfo {author} {\bibfnamefont {A.~F.}\ \bibnamefont
  {Goncharov}}, \bibinfo {author} {\bibfnamefont {P.}~\bibnamefont
  {Postorino}},\ and\ \bibinfo {author} {\bibfnamefont {W.~L.}\ \bibnamefont
  {Mao}},\ }\bibfield  {title} {\bibinfo {title} {Persistence of jahn-teller
  distortion up to the insulator to metal transition in
  ${\mathrm{lamno}}_{3}$},\ }\href
  {https://doi.org/10.1103/PhysRevLett.106.066402} {\bibfield  {journal}
  {\bibinfo  {journal} {Phys. Rev. Lett.}\ }\textbf {\bibinfo {volume} {106}},\
  \bibinfo {pages} {066402} (\bibinfo {year} {2011})}\BibitemShut {NoStop}%
\bibitem [{\citenamefont {Ro\ifmmode~\acute{s}\else \'{s}\fi{}ciszewski}\ and\
  \citenamefont {Ole\ifmmode~\acute{s}\else \'{s}\fi{}}(2019)}]{LMnO_OO}%
  \BibitemOpen
  \bibfield  {author} {\bibinfo {author} {\bibfnamefont {K.}~\bibnamefont
  {Ro\ifmmode~\acute{s}\else \'{s}\fi{}ciszewski}}\ and\ \bibinfo {author}
  {\bibfnamefont {A.~M.}\ \bibnamefont {Ole\ifmmode~\acute{s}\else
  \'{s}\fi{}}},\ }\bibfield  {title} {\bibinfo {title} {Spin-orbital order in
  ${\mathrm{lamno}}_{3}$: $d\ensuremath{-}p$ model study},\ }\href
  {https://doi.org/10.1103/PhysRevB.99.155108} {\bibfield  {journal} {\bibinfo
  {journal} {Phys. Rev. B}\ }\textbf {\bibinfo {volume} {99}},\ \bibinfo
  {pages} {155108} (\bibinfo {year} {2019})}\BibitemShut {NoStop}%
\bibitem [{\citenamefont {Moussa}\ \emph
  {et~al.}(1996{\natexlab{a}})\citenamefont {Moussa}, \citenamefont {Hennion},
  \citenamefont {Rodriguez-Carvajal}, \citenamefont {Moudden}, \citenamefont
  {Pinsard},\ and\ \citenamefont {Revcolevschi}}]{LMO_TN}%
  \BibitemOpen
  \bibfield  {author} {\bibinfo {author} {\bibfnamefont {F.}~\bibnamefont
  {Moussa}}, \bibinfo {author} {\bibfnamefont {M.}~\bibnamefont {Hennion}},
  \bibinfo {author} {\bibfnamefont {J.}~\bibnamefont {Rodriguez-Carvajal}},
  \bibinfo {author} {\bibfnamefont {H.}~\bibnamefont {Moudden}}, \bibinfo
  {author} {\bibfnamefont {L.}~\bibnamefont {Pinsard}},\ and\ \bibinfo {author}
  {\bibfnamefont {A.}~\bibnamefont {Revcolevschi}},\ }\bibfield  {title}
  {\bibinfo {title} {Spin waves in the antiferromagnet perovskite
  lamn${\mathrm{o}}_{3}$: A neutron-scattering study},\ }\href
  {https://doi.org/10.1103/PhysRevB.54.15149} {\bibfield  {journal} {\bibinfo
  {journal} {Phys. Rev. B}\ }\textbf {\bibinfo {volume} {54}},\ \bibinfo
  {pages} {15149} (\bibinfo {year} {1996}{\natexlab{a}})}\BibitemShut {NoStop}%
\bibitem [{\citenamefont {Bl\"ochl}(1994)}]{Bloch1994}%
  \BibitemOpen
  \bibfield  {author} {\bibinfo {author} {\bibfnamefont {P.~E.}\ \bibnamefont
  {Bl\"ochl}},\ }\bibfield  {title} {\bibinfo {title} {Projector augmented-wave
  method},\ }\href {https://doi.org/10.1103/PhysRevB.50.17953} {\bibfield
  {journal} {\bibinfo  {journal} {Phys. Rev. B}\ }\textbf {\bibinfo {volume}
  {50}},\ \bibinfo {pages} {17953} (\bibinfo {year} {1994})}\BibitemShut
  {NoStop}%
\bibitem [{\citenamefont {Kresse}\ and\ \citenamefont
  {Joubert}(1999)}]{Kresse1999}%
  \BibitemOpen
  \bibfield  {author} {\bibinfo {author} {\bibfnamefont {G.}~\bibnamefont
  {Kresse}}\ and\ \bibinfo {author} {\bibfnamefont {D.}~\bibnamefont
  {Joubert}},\ }\bibfield  {title} {\bibinfo {title} {From ultrasoft
  pseudopotentials to the projector augmented-wave method},\ }\href
  {https://doi.org/10.1103/PhysRevB.59.1758} {\bibfield  {journal} {\bibinfo
  {journal} {Phys. Rev. B}\ }\textbf {\bibinfo {volume} {59}},\ \bibinfo
  {pages} {1758} (\bibinfo {year} {1999})}\BibitemShut {NoStop}%
\bibitem [{\citenamefont {Kresse}\ and\ \citenamefont
  {Hafner}(1993)}]{Kresse1993}%
  \BibitemOpen
  \bibfield  {author} {\bibinfo {author} {\bibfnamefont {G.}~\bibnamefont
  {Kresse}}\ and\ \bibinfo {author} {\bibfnamefont {J.}~\bibnamefont
  {Hafner}},\ }\bibfield  {title} {\bibinfo {title} {Ab initio molecular
  dynamics for liquid metals},\ }\href
  {https://doi.org/10.1103/PhysRevB.47.558} {\bibfield  {journal} {\bibinfo
  {journal} {Phys. Rev. B}\ }\textbf {\bibinfo {volume} {47}},\ \bibinfo
  {pages} {558} (\bibinfo {year} {1993})}\BibitemShut {NoStop}%
\bibitem [{\citenamefont {Kresse}\ and\ \citenamefont
  {Furthm\"uller}(1996)}]{Kresse1996}%
  \BibitemOpen
  \bibfield  {author} {\bibinfo {author} {\bibfnamefont {G.}~\bibnamefont
  {Kresse}}\ and\ \bibinfo {author} {\bibfnamefont {J.}~\bibnamefont
  {Furthm\"uller}},\ }\bibfield  {title} {\bibinfo {title} {Efficient iterative
  schemes for ab initio total-energy calculations using a plane-wave basis
  set},\ }\href {https://doi.org/10.1103/PhysRevB.54.11169} {\bibfield
  {journal} {\bibinfo  {journal} {Phys. Rev. B}\ }\textbf {\bibinfo {volume}
  {54}},\ \bibinfo {pages} {11169} (\bibinfo {year} {1996})}\BibitemShut
  {NoStop}%
\bibitem [{\citenamefont {Liechtenstein}\ \emph {et~al.}(1995)\citenamefont
  {Liechtenstein}, \citenamefont {Anisimov},\ and\ \citenamefont
  {Zaanen}}]{Liechtenstein_U}%
  \BibitemOpen
  \bibfield  {author} {\bibinfo {author} {\bibfnamefont {A.~I.}\ \bibnamefont
  {Liechtenstein}}, \bibinfo {author} {\bibfnamefont {V.~I.}\ \bibnamefont
  {Anisimov}},\ and\ \bibinfo {author} {\bibfnamefont {J.}~\bibnamefont
  {Zaanen}},\ }\bibfield  {title} {\bibinfo {title} {Density-functional theory
  and strong interactions: Orbital ordering in mott-hubbard insulators},\
  }\href {https://doi.org/10.1103/PhysRevB.52.R5467} {\bibfield  {journal}
  {\bibinfo  {journal} {Phys. Rev. B}\ }\textbf {\bibinfo {volume} {52}},\
  \bibinfo {pages} {R5467} (\bibinfo {year} {1995})}\BibitemShut {NoStop}%
\bibitem [{\citenamefont {Schmitt}\ \emph {et~al.}(2020)\citenamefont
  {Schmitt}, \citenamefont {Zhang}, \citenamefont {Mercy},\ and\ \citenamefont
  {Ghosez}}]{LaMnO3_marcus}%
  \BibitemOpen
  \bibfield  {author} {\bibinfo {author} {\bibfnamefont {M.~M.}\ \bibnamefont
  {Schmitt}}, \bibinfo {author} {\bibfnamefont {Y.}~\bibnamefont {Zhang}},
  \bibinfo {author} {\bibfnamefont {A.}~\bibnamefont {Mercy}},\ and\ \bibinfo
  {author} {\bibfnamefont {P.}~\bibnamefont {Ghosez}},\ }\bibfield  {title}
  {\bibinfo {title} {Electron-lattice interplay in ${\mathrm{lamno}}_{3}$ from
  canonical jahn-teller distortion notations},\ }\href
  {https://doi.org/10.1103/PhysRevB.101.214304} {\bibfield  {journal} {\bibinfo
   {journal} {Phys. Rev. B}\ }\textbf {\bibinfo {volume} {101}},\ \bibinfo
  {pages} {214304} (\bibinfo {year} {2020})}\BibitemShut {NoStop}%
\bibitem [{\citenamefont {Stokes}\ \emph {et~al.}({\natexlab{a}})\citenamefont
  {Stokes}, \citenamefont {Hatch},\ and\ \citenamefont
  {Campbell}}]{Isodistort}%
  \BibitemOpen
  \bibfield  {author} {\bibinfo {author} {\bibfnamefont {H.~T.}\ \bibnamefont
  {Stokes}}, \bibinfo {author} {\bibfnamefont {D.~M.}\ \bibnamefont {Hatch}},\
  and\ \bibinfo {author} {\bibfnamefont {B.~J.}\ \bibnamefont {Campbell}},\
  }\href@noop {} {\bibinfo {title} {{ISODISTORT, ISOTROPY} software suite}},\
  \bibinfo {howpublished} {\url{https://iso.byu.edu}}
  ({\natexlab{a}})\BibitemShut {NoStop}%
\bibitem [{\citenamefont {Campbell}\ \emph {et~al.}(2006)\citenamefont
  {Campbell}, \citenamefont {Stokes}, \citenamefont {Tanner},\ and\
  \citenamefont {Hatch}}]{Isodistort1}%
  \BibitemOpen
  \bibfield  {author} {\bibinfo {author} {\bibfnamefont {B.~J.}\ \bibnamefont
  {Campbell}}, \bibinfo {author} {\bibfnamefont {H.~T.}\ \bibnamefont
  {Stokes}}, \bibinfo {author} {\bibfnamefont {D.~E.}\ \bibnamefont {Tanner}},\
  and\ \bibinfo {author} {\bibfnamefont {D.~M.}\ \bibnamefont {Hatch}},\
  }\bibfield  {title} {\bibinfo {title} {{{\it ISODISPLACE}: a web-based tool
  for exploring structural distortions}},\ }\href@noop {} {\bibfield  {journal}
  {\bibinfo  {journal} {Journal of Applied Crystallography}\ }\textbf {\bibinfo
  {volume} {39}},\ \bibinfo {pages} {607} (\bibinfo {year} {2006})}\BibitemShut
  {NoStop}%
\bibitem [{\citenamefont {Spaldin}\ \emph {et~al.}(2013)\citenamefont
  {Spaldin}, \citenamefont {Fechner}, \citenamefont {Bousquet}, \citenamefont
  {Balatsky},\ and\ \citenamefont {Nordstr\"om}}]{Spaldin2013}%
  \BibitemOpen
  \bibfield  {author} {\bibinfo {author} {\bibfnamefont {N.~A.}\ \bibnamefont
  {Spaldin}}, \bibinfo {author} {\bibfnamefont {M.}~\bibnamefont {Fechner}},
  \bibinfo {author} {\bibfnamefont {E.}~\bibnamefont {Bousquet}}, \bibinfo
  {author} {\bibfnamefont {A.}~\bibnamefont {Balatsky}},\ and\ \bibinfo
  {author} {\bibfnamefont {L.}~\bibnamefont {Nordstr\"om}},\ }\bibfield
  {title} {\bibinfo {title} {Monopole-based formalism for the diagonal
  magnetoelectric response},\ }\href
  {https://doi.org/10.1103/PhysRevB.88.094429} {\bibfield  {journal} {\bibinfo
  {journal} {Phys. Rev. B}\ }\textbf {\bibinfo {volume} {88}},\ \bibinfo
  {pages} {094429} (\bibinfo {year} {2013})}\BibitemShut {NoStop}%
\bibitem [{\citenamefont {Cricchio}(2010)}]{Cricchio2010}%
  \BibitemOpen
  \bibfield  {author} {\bibinfo {author} {\bibfnamefont {F.}~\bibnamefont
  {Cricchio}},\ }\emph {\bibinfo {title} {Multipoles in {Correlated} {Electron}
  {Materials}}},\ \href
  {http://urn.kb.se/resolve?urn=urn:nbn:se:uu:diva-132068} {Ph.D. thesis},\
  \bibinfo  {school} {Uppsala University} (\bibinfo {year} {2010})\BibitemShut
  {NoStop}%
\bibitem [{\citenamefont {Urru}\ and\ \citenamefont
  {Spaldin}(2022)}]{Urru2022}%
  \BibitemOpen
  \bibfield  {author} {\bibinfo {author} {\bibfnamefont {A.}~\bibnamefont
  {Urru}}\ and\ \bibinfo {author} {\bibfnamefont {N.~A.}\ \bibnamefont
  {Spaldin}},\ }\bibfield  {title} {\bibinfo {title} {Magnetic octupole tensor
  decomposition and second-order magnetoelectric effect},\ }\href
  {https://doi.org/https://doi.org/10.1016/j.aop.2022.168964} {\bibfield
  {journal} {\bibinfo  {journal} {Ann. Phys.}\ }\textbf {\bibinfo {volume}
  {447}},\ \bibinfo {pages} {168964} (\bibinfo {year} {2022})}\BibitemShut
  {NoStop}%
\bibitem [{\citenamefont {Moussa}\ \emph
  {et~al.}(1996{\natexlab{b}})\citenamefont {Moussa}, \citenamefont {Hennion},
  \citenamefont {Rodriguez-Carvajal}, \citenamefont {Moudden}, \citenamefont
  {Pinsard},\ and\ \citenamefont {Revcolevschi}}]{LMO_dist1}%
  \BibitemOpen
  \bibfield  {author} {\bibinfo {author} {\bibfnamefont {F.}~\bibnamefont
  {Moussa}}, \bibinfo {author} {\bibfnamefont {M.}~\bibnamefont {Hennion}},
  \bibinfo {author} {\bibfnamefont {J.}~\bibnamefont {Rodriguez-Carvajal}},
  \bibinfo {author} {\bibfnamefont {H.}~\bibnamefont {Moudden}}, \bibinfo
  {author} {\bibfnamefont {L.}~\bibnamefont {Pinsard}},\ and\ \bibinfo {author}
  {\bibfnamefont {A.}~\bibnamefont {Revcolevschi}},\ }\bibfield  {title}
  {\bibinfo {title} {Spin waves in the antiferromagnet perovskite
  lamn${\mathrm{o}}_{3}$: A neutron-scattering study},\ }\href
  {https://doi.org/10.1103/PhysRevB.54.15149} {\bibfield  {journal} {\bibinfo
  {journal} {Phys. Rev. B}\ }\textbf {\bibinfo {volume} {54}},\ \bibinfo
  {pages} {15149} (\bibinfo {year} {1996}{\natexlab{b}})}\BibitemShut {NoStop}%
\bibitem [{\citenamefont {Elemans}\ \emph {et~al.}(1971)\citenamefont
  {Elemans}, \citenamefont {{Van Laar}}, \citenamefont {{Van Der Veen}},\ and\
  \citenamefont {Loopstra}}]{LMO_dist2}%
  \BibitemOpen
  \bibfield  {author} {\bibinfo {author} {\bibfnamefont {J.~B.}\ \bibnamefont
  {Elemans}}, \bibinfo {author} {\bibfnamefont {B.}~\bibnamefont {{Van Laar}}},
  \bibinfo {author} {\bibfnamefont {K.}~\bibnamefont {{Van Der Veen}}},\ and\
  \bibinfo {author} {\bibfnamefont {B.}~\bibnamefont {Loopstra}},\ }\bibfield
  {title} {\bibinfo {title} {The crystallographic and magnetic structures of
  {La$_{1-x}$Ba$_x$Mn$_{1-x}$Me$_x$O$_3$} ({M}e = {M}n or {T}i)},\ }\href
  {https://doi.org/https://doi.org/10.1016/0022-4596(71)90034-X} {\bibfield
  {journal} {\bibinfo  {journal} {Journal of Solid State Chemistry}\ }\textbf
  {\bibinfo {volume} {3}},\ \bibinfo {pages} {238} (\bibinfo {year}
  {1971})}\BibitemShut {NoStop}%
\bibitem [{\citenamefont {Matar}(2003)}]{CMO_matar}%
  \BibitemOpen
  \bibfield  {author} {\bibinfo {author} {\bibfnamefont {S.~F.}\ \bibnamefont
  {Matar}},\ }\bibfield  {title} {\bibinfo {title} {Ab initio investigations in
  magnetic oxides},\ }\href
  {https://doi.org/https://doi.org/10.1016/j.progsolidstchem.2004.01.001}
  {\bibfield  {journal} {\bibinfo  {journal} {Progress in Solid State
  Chemistry}\ }\textbf {\bibinfo {volume} {31}},\ \bibinfo {pages} {239}
  (\bibinfo {year} {2003})}\BibitemShut {NoStop}%
\bibitem [{\citenamefont {Geller}\ and\ \citenamefont {Wood}(1956)}]{LFO_wood}%
  \BibitemOpen
  \bibfield  {author} {\bibinfo {author} {\bibfnamefont {S.}~\bibnamefont
  {Geller}}\ and\ \bibinfo {author} {\bibfnamefont {E.~A.}\ \bibnamefont
  {Wood}},\ }\bibfield  {title} {\bibinfo {title} {{Crystallographic studies of
  perovskite-like compounds. I. Rare earth orthoferrites and YFeO${\sb 3}$,
  YCrO${\sb 3}$, YAlO${\sb 3}$}},\ }\href
  {https://doi.org/10.1107/S0365110X56001571} {\bibfield  {journal} {\bibinfo
  {journal} {Acta Crystallographica}\ }\textbf {\bibinfo {volume} {9}},\
  \bibinfo {pages} {563} (\bibinfo {year} {1956})}\BibitemShut {NoStop}%
\bibitem [{\citenamefont {Sławiński}\ \emph {et~al.}(2005)\citenamefont
  {Sławiński}, \citenamefont {Przeniosło}, \citenamefont {Sosnowska},\ and\
  \citenamefont {Suard}}]{NdFeO_2005}%
  \BibitemOpen
  \bibfield  {author} {\bibinfo {author} {\bibfnamefont {W.}~\bibnamefont
  {Sławiński}}, \bibinfo {author} {\bibfnamefont {R.}~\bibnamefont
  {Przeniosło}}, \bibinfo {author} {\bibfnamefont {I.}~\bibnamefont
  {Sosnowska}},\ and\ \bibinfo {author} {\bibfnamefont {E.}~\bibnamefont
  {Suard}},\ }\bibfield  {title} {\bibinfo {title} {Spin reorientation and
  structural changes in ndfeo3},\ }\href
  {https://doi.org/10.1088/0953-8984/17/29/002} {\bibfield  {journal} {\bibinfo
   {journal} {Journal of Physics: Condensed Matter}\ }\textbf {\bibinfo
  {volume} {17}},\ \bibinfo {pages} {4605} (\bibinfo {year}
  {2005})}\BibitemShut {NoStop}%
\bibitem [{\citenamefont {Bandyopadhyay}\ and\ \citenamefont
  {Ghosez}(2024{\natexlab{a}})}]{Subhadeep_BiNiO3}%
  \BibitemOpen
  \bibfield  {author} {\bibinfo {author} {\bibfnamefont {S.}~\bibnamefont
  {Bandyopadhyay}}\ and\ \bibinfo {author} {\bibfnamefont {P.}~\bibnamefont
  {Ghosez}},\ }\bibfield  {title} {\bibinfo {title} {Latent electronic
  (anti-)ferroelectricity in {BiNiO$_3$}},\ }\href
  {https://doi.org/10.1103/PhysRevLett.133.146801} {\bibfield  {journal}
  {\bibinfo  {journal} {Phys. Rev. Lett.}\ }\textbf {\bibinfo {volume} {133}},\
  \bibinfo {pages} {146801} (\bibinfo {year} {2024}{\natexlab{a}})}\BibitemShut
  {NoStop}%
\bibitem [{\citenamefont {Komarek}\ \emph {et~al.}(2008)\citenamefont
  {Komarek}, \citenamefont {Streltsov}, \citenamefont {Isobe}, \citenamefont
  {M\"oller}, \citenamefont {Hoelzel}, \citenamefont {Senyshyn}, \citenamefont
  {Trots}, \citenamefont {Fern\'andez-D\'{\i}az}, \citenamefont {Hansen},
  \citenamefont {Gotou}, \citenamefont {Yagi}, \citenamefont {Ueda},
  \citenamefont {Anisimov}, \citenamefont {Gr\"uninger}, \citenamefont
  {Khomskii},\ and\ \citenamefont {Braden}}]{CaCrO3_Khomskii}%
  \BibitemOpen
  \bibfield  {author} {\bibinfo {author} {\bibfnamefont {A.~C.}\ \bibnamefont
  {Komarek}}, \bibinfo {author} {\bibfnamefont {S.~V.}\ \bibnamefont
  {Streltsov}}, \bibinfo {author} {\bibfnamefont {M.}~\bibnamefont {Isobe}},
  \bibinfo {author} {\bibfnamefont {T.}~\bibnamefont {M\"oller}}, \bibinfo
  {author} {\bibfnamefont {M.}~\bibnamefont {Hoelzel}}, \bibinfo {author}
  {\bibfnamefont {A.}~\bibnamefont {Senyshyn}}, \bibinfo {author}
  {\bibfnamefont {D.}~\bibnamefont {Trots}}, \bibinfo {author} {\bibfnamefont
  {M.~T.}\ \bibnamefont {Fern\'andez-D\'{\i}az}}, \bibinfo {author}
  {\bibfnamefont {T.}~\bibnamefont {Hansen}}, \bibinfo {author} {\bibfnamefont
  {H.}~\bibnamefont {Gotou}}, \bibinfo {author} {\bibfnamefont
  {T.}~\bibnamefont {Yagi}}, \bibinfo {author} {\bibfnamefont {Y.}~\bibnamefont
  {Ueda}}, \bibinfo {author} {\bibfnamefont {V.~I.}\ \bibnamefont {Anisimov}},
  \bibinfo {author} {\bibfnamefont {M.}~\bibnamefont {Gr\"uninger}}, \bibinfo
  {author} {\bibfnamefont {D.~I.}\ \bibnamefont {Khomskii}},\ and\ \bibinfo
  {author} {\bibfnamefont {M.}~\bibnamefont {Braden}},\ }\bibfield  {title}
  {\bibinfo {title} {${\mathrm{cacro}}_{3}$: An anomalous antiferromagnetic
  metallic oxide},\ }\href {https://doi.org/10.1103/PhysRevLett.101.167204}
  {\bibfield  {journal} {\bibinfo  {journal} {Phys. Rev. Lett.}\ }\textbf
  {\bibinfo {volume} {101}},\ \bibinfo {pages} {167204} (\bibinfo {year}
  {2008})}\BibitemShut {NoStop}%
\bibitem [{\citenamefont {Bordet}\ \emph {et~al.}(1993)\citenamefont {Bordet},
  \citenamefont {Chaillout}, \citenamefont {Marezio}, \citenamefont {Huang},
  \citenamefont {Santoro}, \citenamefont {Cheong}, \citenamefont {Takagi},
  \citenamefont {Oglesby},\ and\ \citenamefont {Batlogg}}]{AVO_1}%
  \BibitemOpen
  \bibfield  {author} {\bibinfo {author} {\bibfnamefont {P.}~\bibnamefont
  {Bordet}}, \bibinfo {author} {\bibfnamefont {C.}~\bibnamefont {Chaillout}},
  \bibinfo {author} {\bibfnamefont {M.}~\bibnamefont {Marezio}}, \bibinfo
  {author} {\bibfnamefont {Q.}~\bibnamefont {Huang}}, \bibinfo {author}
  {\bibfnamefont {A.}~\bibnamefont {Santoro}}, \bibinfo {author} {\bibfnamefont
  {S.-W.}\ \bibnamefont {Cheong}}, \bibinfo {author} {\bibfnamefont
  {H.}~\bibnamefont {Takagi}}, \bibinfo {author} {\bibfnamefont
  {C.}~\bibnamefont {Oglesby}},\ and\ \bibinfo {author} {\bibfnamefont
  {B.}~\bibnamefont {Batlogg}},\ }\bibfield  {title} {\bibinfo {title}
  {Structural aspects of the crystallographic-magnetic transition in lavo3
  around 140 k},\ }\href
  {https://doi.org/https://doi.org/10.1006/jssc.1993.1285} {\bibfield
  {journal} {\bibinfo  {journal} {Journal of Solid State Chemistry}\ }\textbf
  {\bibinfo {volume} {106}},\ \bibinfo {pages} {253} (\bibinfo {year}
  {1993})}\BibitemShut {NoStop}%
\bibitem [{\citenamefont {Miyasaka}\ \emph {et~al.}(2003)\citenamefont
  {Miyasaka}, \citenamefont {Okimoto}, \citenamefont {Iwama},\ and\
  \citenamefont {Tokura}}]{AVO_2}%
  \BibitemOpen
  \bibfield  {author} {\bibinfo {author} {\bibfnamefont {S.}~\bibnamefont
  {Miyasaka}}, \bibinfo {author} {\bibfnamefont {Y.}~\bibnamefont {Okimoto}},
  \bibinfo {author} {\bibfnamefont {M.}~\bibnamefont {Iwama}},\ and\ \bibinfo
  {author} {\bibfnamefont {Y.}~\bibnamefont {Tokura}},\ }\bibfield  {title}
  {\bibinfo {title} {Spin-orbital phase diagram of perovskite-type
  $r{\mathrm{vo}}_{3}$ $(r=\mathrm{rare}$-earth ion or y)},\ }\href
  {https://doi.org/10.1103/PhysRevB.68.100406} {\bibfield  {journal} {\bibinfo
  {journal} {Phys. Rev. B}\ }\textbf {\bibinfo {volume} {68}},\ \bibinfo
  {pages} {100406} (\bibinfo {year} {2003})}\BibitemShut {NoStop}%
\bibitem [{\citenamefont {Varignon}\ \emph
  {et~al.}(2017{\natexlab{a}})\citenamefont {Varignon}, \citenamefont
  {Grisolia}, \citenamefont {Preziosi}, \citenamefont {Ghosez},\ and\
  \citenamefont {Bibes}}]{Varignon_titanates}%
  \BibitemOpen
  \bibfield  {author} {\bibinfo {author} {\bibfnamefont {J.}~\bibnamefont
  {Varignon}}, \bibinfo {author} {\bibfnamefont {M.~N.}\ \bibnamefont
  {Grisolia}}, \bibinfo {author} {\bibfnamefont {D.}~\bibnamefont {Preziosi}},
  \bibinfo {author} {\bibfnamefont {P.}~\bibnamefont {Ghosez}},\ and\ \bibinfo
  {author} {\bibfnamefont {M.}~\bibnamefont {Bibes}},\ }\bibfield  {title}
  {\bibinfo {title} {Origin of the orbital and spin ordering in rare-earth
  titanates},\ }\href {https://doi.org/10.1103/PhysRevB.96.235106} {\bibfield
  {journal} {\bibinfo  {journal} {Phys. Rev. B}\ }\textbf {\bibinfo {volume}
  {96}},\ \bibinfo {pages} {235106} (\bibinfo {year}
  {2017}{\natexlab{a}})}\BibitemShut {NoStop}%
\bibitem [{\citenamefont {Bandyopadhyay}\ and\ \citenamefont
  {Ghosez}(2024{\natexlab{b}})}]{TMO_subhadeep}%
  \BibitemOpen
  \bibfield  {author} {\bibinfo {author} {\bibfnamefont {S.}~\bibnamefont
  {Bandyopadhyay}}\ and\ \bibinfo {author} {\bibfnamefont {P.}~\bibnamefont
  {Ghosez}},\ }\href {https://arxiv.org/abs/2407.21406} {\bibinfo {title}
  {Structurally triggered orbital and charge orderings in {TlMnO$_3$} and
  related compounds}} (\bibinfo {year} {2024}{\natexlab{b}}),\ \Eprint
  {https://arxiv.org/abs/2407.21406} {arXiv:2407.21406 [cond-mat.str-el]}
  \BibitemShut {NoStop}%
\bibitem [{\citenamefont {Goldschmidt}(1926)}]{Goldschmidt_1926}%
  \BibitemOpen
  \bibfield  {author} {\bibinfo {author} {\bibfnamefont {V.~M.}\ \bibnamefont
  {Goldschmidt}},\ }\bibfield  {title} {\bibinfo {title} {Die gesetze der
  krystallochemie},\ }\href {https://doi.org/10.1007/BF01507527} {\bibfield
  {journal} {\bibinfo  {journal} {Naturwissenschaften}\ }\textbf {\bibinfo
  {volume} {14}},\ \bibinfo {pages} {477} (\bibinfo {year} {1926})}\BibitemShut
  {NoStop}%
\bibitem [{\citenamefont {Benedek}\ and\ \citenamefont
  {Fennie}(2013)}]{Benedek_2013}%
  \BibitemOpen
  \bibfield  {author} {\bibinfo {author} {\bibfnamefont {N.~A.}\ \bibnamefont
  {Benedek}}\ and\ \bibinfo {author} {\bibfnamefont {C.~J.}\ \bibnamefont
  {Fennie}},\ }\bibfield  {title} {\bibinfo {title} {Why are there so few
  perovskite ferroelectrics?},\ }\href {https://doi.org/10.1021/jp402046t}
  {\bibfield  {journal} {\bibinfo  {journal} {The Journal of Physical Chemistry
  C}\ }\textbf {\bibinfo {volume} {117}},\ \bibinfo {pages} {13339} (\bibinfo
  {year} {2013})}\BibitemShut {NoStop}%
\bibitem [{\citenamefont {Mercy}\ \emph {et~al.}(2017)\citenamefont {Mercy},
  \citenamefont {Bieder}, \citenamefont {{\'I}{\~{n}}iguez},\ and\
  \citenamefont {Ghosez}}]{YNO_Mercy}%
  \BibitemOpen
  \bibfield  {author} {\bibinfo {author} {\bibfnamefont {A.}~\bibnamefont
  {Mercy}}, \bibinfo {author} {\bibfnamefont {J.}~\bibnamefont {Bieder}},
  \bibinfo {author} {\bibfnamefont {J.}~\bibnamefont {{\'I}{\~{n}}iguez}},\
  and\ \bibinfo {author} {\bibfnamefont {P.}~\bibnamefont {Ghosez}},\
  }\bibfield  {title} {\bibinfo {title} {Structurally triggered metal-insulator
  transition in rare-earth nickelates},\ }\href
  {https://doi.org/10.1038/s41467-017-01811-x} {\bibfield  {journal} {\bibinfo
  {journal} {Nature Communications}\ }\textbf {\bibinfo {volume} {8}},\
  \bibinfo {pages} {1677} (\bibinfo {year} {2017})}\BibitemShut {NoStop}%
\bibitem [{\citenamefont {Varignon}\ \emph {et~al.}(2015)\citenamefont
  {Varignon}, \citenamefont {Bristowe}, \citenamefont {Bousquet},\ and\
  \citenamefont {Ghosez}}]{varignon_AVO}%
  \BibitemOpen
  \bibfield  {author} {\bibinfo {author} {\bibfnamefont {J.}~\bibnamefont
  {Varignon}}, \bibinfo {author} {\bibfnamefont {N.~C.}\ \bibnamefont
  {Bristowe}}, \bibinfo {author} {\bibfnamefont {E.}~\bibnamefont {Bousquet}},\
  and\ \bibinfo {author} {\bibfnamefont {P.}~\bibnamefont {Ghosez}},\
  }\bibfield  {title} {\bibinfo {title} {Coupling and electrical control of
  structural, orbital and magnetic orders in perovskites},\ }\href
  {https://doi.org/10.1038/srep15364} {\bibfield  {journal} {\bibinfo
  {journal} {Scientific Reports}\ }\textbf {\bibinfo {volume} {5}},\ \bibinfo
  {pages} {15364} (\bibinfo {year} {2015})}\BibitemShut {NoStop}%
\bibitem [{\citenamefont {Pecharsky}\ and\ \citenamefont {Jr.}(2024)}]{Radii}%
  \BibitemOpen
  \bibfield  {author} {\bibinfo {author} {\bibfnamefont {K.~A.}\ \bibnamefont
  {Pecharsky}, \bibfnamefont {Vitalij K.~Gschneidner}}\ and\ \bibinfo {author}
  {\bibnamefont {Jr.}},\ }\bibfield  {title} {\bibinfo {title} {rare-earth
  element},\ }\href {https://www.britannica.com/science/rare-earth-element}
  {\bibfield  {journal} {\bibinfo  {journal} {Encyclopedia Britannica, 26 Sep.
  2024}\ } (\bibinfo {year} {2024})}\BibitemShut {NoStop}%
\bibitem [{\citenamefont {Stokes}\ \emph {et~al.}({\natexlab{b}})\citenamefont
  {Stokes}, \citenamefont {Hatch},\ and\ \citenamefont
  {Campbell}}]{Invariants}%
  \BibitemOpen
  \bibfield  {author} {\bibinfo {author} {\bibfnamefont {H.~T.}\ \bibnamefont
  {Stokes}}, \bibinfo {author} {\bibfnamefont {D.~M.}\ \bibnamefont {Hatch}},\
  and\ \bibinfo {author} {\bibfnamefont {B.~J.}\ \bibnamefont {Campbell}},\
  }\href {https://iso.byu.edu} {\bibfield  {journal} {\bibinfo  {journal}
  {INVARIANTS, ISOTROPY Software Suite}\ } ({\natexlab{b}})}\BibitemShut
  {NoStop}%
\bibitem [{\citenamefont {Hatch}\ and\ \citenamefont
  {Stokes}(2003)}]{Invariants2}%
  \BibitemOpen
  \bibfield  {author} {\bibinfo {author} {\bibfnamefont {D.~M.}\ \bibnamefont
  {Hatch}}\ and\ \bibinfo {author} {\bibfnamefont {H.~T.}\ \bibnamefont
  {Stokes}},\ }\bibfield  {title} {\bibinfo {title} {{{\it INVARIANTS}: program
  for obtaining a list of invariant polynomials of the order-parameter
  components associated with irreducible representations of a space group}},\
  }\href {https://doi.org/10.1107/S0021889803005946} {\bibfield  {journal}
  {\bibinfo  {journal} {Journal of Applied Crystallography}\ }\textbf {\bibinfo
  {volume} {36}},\ \bibinfo {pages} {951} (\bibinfo {year} {2003})}\BibitemShut
  {NoStop}%
\bibitem [{\citenamefont {Vukmirović}\ \emph {et~al.}(2023)\citenamefont
  {Vukmirović}, \citenamefont {Joksović}, \citenamefont {Piper},
  \citenamefont {Nesterović}, \citenamefont {Novaković}, \citenamefont
  {Rakić}, \citenamefont {Milanović},\ and\ \citenamefont
  {Srdić}}]{LaMnO3_ceramics_2023}%
  \BibitemOpen
  \bibfield  {author} {\bibinfo {author} {\bibfnamefont {J.}~\bibnamefont
  {Vukmirović}}, \bibinfo {author} {\bibfnamefont {S.}~\bibnamefont
  {Joksović}}, \bibinfo {author} {\bibfnamefont {D.}~\bibnamefont {Piper}},
  \bibinfo {author} {\bibfnamefont {A.}~\bibnamefont {Nesterović}}, \bibinfo
  {author} {\bibfnamefont {M.}~\bibnamefont {Novaković}}, \bibinfo {author}
  {\bibfnamefont {S.}~\bibnamefont {Rakić}}, \bibinfo {author} {\bibfnamefont
  {M.}~\bibnamefont {Milanović}},\ and\ \bibinfo {author} {\bibfnamefont
  {V.~V.}\ \bibnamefont {Srdić}},\ }\bibfield  {title} {\bibinfo {title}
  {Epitaxial growth of lamno3 thin films on different single crystal substrates
  by polymer assisted deposition},\ }\href
  {https://doi.org/https://doi.org/10.1016/j.ceramint.2022.09.207} {\bibfield
  {journal} {\bibinfo  {journal} {Ceramics International}\ }\textbf {\bibinfo
  {volume} {49}},\ \bibinfo {pages} {2366} (\bibinfo {year}
  {2023})}\BibitemShut {NoStop}%
\bibitem [{\citenamefont {Aruta}\ \emph {et~al.}(2006)\citenamefont {Aruta},
  \citenamefont {Angeloni}, \citenamefont {Balestrino}, \citenamefont {Boggio},
  \citenamefont {Medaglia}, \citenamefont {Tebano}, \citenamefont {Davidson},
  \citenamefont {Baldini}, \citenamefont {Di~Castro}, \citenamefont
  {Postorino}, \citenamefont {Dore}, \citenamefont {Sidorenko}, \citenamefont
  {Allodi},\ and\ \citenamefont {De~Renzi}}]{LaMnO3_on_LaAlO3}%
  \BibitemOpen
  \bibfield  {author} {\bibinfo {author} {\bibfnamefont {C.}~\bibnamefont
  {Aruta}}, \bibinfo {author} {\bibfnamefont {M.}~\bibnamefont {Angeloni}},
  \bibinfo {author} {\bibfnamefont {G.}~\bibnamefont {Balestrino}}, \bibinfo
  {author} {\bibfnamefont {N.~G.}\ \bibnamefont {Boggio}}, \bibinfo {author}
  {\bibfnamefont {P.~G.}\ \bibnamefont {Medaglia}}, \bibinfo {author}
  {\bibfnamefont {A.}~\bibnamefont {Tebano}}, \bibinfo {author} {\bibfnamefont
  {B.}~\bibnamefont {Davidson}}, \bibinfo {author} {\bibfnamefont
  {M.}~\bibnamefont {Baldini}}, \bibinfo {author} {\bibfnamefont
  {D.}~\bibnamefont {Di~Castro}}, \bibinfo {author} {\bibfnamefont
  {P.}~\bibnamefont {Postorino}}, \bibinfo {author} {\bibfnamefont
  {P.}~\bibnamefont {Dore}}, \bibinfo {author} {\bibfnamefont {A.}~\bibnamefont
  {Sidorenko}}, \bibinfo {author} {\bibfnamefont {G.}~\bibnamefont {Allodi}},\
  and\ \bibinfo {author} {\bibfnamefont {R.}~\bibnamefont {De~Renzi}},\
  }\bibfield  {title} {\bibinfo {title} {Preparation and characterization of
  lamno3 thin films grown by pulsed laser deposition},\ }\href
  {https://doi.org/10.1063/1.2217983} {\bibfield  {journal} {\bibinfo
  {journal} {Journal of Applied Physics}\ }\textbf {\bibinfo {volume} {100}},\
  \bibinfo {pages} {023910} (\bibinfo {year} {2006})}\BibitemShut {NoStop}%
\bibitem [{\citenamefont {Marton}\ \emph {et~al.}(2010)\citenamefont {Marton},
  \citenamefont {{A. Seo}}, \citenamefont {Egami},\ and\ \citenamefont
  {Lee}}]{LaMnO3_marton}%
  \BibitemOpen
  \bibfield  {author} {\bibinfo {author} {\bibfnamefont {Z.}~\bibnamefont
  {Marton}}, \bibinfo {author} {\bibfnamefont {S.~S.}\ \bibnamefont {{A.
  Seo}}}, \bibinfo {author} {\bibfnamefont {T.}~\bibnamefont {Egami}},\ and\
  \bibinfo {author} {\bibfnamefont {H.~N.}\ \bibnamefont {Lee}},\ }\bibfield
  {title} {\bibinfo {title} {Growth control of stoichiometry in lamno3
  epitaxial thin films by pulsed laser deposition},\ }\href
  {https://doi.org/https://doi.org/10.1016/j.jcrysgro.2010.07.013} {\bibfield
  {journal} {\bibinfo  {journal} {Journal of Crystal Growth}\ }\textbf
  {\bibinfo {volume} {312}},\ \bibinfo {pages} {2923} (\bibinfo {year}
  {2010})}\BibitemShut {NoStop}%
\bibitem [{\citenamefont {Roqueta}\ \emph {et~al.}(2015)\citenamefont
  {Roqueta}, \citenamefont {Pomar}, \citenamefont {Balcells}, \citenamefont
  {Frontera}, \citenamefont {Valencia}, \citenamefont {Abrudan}, \citenamefont
  {Bozzo}, \citenamefont {Konstantinović}, \citenamefont {Santiso},\ and\
  \citenamefont {Martínez}}]{LaMnO3_Roqueta}%
  \BibitemOpen
  \bibfield  {author} {\bibinfo {author} {\bibfnamefont {J.}~\bibnamefont
  {Roqueta}}, \bibinfo {author} {\bibfnamefont {A.}~\bibnamefont {Pomar}},
  \bibinfo {author} {\bibfnamefont {L.}~\bibnamefont {Balcells}}, \bibinfo
  {author} {\bibfnamefont {C.}~\bibnamefont {Frontera}}, \bibinfo {author}
  {\bibfnamefont {S.}~\bibnamefont {Valencia}}, \bibinfo {author}
  {\bibfnamefont {R.}~\bibnamefont {Abrudan}}, \bibinfo {author} {\bibfnamefont
  {B.}~\bibnamefont {Bozzo}}, \bibinfo {author} {\bibfnamefont
  {Z.}~\bibnamefont {Konstantinović}}, \bibinfo {author} {\bibfnamefont
  {J.}~\bibnamefont {Santiso}},\ and\ \bibinfo {author} {\bibfnamefont
  {B.}~\bibnamefont {Martínez}},\ }\bibfield  {title} {\bibinfo {title}
  {Strain-engineered ferromagnetism in lamno3 thin films},\ }\href
  {https://doi.org/10.1021/acs.cgd.5b00884} {\bibfield  {journal} {\bibinfo
  {journal} {Crystal Growth \& Design}\ }\textbf {\bibinfo {volume} {15}},\
  \bibinfo {pages} {5332} (\bibinfo {year} {2015})}\BibitemShut {NoStop}%
\bibitem [{\citenamefont {Rondinelli}\ and\ \citenamefont
  {Fennie}(2012)}]{HIF_Pnma_adv_mat_2012}%
  \BibitemOpen
  \bibfield  {author} {\bibinfo {author} {\bibfnamefont {J.~M.}\ \bibnamefont
  {Rondinelli}}\ and\ \bibinfo {author} {\bibfnamefont {C.~J.}\ \bibnamefont
  {Fennie}},\ }\bibfield  {title} {\bibinfo {title} {Octahedral
  rotation-induced ferroelectricity in cation ordered perovskites},\ }\href
  {https://doi.org/https://doi.org/10.1002/adma.201104674} {\bibfield
  {journal} {\bibinfo  {journal} {Advanced Materials}\ }\textbf {\bibinfo
  {volume} {24}},\ \bibinfo {pages} {1961} (\bibinfo {year}
  {2012})}\BibitemShut {NoStop}%
\bibitem [{\citenamefont {Bousquet}\ \emph {et~al.}(2008)\citenamefont
  {Bousquet}, \citenamefont {Dawber}, \citenamefont {Stucki}, \citenamefont
  {Lichtensteiger}, \citenamefont {Hermet}, \citenamefont {Gariglio},
  \citenamefont {Triscone},\ and\ \citenamefont {Ghosez}}]{Eric_2008}%
  \BibitemOpen
  \bibfield  {author} {\bibinfo {author} {\bibfnamefont {E.}~\bibnamefont
  {Bousquet}}, \bibinfo {author} {\bibfnamefont {M.}~\bibnamefont {Dawber}},
  \bibinfo {author} {\bibfnamefont {N.}~\bibnamefont {Stucki}}, \bibinfo
  {author} {\bibfnamefont {C.}~\bibnamefont {Lichtensteiger}}, \bibinfo
  {author} {\bibfnamefont {P.}~\bibnamefont {Hermet}}, \bibinfo {author}
  {\bibfnamefont {S.}~\bibnamefont {Gariglio}}, \bibinfo {author}
  {\bibfnamefont {J.-M.}\ \bibnamefont {Triscone}},\ and\ \bibinfo {author}
  {\bibfnamefont {P.}~\bibnamefont {Ghosez}},\ }\bibfield  {title} {\bibinfo
  {title} {Improper ferroelectricity in perovskite oxide artificial
  superlattices},\ }\href {https://doi.org/10.1038/nature06817} {\bibfield
  {journal} {\bibinfo  {journal} {Nature}\ }\textbf {\bibinfo {volume} {452}},\
  \bibinfo {pages} {732} (\bibinfo {year} {2008})}\BibitemShut {NoStop}%
\bibitem [{\citenamefont {Benedek}\ \emph {et~al.}(2012)\citenamefont
  {Benedek}, \citenamefont {Mulder},\ and\ \citenamefont
  {Fennie}}]{BENEDEK2012}%
  \BibitemOpen
  \bibfield  {author} {\bibinfo {author} {\bibfnamefont {N.~A.}\ \bibnamefont
  {Benedek}}, \bibinfo {author} {\bibfnamefont {A.~T.}\ \bibnamefont
  {Mulder}},\ and\ \bibinfo {author} {\bibfnamefont {C.~J.}\ \bibnamefont
  {Fennie}},\ }\bibfield  {title} {\bibinfo {title} {Polar octahedral
  rotations: A path to new multifunctional materials},\ }\href
  {https://doi.org/https://doi.org/10.1016/j.jssc.2012.04.012} {\bibfield
  {journal} {\bibinfo  {journal} {Journal of Solid State Chemistry}\ }\textbf
  {\bibinfo {volume} {195}},\ \bibinfo {pages} {11} (\bibinfo {year} {2012})},\
  \bibinfo {note} {polar Inorganic Materials: Design Strategies and Functional
  Properties}\BibitemShut {NoStop}%
\bibitem [{\citenamefont {Varbaro}\ \emph {et~al.}(2024)\citenamefont
  {Varbaro}, \citenamefont {Mundet}, \citenamefont {Bandyopadhyay},
  \citenamefont {Domínguez}, \citenamefont {Fowlie}, \citenamefont {Korosec},
  \citenamefont {Hsu}, \citenamefont {Alexander}, \citenamefont {Ghosez},\ and\
  \citenamefont {Triscone}}]{Subhdeep_APL}%
  \BibitemOpen
  \bibfield  {author} {\bibinfo {author} {\bibfnamefont {L.}~\bibnamefont
  {Varbaro}}, \bibinfo {author} {\bibfnamefont {B.}~\bibnamefont {Mundet}},
  \bibinfo {author} {\bibfnamefont {S.}~\bibnamefont {Bandyopadhyay}}, \bibinfo
  {author} {\bibfnamefont {C.}~\bibnamefont {Domínguez}}, \bibinfo {author}
  {\bibfnamefont {J.}~\bibnamefont {Fowlie}}, \bibinfo {author} {\bibfnamefont
  {L.}~\bibnamefont {Korosec}}, \bibinfo {author} {\bibfnamefont {C.-Y.}\
  \bibnamefont {Hsu}}, \bibinfo {author} {\bibfnamefont {D.~T.~L.}\
  \bibnamefont {Alexander}}, \bibinfo {author} {\bibfnamefont {P.}~\bibnamefont
  {Ghosez}},\ and\ \bibinfo {author} {\bibfnamefont {J.-M.}\ \bibnamefont
  {Triscone}},\ }\bibfield  {title} {\bibinfo {title} {Structural study of
  nickelate based heterostructures},\ }\href
  {https://doi.org/10.1063/5.0184306} {\bibfield  {journal} {\bibinfo
  {journal} {APL Materials}\ }\textbf {\bibinfo {volume} {12}},\ \bibinfo
  {pages} {031104} (\bibinfo {year} {2024})}\BibitemShut {NoStop}%
\bibitem [{Note1()}]{Note1}%
  \BibitemOpen
  \bibinfo {note} {E and E'-type AFM describe an in-plane up-up-down-down spin
  order that couples ferroically and antiferroically along the out of plane
  direction respectively.}\BibitemShut {Stop}%
\bibitem [{\citenamefont {Sergienko}\ \emph {et~al.}(2006)\citenamefont
  {Sergienko}, \citenamefont {\ifmmode~\mbox{\c{S}}\else \c{S}\fi{}en},\ and\
  \citenamefont {Dagotto}}]{Dagotto_PRL}%
  \BibitemOpen
  \bibfield  {author} {\bibinfo {author} {\bibfnamefont {I.~A.}\ \bibnamefont
  {Sergienko}}, \bibinfo {author} {\bibfnamefont {C.}~\bibnamefont
  {\ifmmode~\mbox{\c{S}}\else \c{S}\fi{}en}},\ and\ \bibinfo {author}
  {\bibfnamefont {E.}~\bibnamefont {Dagotto}},\ }\bibfield  {title} {\bibinfo
  {title} {Ferroelectricity in the magnetic $e$-phase of orthorhombic
  perovskites},\ }\href {https://doi.org/10.1103/PhysRevLett.97.227204}
  {\bibfield  {journal} {\bibinfo  {journal} {Phys. Rev. Lett.}\ }\textbf
  {\bibinfo {volume} {97}},\ \bibinfo {pages} {227204} (\bibinfo {year}
  {2006})}\BibitemShut {NoStop}%
\bibitem [{\citenamefont {Picozzi}\ \emph {et~al.}(2007)\citenamefont
  {Picozzi}, \citenamefont {Yamauchi}, \citenamefont {Sanyal}, \citenamefont
  {Sergienko},\ and\ \citenamefont {Dagotto}}]{Silvia_HMnO3}%
  \BibitemOpen
  \bibfield  {author} {\bibinfo {author} {\bibfnamefont {S.}~\bibnamefont
  {Picozzi}}, \bibinfo {author} {\bibfnamefont {K.}~\bibnamefont {Yamauchi}},
  \bibinfo {author} {\bibfnamefont {B.}~\bibnamefont {Sanyal}}, \bibinfo
  {author} {\bibfnamefont {I.~A.}\ \bibnamefont {Sergienko}},\ and\ \bibinfo
  {author} {\bibfnamefont {E.}~\bibnamefont {Dagotto}},\ }\bibfield  {title}
  {\bibinfo {title} {Dual nature of improper ferroelectricity in a
  magnetoelectric multiferroic},\ }\href
  {https://doi.org/10.1103/PhysRevLett.99.227201} {\bibfield  {journal}
  {\bibinfo  {journal} {Phys. Rev. Lett.}\ }\textbf {\bibinfo {volume} {99}},\
  \bibinfo {pages} {227201} (\bibinfo {year} {2007})}\BibitemShut {NoStop}%
\bibitem [{\citenamefont {Varignon}\ \emph
  {et~al.}(2017{\natexlab{b}})\citenamefont {Varignon}, \citenamefont
  {Grisolia}, \citenamefont {{\'I}{\~n}iguez}, \citenamefont
  {Barth{\'e}l{\'e}my},\ and\ \citenamefont {Bibes}}]{Varignon_2017}%
  \BibitemOpen
  \bibfield  {author} {\bibinfo {author} {\bibfnamefont {J.}~\bibnamefont
  {Varignon}}, \bibinfo {author} {\bibfnamefont {M.~N.}\ \bibnamefont
  {Grisolia}}, \bibinfo {author} {\bibfnamefont {J.}~\bibnamefont
  {{\'I}{\~n}iguez}}, \bibinfo {author} {\bibfnamefont {A.}~\bibnamefont
  {Barth{\'e}l{\'e}my}},\ and\ \bibinfo {author} {\bibfnamefont
  {M.}~\bibnamefont {Bibes}},\ }\bibfield  {title} {\bibinfo {title} {Complete
  phase diagram of rare-earth nickelates from first-principles},\ }\href
  {https://doi.org/10.1038/s41535-017-0024-9} {\bibfield  {journal} {\bibinfo
  {journal} {npj Quantum Materials}\ }\textbf {\bibinfo {volume} {2}},\
  \bibinfo {pages} {21} (\bibinfo {year} {2017}{\natexlab{b}})}\BibitemShut
  {NoStop}%
\bibitem [{\citenamefont {Meier}\ \emph {et~al.}(2025)\citenamefont {Meier},
  \citenamefont {Carta}, \citenamefont {Ederer},\ and\ \citenamefont
  {Cano}}]{Cano_claude_2025}%
  \BibitemOpen
  \bibfield  {author} {\bibinfo {author} {\bibfnamefont {Q.~N.}\ \bibnamefont
  {Meier}}, \bibinfo {author} {\bibfnamefont {A.}~\bibnamefont {Carta}},
  \bibinfo {author} {\bibfnamefont {C.}~\bibnamefont {Ederer}},\ and\ \bibinfo
  {author} {\bibfnamefont {A.}~\bibnamefont {Cano}},\ }\href
  {https://arxiv.org/abs/2502.01515} {\bibinfo {title} {(anti-)altermagnetism
  from orbital ordering in the ruddlesden-popper chromates
  sr$_{n+1}$cr$_n$o$_{3n+1}$}} (\bibinfo {year} {2025}),\ \Eprint
  {https://arxiv.org/abs/2502.01515} {arXiv:2502.01515 [cond-mat.mtrl-sci]}
  \BibitemShut {NoStop}%
\bibitem [{\citenamefont {Šmejkal}(2024{\natexlab{b}})}]{Libor_2025_BaCuF4}%
  \BibitemOpen
  \bibfield  {author} {\bibinfo {author} {\bibfnamefont {L.}~\bibnamefont
  {Šmejkal}},\ }\href {https://arxiv.org/abs/2411.19928} {\bibinfo {title}
  {Altermagnetic multiferroics and altermagnetoelectric effect}} (\bibinfo
  {year} {2024}{\natexlab{b}}),\ \Eprint {https://arxiv.org/abs/2411.19928}
  {arXiv:2411.19928 [cond-mat.mtrl-sci]} \BibitemShut {NoStop}%
\bibitem [{\citenamefont {Gu}\ \emph {et~al.}(2025{\natexlab{b}})\citenamefont
  {Gu}, \citenamefont {Liu}, \citenamefont {Zhu}, \citenamefont {Yananose},
  \citenamefont {Chen}, \citenamefont {Hu}, \citenamefont {Stroppa},\ and\
  \citenamefont {Liu}}]{Stroppa_2025}%
  \BibitemOpen
  \bibfield  {author} {\bibinfo {author} {\bibfnamefont {M.}~\bibnamefont
  {Gu}}, \bibinfo {author} {\bibfnamefont {Y.}~\bibnamefont {Liu}}, \bibinfo
  {author} {\bibfnamefont {H.}~\bibnamefont {Zhu}}, \bibinfo {author}
  {\bibfnamefont {K.}~\bibnamefont {Yananose}}, \bibinfo {author}
  {\bibfnamefont {X.}~\bibnamefont {Chen}}, \bibinfo {author} {\bibfnamefont
  {Y.}~\bibnamefont {Hu}}, \bibinfo {author} {\bibfnamefont {A.}~\bibnamefont
  {Stroppa}},\ and\ \bibinfo {author} {\bibfnamefont {Q.}~\bibnamefont {Liu}},\
  }\href {https://arxiv.org/abs/2411.14216} {\bibinfo {title} {Ferroelectric
  switchable altermagnetism}} (\bibinfo {year} {2025}{\natexlab{b}}),\ \Eprint
  {https://arxiv.org/abs/2411.14216} {arXiv:2411.14216 [cond-mat.mtrl-sci]}
  \BibitemShut {NoStop}%
\bibitem [{\citenamefont {Duan}\ \emph
  {et~al.}(2025{\natexlab{b}})\citenamefont {Duan}, \citenamefont {Zhang},
  \citenamefont {Zhu}, \citenamefont {Liu}, \citenamefont {Zhang},
  \citenamefont {Zutic},\ and\ \citenamefont {Zhou}}]{Duan_2025}%
  \BibitemOpen
  \bibfield  {author} {\bibinfo {author} {\bibfnamefont {X.}~\bibnamefont
  {Duan}}, \bibinfo {author} {\bibfnamefont {J.}~\bibnamefont {Zhang}},
  \bibinfo {author} {\bibfnamefont {Z.}~\bibnamefont {Zhu}}, \bibinfo {author}
  {\bibfnamefont {Y.}~\bibnamefont {Liu}}, \bibinfo {author} {\bibfnamefont
  {Z.}~\bibnamefont {Zhang}}, \bibinfo {author} {\bibfnamefont
  {I.}~\bibnamefont {Zutic}},\ and\ \bibinfo {author} {\bibfnamefont
  {T.}~\bibnamefont {Zhou}},\ }\href {https://arxiv.org/abs/2410.06071}
  {\bibinfo {title} {Antiferroelectric altermagnets: Antiferroelectricity
  alters magnets}} (\bibinfo {year} {2025}{\natexlab{b}}),\ \Eprint
  {https://arxiv.org/abs/2410.06071} {arXiv:2410.06071 [cond-mat.mtrl-sci]}
  \BibitemShut {NoStop}%
\bibitem [{\citenamefont {Dong}\ \emph {et~al.}(2025)\citenamefont {Dong},
  \citenamefont {Wu}, \citenamefont {Zhu}, \citenamefont {Zheng}, \citenamefont
  {Li},\ and\ \citenamefont {Zhang}}]{Dong2025}%
  \BibitemOpen
  \bibfield  {author} {\bibinfo {author} {\bibfnamefont {J.}~\bibnamefont
  {Dong}}, \bibinfo {author} {\bibfnamefont {K.}~\bibnamefont {Wu}}, \bibinfo
  {author} {\bibfnamefont {M.}~\bibnamefont {Zhu}}, \bibinfo {author}
  {\bibfnamefont {F.}~\bibnamefont {Zheng}}, \bibinfo {author} {\bibfnamefont
  {X.}~\bibnamefont {Li}},\ and\ \bibinfo {author} {\bibfnamefont
  {J.}~\bibnamefont {Zhang}},\ }\href {https://arxiv.org/abs/2501.02914}
  {\bibinfo {title} {Nonrelativistic spin-splitting multiferroic
  antiferromagnet and compensated ferrimagnet with zero net magnetization}}
  (\bibinfo {year} {2025}),\ \Eprint {https://arxiv.org/abs/2501.02914}
  {arXiv:2501.02914 [cond-mat.mtrl-sci]} \BibitemShut {NoStop}%
\end{thebibliography}%
\end{document}